\documentclass[11pt,a4paper]{article}
\usepackage[T1]{fontenc}
\usepackage[utf8]{inputenc}
\usepackage{tgtermes}
\usepackage[margin=24mm]{geometry}
\usepackage[expansion=false]{microtype}
\usepackage{amsmath,amssymb,amsthm}
\usepackage{booktabs,tabularx,array}
\usepackage{graphicx,xcolor,tikz,pgfplots}
\usetikzlibrary{arrows.meta,positioning,calc,fit,shapes.geometric}
\usepgfplotslibrary{groupplots,fillbetween}
\pgfplotsset{compat=1.18}
\usepackage[font=small,labelfont=bf]{caption}
\usepackage{enumitem}
\usepackage[superscript]{cite}
\usepackage{pdflscape,float}
\definecolor{citationcolor}{HTML}{9B4A12}
\usepackage[colorlinks=true,linkcolor=black,citecolor=citationcolor,urlcolor=blue!50!black]{hyperref}
\definecolor{blue}{HTML}{0072B2}
\definecolor{orange}{HTML}{D55E00}
\definecolor{green}{HTML}{009E73}
\definecolor{purple}{HTML}{9D5B91}
\definecolor{lightgray}{HTML}{F4F5F6}
\renewcommand{\arraystretch}{1.16}
\setlist{nosep,leftmargin=*}
\newenvironment{eqvars}{\par\smallskip\noindent{\small\textbf{Variables and model roles:}}\nopagebreak\begin{itemize}[label=\textbullet,leftmargin=1.45em,topsep=3pt,itemsep=2pt,parsep=0pt]\small}{\end{itemize}}
\newcolumntype{Y}{>{\raggedright\arraybackslash}X}
\newcommand{\HT}{5-HT\textsubscript{2A}}
\newcommand{\sig}{\operatorname{sigmoid}}
\newcommand{\deff}{D_{\mathrm{eff}}}
\newcommand{\figinput}[3]{\begin{figure}[H]\centering#1\caption{#2}\label{fig:#3}\end{figure}}
\usepackage{fontawesome5}
\usepackage{tgheros}
\usetikzlibrary{patterns,shapes.misc,decorations.pathmorphing,backgrounds,shadows.blur,shadows,3d,decorations.text,decorations.pathreplacing,decorations.markings,fadings}
\definecolor{dopaminecolor}{RGB}{230,159,0}
\definecolor{serotonincolor}{RGB}{86,180,233}
\definecolor{consciouscolor}{RGB}{0,158,115}
\definecolor{psychosiscolor}{RGB}{175,155,0}
\definecolor{deathcolor}{RGB}{88,88,88}
\definecolor{sleepcolor}{RGB}{0,114,178}
\definecolor{wakecolor}{RGB}{213,94,0}
\definecolor{flowcolor}{RGB}{204,121,167}
\definecolor{seizurecolor}{RGB}{230,159,0}
\definecolor{mygray10}{gray}{0.10}
\definecolor{mygray40}{gray}{0.40}
\definecolor{mygray70}{gray}{0.70}
\newsavebox{\artworkbox}
\newcommand{\inputartwork}[1]{\begingroup
\normalfont\normalsize
\setlength{\parskip}{0pt}\setlength{\parindent}{0pt}
\renewcommand{\arraystretch}{1}\setlength{\tabcolsep}{6pt}
#1\endgroup}
\newcommand{\fitartwork}[1]{%
\sbox{\artworkbox}{\inputartwork{#1}}%
\ifdim\wd\artworkbox>\linewidth
  \sbox{\artworkbox}{\resizebox{\linewidth}{!}{\usebox{\artworkbox}}}%
\fi
\ifdim\dimexpr\ht\artworkbox+\dp\artworkbox\relax>0.77\textheight
  \resizebox{!}{0.77\textheight}{\usebox{\artworkbox}}%
\else\usebox{\artworkbox}\fi}
\renewcommand{\figinput}[3]{\begin{figure}[H]\centering\fitartwork{#1}\caption{#2}\label{fig:#3}\end{figure}}
\newcommand{\landfiginput}[3]{\begin{landscape}\begin{figure}[H]\centering
\sbox{\artworkbox}{\inputartwork{#1}}%
\sbox{\artworkbox}{\resizebox{\linewidth}{!}{\usebox{\artworkbox}}}%
\ifdim\dimexpr\ht\artworkbox+\dp\artworkbox\relax>0.80\textwidth
\resizebox{!}{0.80\textwidth}{\usebox{\artworkbox}}%
\else\usebox{\artworkbox}\fi
\caption{#2}\label{fig:#3}\end{figure}\end{landscape}}

\newcommand{\fitlandart}[1]{%
\sbox{\artworkbox}{\inputartwork{#1}}%
\sbox{\artworkbox}{\resizebox{\linewidth}{!}{\usebox{\artworkbox}}}%
\ifdim\dimexpr\ht\artworkbox+\dp\artworkbox\relax>0.80\textwidth
\resizebox{!}{0.80\textwidth}{\usebox{\artworkbox}}%
\else\usebox{\artworkbox}\fi}

\theoremstyle{definition}
\newtheorem{hypothesis}{Hypothesis}
\newtheorem{prediction}{Prediction}
\hypersetup{pdftitle={A regional dopamine-serotonin control model of conscious state},pdfauthor={Diogo Sousa}}
\makeatletter
\renewcommand{\@maketitle}{%
  \begingroup
  \setlength{\parskip}{0pt}
  \vspace*{-1cm}
  \hrule height 2pt
  \vskip 0.22in
  \centering
  {\fontsize{20}{24}\selectfont\bfseries\scshape\@title\par}
  \vskip 0.22in
  \hrule height 2pt
  \vskip 0.30in
  {\small\renewcommand{\arraystretch}{1}%
   \begin{tabular}{c}\@author\end{tabular}\par}
  \vskip 0.30in
  {\small\@date\par}
  \vskip 0.18in
  \endgroup
}
\makeatother

\title{A regional dopamine-serotonin control model of conscious state}
\author{\textbf{Diogo Sousa}\hspace{1mm}\href{https://orcid.org/0009-0007-1067-9343}{\includegraphics[width=0.3cm]{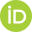}\textbf{0009-0007-1067-9343}}\\
\textit{Biomechanics and Health Unit}\\
\textit{INEGI - Institute of Science and Innovation}\\
\textit{in Mechanical and Industrial Engineering}\\
Porto, Portugal\\
\href{mailto:dssousa@inegi.up.pt}{dssousa@inegi.up.pt}}
\date{September 15, 2026}
\begin{document}
\maketitle
\begin{abstract}
This article proposes a dopamine-serotonin theory of conscious state control.
Regional dopamine signaling controls a candidate intensity coordinate.
Serotonergic signaling changes a candidate coordinate of felt complexity and significance within the capacity of the neural substrate.
A receptor model, a regional bottleneck, and a cumulative intensity response with an upper collapse gate connect these coordinates.
Extensions cover sleep, seizures, regional demand, species differences, development, terminal collapse, and receptor priming.
A fixed regional benchmark and a weight-cancelling sensitivity identity provide quantitative tests across perturbations.

Exploratory PPMI analyses examine REM timing and regional motor response.
Restricted waiting for sustained REM uses 10,537 nights from 137 participants, with formulation variation in 36 participants.
A formulation contrast gives $+1.220$ minutes per 100 mg/day LEDD substitution at fixed total levodopa LEDD.
Its 95\% interval is $0.425$ to $2.014$, with $p=0.0546$ after Holm adjustment across six tests.
Individual trends reduce the estimate to $+0.905$ minutes (95\% interval $-0.037$ to $1.847$).

Removing the largest information contributor reduces the trend-adjusted estimate to $+0.482$ minutes (95\% interval $-0.308$ to $1.272$).
Both temporal models give positive estimates after deleting their two or three largest contributors.
All four intervals span zero.
Episode persistence and the direct REM-versus-deep comparison have intervals spanning zero.
Conventional REM phase and latency coefficients are positive in both temporal models.

A regional motor comparison uses an earlier OFF examination in 254 participants.
An ordinal model allows imaging to explain baseline burden under a common medication response.
The incremental prediction gain is uncertain (log-score difference $-7.38\times10^{-5}$, 95\% interval $[-3.72,3.68]\times10^{-3}$).
These observational outcomes measure sleep and motor behavior, rather than effective receptor signaling or conscious experience.

The REM pattern motivates a prospective entry-mechanism test.
Regional interventions, transition dynamics, receptor assays, and direct reports of experience can assess the central control law.
\end{abstract}

\section{Introduction}
Conscious experience changes across wakefulness, sleep, anesthesia, seizures, and drug exposure.
A useful theory must connect these changes to mechanisms and measurable predictions.
This article proposes that regional dopamine and serotonin signals control two important dimensions of these changes.
The proposal is a candidate control theory, with explicit equations and conditions for failure.

The central claim concerns effective signaling, rather than total transmitter concentration.
Receptor density, occupancy, efficacy, circuit location, and clearance can change the same chemical input into different regional effects.
A regional bottleneck can thus reduce conscious access during a rise in average neural activity.
Serotonergic signaling can change the structure of experience at a similar level of the dopamine coordinate.
Both claims require tests that separate experience from movement, sleep classification, and task performance.

Several experimental findings motivate the regional architecture.
Activation of ventral tegmental dopamine neurons can cause emergence from anesthesia in mice \cite{Taylor2016}.
Transient dopamine signaling in the basolateral amygdala can initiate REM sleep in mice \cite{Hasegawa2022}.
These findings require a regional account of dopamine function.

Human studies link psilocin exposure and \HT{} occupancy to subjective drug intensity \cite{Madsen2019}.
Some psychoactive drugs increase measures of spontaneous neural signal diversity \cite{Schartner2017}.
These measurements are relevant to a complexity hypothesis, but they are not interchangeable.
Reports of experience during both REM and non-REM sleep also exclude a direct equivalence between REM and consciousness \cite{Siclari2017}.

The article first defines the core model, then examines regional and sleep mechanisms.
Exploratory PPMI analyses constrain a subset of its predictions.
Later sections extend the model to collapse, comparative biology, development, intracellular pathways, and receptor priming.
Each extension identifies its assumptions and a result that could contradict it.

\subsection{Five quantitative commitments}\label{sec:commitments}
The theory is organized around five commitments that can be evaluated separately (Table~\ref{tab:commitments}).
W0, X0, and X1 specify fixed numerical benchmarks.
The receptor-gain inequality and sensitivity-transfer identity restrict broader equation families.
Failure of a benchmark rejects that benchmark.
Failure of an identity challenges the corresponding mechanism.

\begin{table}[H]\centering\small
\caption{Quantitative commitments and outcomes that would reject them. Regional signals use the waking reference defined in Section~\ref{sec:definitions}.}
\label{tab:commitments}
\begin{tabularx}{\linewidth}{@{}p{1.3cm}Y Y@{}}
\toprule
Law & Fixed prediction & Contradictory outcome \\
\midrule
W0 & Weights $(4,3,2,1,1)/11$ and $\kappa=4$. At $(0.7,1.4,1,1,1)$, the PFC-to-motor sensitivity ratio is $21.926$. & A regional outcome-slope ratio excluding this value under the shared-readout conditions. \\
X0 & Quadratic regional demand with W0 weights. The matched-bottleneck comparison predicts $\Delta\dot E/c_Q=-5.216$. & A reserve-change contrast excluding this value after independent cost calibration (Section~\ref{sec:regionalexcess}). \\
X1 & With $\alpha_E=1/4$ and $\epsilon_E=1/20$, the demand derivative changes sign at relative reserve $e=0.45$. & A derivative reversal excluding $0.45$, or precise derivatives with the opposite signs in the declared scalar comparison. \\
RPR & Gain exceeds one exactly when $u_N>(1+u_AA)/(1+A)$. At $u_N=1.4$, $u_A=2$, $A=1$, gain is $0.933$. & A gain above one at these measured inputs, with residual occupancy, efficacy, and ligand input controlled. \\
Transfer & Equation~\ref{eq:ratiotransfer} fixes a linear log-ratio relation. For $\kappa=4$ and a regional-difference change of $+0.7$, its value is $-2.800$. & A held-out log ratio excluding $-2.800$, or failure of one calibrated $\kappa$ to predict additional contrasts. \\
\bottomrule
\end{tabularx}
\end{table}

The comparison fixes anatomy, signal normalization, temporal filtering, physiological context, and a locally nonzero shared observation slope before evaluation.
Measured input error must propagate into each prediction interval.
A study rejects a target when its prespecified simultaneous uncertainty interval excludes the prediction beyond independently validated measurement tolerance.
This rule prevents uncertain measurements from becoming either apparent confirmations or automatic rejections.
The transfer identity removes unknown regional weights, but still needs measured signal differences and a fixed or independently calibrated $\kappa$.

The main technological need is simultaneous regional signaling, selective perturbation, and a validated readout of conscious state.
Parts of this program can already be tested in receptor and circuit preparations.

\subsection{Related theories and the proposed contribution}\label{sec:related}
The proposal builds on established accounts of affect, conscious access, and psychedelic experience.
Its central contribution is a regional control law that imposes shared quantitative restrictions across perturbations.
The soft minimum is a standard mathematical operator.
The proposed biological use of that operator, together with its transferable response ratios, is the specific hypothesis advanced here.

\paragraph{Affective and subcortical accounts:}
Solms and Panksepp place primary affective consciousness in subcortical systems and describe its contribution to later cognitive forms \cite{SolmsPanksepp2012}.
This establishes an important precedent for a theory that gives neuromodulatory and subcortical systems a central role.
Solms also proposed a dopamine-related forebrain mechanism for dreaming that can dissociate from REM generation \cite{Solms2000}.
That account precedes the separation of sleep stage and dream experience developed in Section~\ref{sec:rementrytheory}.

The present proposal adds an explicit transformation from regional receptor signaling to a coordinated state input.
Its cortical and subcortical weights are fixed before evaluation and must apply unchanged across perturbations.
A prefrontal deficit is neither universally necessary nor sufficient for state loss, because its effect depends on its size and the other regional signals (Section~\ref{sec:seizure}).
The decisive question is whether measured regional changes obey the proposed control law.

\paragraph{Global neuronal workspace:}
The global neuronal workspace account associates conscious access with amplification and widespread availability of selected information \cite{DehaeneChangeux2011}.
The regional dopamine model instead specifies a proposed control input to the operating state of such networks.
These accounts can coexist if dopamine controls access through workspace dynamics.
However, a neuromodulatory influence alone would not establish the conditional sufficiency claim in Hypothesis~1.
Different intensity outcomes at matched modeled inputs, histories, and physiological context would contradict that stronger claim.
Simultaneous regional signaling and network measurements can distinguish a common control input from an additional independent network determinant.

\paragraph{Integrated information theory:}
Integrated information theory identifies experience with the intrinsic causal structure of a physical substrate \cite{Albantakis2023}.
The present equations specify state control and substrate constraints, rather than an identity between experience and a causal structure.
The substrate-capacity term is not an estimate of integrated information.
Likewise, \(\Psi\) denotes the proposed complexity of state contents, not an IIT quantity.
A perturbation could change a control coordinate, a substrate property, or both.
Joint measurements must establish which change explains the observed state difference.

\paragraph{Entropic brain and REBUS:}
The entropic brain hypothesis links neural entropy with the richness and organization of conscious states \cite{CarhartHarris2014}.
REBUS connects psychedelic action at \HT{} receptors with reduced precision of high-level priors and increased influence of bottom-up signals \cite{CarhartHarrisFriston2019}.
These are direct precedents for the proposed serotonergic complexity axis.
The association between psychedelic signaling and complex experience is therefore not claimed as a new discovery.
The present model adds a dopamine-dependent regional bottleneck and an explicit separation between intensity, complexity drive, and substrate capacity.

The models make overlapping predictions about psychedelics, so increased signal diversity alone cannot distinguish them.
A useful comparison varies regional dopamine balance while holding measured effective \HT{} drive, physiological context, and substrate conditions constant.
Small regional perturbations then test the sensitivity-transfer identity in Equation~\ref{eq:ratiotransfer}.
REBUS does not itself impose this exponential relation between regional signal differences and outcome slopes.
Satisfying that restriction would support the added regional control law, while remaining compatible with REBUS as a content-level mechanism.

A complementary experiment combines receptor assays, reports of experiential complexity, and independent estimates of prior precision.
It varies contextual expectations at matched receptor drive and intensity, with the fixed content measurements in Section~\ref{sec:experiencemeasurement}.
Expectations are not an input to Equation~\ref{eq:complexity}, so the model predicts equal complexity across these contexts.
Reliable differences in the prespecified content score would identify a missing causal input, rather than a different point on the receptor curve.
These experiments test the proposed separations between mechanisms without assuming that the theories are mutually exclusive.

\section{Mathematical framework}
\subsection{Definitions and scope}\label{sec:definitions}
The model has two latent coordinates: intensity $C$ and dynamic complexity $\Psi$.
Intensity denotes the proposed strength with which experience is subjectively present.
Complexity denotes how complex and significant the contents feel to the person who experiences them.
The model permits a simple experience with strong subjective presence, or a diverse experience with weaker presence.
Neither coordinate has a validated universal measurement scale.
Future measurements can provide estimates of the candidate relationships specified by these equations.

The model follows a defined sequence.
Local receptor responses supply regional signals.
Temporal integration accounts for recent signal history, and spatial aggregation gives the bottleneck coordinate.
The state function then maps that coordinate and its physiological context to conscious intensity.
The regional demand extension also lets excessive local signaling affect reserve and injury (Section~\ref{sec:regionalexcess}).

\begin{eqvars}
\item $r$ (\textbf{Regional index}): Selects one region or circuit at a fixed spatial resolution.
\item $t$ (\textbf{Time}): Locates each input and state on the same time axis.
\item $D_r(t)$ (\textbf{Regional dopamine signal}): Gives the dopamine signal after regional normalization and temporal integration.
\item $H_r(t)$ (\textbf{Regional serotonin signal}): Gives the regional serotonergic signal relative to its specified reference.
\item $C$ (\textbf{Conscious intensity}): Represents the proposed strength of subjective experience as a whole. Content diversity, emotional intensity, motor output, and task accuracy are different quantities.
\item $\Psi$ (\textbf{Experiential complexity}): Summarizes felt complexity and personal significance. It concerns the person's experience, rather than an external judgment of its complexity or truth.
\end{eqvars}

Felt complexity concerns the richness and connections experienced within the content.
Felt significance concerns how meaningful or personally important that content seems.
Neither implies factual accuracy, useful insight, or flexible thought.
A rigid belief can carry strong significance, while diverse mental activity can feel unimportant.
Section~\ref{sec:experiencemeasurement} specifies separate ratings and a fixed summary score for testing $\Psi$.
Diversity, flexibility, and neural entropy provide additional comparisons, rather than interchangeable definitions of this coordinate.

The reference is stable, quiet wakefulness in the same system, with declared physiological context and no acute perturbation.
Each regional signal is divided by its positive time-averaged value in that reference window.
Thus, every regional reference signal equals one, and the normalized bottleneck also equals one.
The window duration, conditions, and measurement protocol must be fixed before the experimental comparison.
A value of one means this waking reference, not full receptor occupancy.

Figures~\ref{fig:continuum} and~\ref{fig:sleep} use 100\% for the global waking reference.
Figure~\ref{fig:statemap} instead uses ordinal axes.
Distances between its labels do not represent measured signal ratios.
Medication dose, extracellular concentration, receptor signaling, and conscious experience are separate quantities.

A dynamic state model can retain different states after different histories.
Comparisons must therefore specify the operating branch and initial transition state (Section~\ref{sec:transition}).

\begin{hypothesis}[Regional control]
Within a specified viable domain, regional dopamine signaling determines the intensity coordinate through a common bottleneck and state function.
Matched bottleneck inputs and histories imply matched intensity outputs when serotonergic drive, reserve, injury, substrate capacity, and declared physiological context are fixed.
The physiological context includes oxygen availability and temperature when these enter the viability mapping.
For the dynamic model, the comparison also holds the operating branch and initial transition state constant.
Effective \HT{} signaling sets the complexity drive within the substrate ceiling.
The same regional weights and control functions must apply across the selected perturbations.
\end{hypothesis}

This is a conditional sufficiency claim, rather than an association with arousal.
Its experimental test also requires a fixed observation model and controls for motor output, report ability, and measurement history.
Reproducible differences in conditional outcome distributions at matched modeled inputs would contradict this claim within the declared domain.

\subsection{Effective dopamine signaling}
The receptor architecture has three components:
\begin{samepage}
\begin{align}
N_r(t)&=\left(\sum_{j\in D_1}w_{jr}\rho_{jr}O_{jr}S_{jr}\right)
 +\theta_r\left(\sum_{j\in D_{2,\mathrm{post}}}w_{jr}\rho_{jr}O_{jr}S_{jr}\right),\label{eq:numerator}\\
A_r(t)&=\sum_{j\in D_{2,\mathrm{auto}}}v_{jr}\rho_{jr}O_{jr}S_{jr},\\
\deff(r,t)&=k_r\frac{N_r(t)}{1+A_r(t)}.\label{eq:deff}
\end{align}
\begin{eqvars}
\item $j$ (\textbf{Receptor index}): Receptor population within region $r$.
\item $N_r$ (\textbf{Postsynaptic contribution}): Combines the weighted postsynaptic contributions. The positive-signal model requires a nonnegative total.
\item $A_r$ (\textbf{Autoreceptor feedback}): Summarizes nonnegative autoreceptor feedback. A larger value reduces effective signaling at fixed postsynaptic input.
\item $\rho_{jr}$ (\textbf{Relative receptor density}): Gives density relative to the reference density. This ratio separates receptor availability from occupancy.
\item $O_{jr}$ (\textbf{Receptor occupancy}): Gives the occupied fraction, between zero and one. It depends on local ligand concentration and receptor affinity.
\item $S_{jr}$ (\textbf{Signaling efficacy}): Gives signaling efficacy relative to its reference. Equal occupancy need not produce equal downstream signaling.
\item $w_{jr},v_{jr}$ (\textbf{Receptor weights}): Assign dimensionless contributions within the receptor model. They differ from the spatial weights $p_r$.
\item $\theta_r$ (\textbf{Postsynaptic D2 circuit gain}): Dimensionless coefficient for the net effect of postsynaptic D$_2$ signaling on the selected regional output. A positive value increases this output, a negative value reduces it, and zero removes this contribution.
\item $k_r$ (\textbf{Regional gain}): Sets the regional reference scale. It summarizes gain factors that are not explicit in the receptor terms.
\item $\deff(r,t)$ (\textbf{Effective dopamine signal}): Combines postsynaptic signaling and autoreceptor feedback before temporal integration.
\item $D_1,D_{2,\mathrm{post}},D_{2,\mathrm{auto}}$ (\textbf{Receptor compartments}): Separate D$_1$, postsynaptic D$_2$, and autoreceptor D$_2$ populations. Their model contributions are not interchangeable.
\end{eqvars}
\end{samepage}

The coefficient $\theta_r$ translates a cellular receptor effect into a net circuit effect.
D$_2$ activation can inhibit a neuron while facilitating a downstream output by reducing an inhibitory projection \cite{Gallo2018}.
Within each receptor pool, weights and efficacy magnitudes are nonnegative.
The coefficient $\theta_r$ carries the postsynaptic D$_2$ circuit sign.
Thus, its sign depends on the circuit and the output being modeled.
It is not a universal receptor constant or a direct measure of consciousness.

Autoreceptor feedback instead enters $A_r$, where it reduces the effective signal at fixed postsynaptic input \cite{Bello2011}.
The regional coefficients and receptor weights must be specified independently of the outcome used to evaluate the theory.
Only the products $\theta_r w_{jr}$ identify the postsynaptic D$_2$ contribution when both factors are freely estimated.
Fixing or normalizing the within-pool weights is therefore necessary before estimating $\theta_r$ separately.
For negative $\theta_r$, the positive-signal model applies only where the total $N_r\geq0$.

For one ligand at equilibrium, occupancy has the form
\begin{samepage}
\begin{equation}
O_{jr}=\frac{L_r}{K_{jr}+L_r},\qquad
\frac{dL_r}{dt}=q_r(t)-\{k_{\mathrm{DAT},r}+k_{\mathrm{other},r}\}L_r.
\label{eq:occupancy}
\end{equation}
\begin{eqvars}
\item $L_r$ (\textbf{Ligand concentration}): Local ligand concentration.
\item $K_{jr}$ (\textbf{Occupancy constant}): Sets the concentration for half occupancy in this equilibrium model. Lower values give greater occupancy at matched concentration.
\item $q_r(t)$ (\textbf{Release rate}): Adds local ligand in concentration per unit time.
\item $k_{\mathrm{DAT},r}$ (\textbf{Transporter clearance}): Sets transporter-dependent ligand removal in inverse time. It acts through concentration dynamics.
\item $k_{\mathrm{other},r}$ (\textbf{Other clearance}): Sets removal through the other modeled routes, also in inverse time.
\end{eqvars}
\end{samepage}

This form locates transporter effects in ligand dynamics.
It avoids a second clearance correction when occupancy already uses the resulting concentration.

A static transporter factor can serve as a reduced approximation:
\begin{samepage}
\begin{equation}
\deff^{\mathrm{static}}(r,t)=k_r\frac{N_r(t)}{1+A_r(t)}
\underbrace{\left(1-\frac{1}{1+\eta_r^{-1}}\right)}_{1/(1+\eta_r)},
\qquad \eta_r=\frac{k_{\mathrm{DAT},r}}{k_{\mathrm{ref},r}}>0.
\label{eq:staticdat}
\end{equation}
\begin{eqvars}
\item $\eta_r$ (\textbf{Relative transporter clearance}): Dimensionless ratio of transporter clearance to reference clearance.
\item $k_{\mathrm{ref},r}$ (\textbf{Reference clearance}): Positive reference clearance rate.
\end{eqvars}
\end{samepage}
This factor requires an occupancy input that does not already include the same clearance effect.
Equations~\ref{eq:occupancy} and \ref{eq:staticdat} are alternative implementations of clearance.
They are not two corrections to apply automatically to the same measurement.

The receptor function converts local ligand input into effective regional signaling.
It need not have an inverted U.
With shared occupancy, $aO/(1+bO)$ increases for positive $a$ and $b$.
The upper decline enters through the viability gate in the signal-to-state relation of Equation~\ref{eq:multistate}.

A receptor-level peak is an optional case under different occupancy constants.
Appendix~\ref{app:receptorpeak} gives its condition and proof.
The state theory does not require that condition.

\subsection{Regional signals, temporal integration, and the bottleneck}
Figure~\ref{fig:regions} shows the candidate regional organization.
Each box supplies its own signal through the receptor model.
A larger signal elsewhere need not fully offset a reduction in a functionally limiting region.
The equations compare this bottleneck hypothesis with a weighted regional mean.

Temporal integration adds a causal memory term:
\begin{samepage}
\begin{equation}
\widetilde D_r(t)=\int_0^\infty\frac{e^{-u/\tau_m}}{\tau_m}\deff(r,t-u)\,du,
\qquad \tau_m\dot{\widetilde D}_r=\deff-\widetilde D_r.
\label{eq:memory}
\end{equation}
\begin{eqvars}
\item $\widetilde D_r(t)$ (\textbf{Integrated regional signal}): Combines recent local inputs with greater weight on more recent signaling.
\item $u$ (\textbf{Input age}): Time elapsed since a past input.
\item $\tau_m$ (\textbf{Memory time}): Sets the duration of regional memory. A larger positive value gives a slower response to changing input.
\item $\dot{\widetilde D}_r$ (\textbf{Signal change}): Rate of change of the integrated signal.
\end{eqvars}
\end{samepage}
The kernel preserves a constant input.
It creates lag, which can produce loops under changing input.
Persistent hysteresis at slow input changes needs an additional feedback mechanism.
Section~\ref{sec:transition} gives a candidate that permits this stronger behavior.

For the spatial equations, set $D_r(t)=\widetilde D_r(t)$ on the regional reference scale.
Thus, the temporal filter acts within each region before the signals are combined.
Filtering the final bottleneck instead would generally give a different result because spatial aggregation is nonlinear.

The spatial comparison is
\begin{samepage}
\begin{align}
\overline D(t)&=\sum_r p_rD_r(t),\qquad p_r\geq0,\quad\sum_r p_r=1,\\
D_{\min}(t)&=\min_{r\in\mathcal R_{\mathrm{crit}}}D_r(t),\\
D_\kappa(t)&=-\frac{1}{\kappa}\log\sum_r p_r e^{-\kappa D_r(t)},\qquad\kappa>0.
\label{eq:softmin}
\end{align}
\begin{eqvars}
\item $p_r$ (\textbf{Prior regional weight}): Assigns a nonnegative regional weight before the test. The weights sum to one and stay fixed across test conditions.
\item $\overline D$ (\textbf{Mean signal}): Combines regions by a weighted arithmetic mean. It is the reference model for spatial compensation.
\item $\mathcal R_{\mathrm{crit}}$ (\textbf{Critical set}): Specifies the hypothesized necessary regions for a selected function. Membership must be fixed before outcome evaluation.
\item $D_{\min}$ (\textbf{Critical minimum}): Lowest signal within that critical set.
\item $D_\kappa$ (\textbf{Bottleneck signal}): Combines the integrated regional signals with greater sensitivity to low-signal regions.
\item $\kappa$ (\textbf{Bottleneck strength}): Sets sensitivity to regional deficits, in inverse signal units. It must use the same signal scale across conditions.
\end{eqvars}
\end{samepage}

The dots and color scale in Figure~\ref{fig:regions} illustrate candidate prior weights $p_r$.
They are not receptor counts, measured demand, or fitted numerical coefficients.
Section~\ref{sec:fixedbenchmark} gives a fixed numerical benchmark for this drawing.
The arrows identify candidate circuit relations, but the scalar aggregation does not specify their connection strengths.
A network model of information exchange would require additional variables and measurements.

The critical set is a hypothesis about necessary regional functions.
Candidate regions include thalamic, cortical, cingulate, and brainstem systems.
Membership must depend on the outcome and experimental evidence.
The model does not establish universal prefrontal necessity by definition.

Normalization gives $D_\kappa=d$ when every region has signal $d$.
For a continuous field, $p_r$ becomes $\omega(r)/\int\omega(r)\,dr$, where $\omega(r)$ is nonnegative regional weight density.
The soft minimum lies between the minimum and the weighted mean.
Its regional sensitivity is
\begin{samepage}
\begin{equation}
\frac{\partial D_\kappa}{\partial D_r}
=\frac{p_r e^{-\kappa D_r}}{\sum_s p_s e^{-\kappa D_s}}.
\label{eq:softweight}
\end{equation}
\begin{eqvars}
\item $\partial D_\kappa/\partial D_r$ (\textbf{Current regional influence}): Gives the change in the bottleneck per small regional signal change. It depends on both $p_r$ and current signals.
\item $s$ (\textbf{Summation index}): Index over the included regions in the denominator.
\end{eqvars}
\end{samepage}

The prior weight and the current influence are different quantities.
At equal regional signals, the current influence equals $p_r$.
A sufficiently low signal increases that region's influence relative to the others.
The limiting region can therefore change over time without changing the prior weights.

A region with a small prior weight can dominate after a sufficiently large signal reduction.
As $\kappa$ increases without bound, $D_\kappa$ approaches the minimum over all regions with positive weight.
The limit equals a restricted critical minimum only when excluded regions have zero weight.
These properties prevent an automatic claim that striatal depletion has only a small effect.

Regional subdivision must preserve weight mass.
If one region is split into two identical-signal subregions, their weights must sum to the original weight.
Otherwise, a change of atlas alone could change the predicted state.

A two-moment form gives a local approximation:
\begin{samepage}
\begin{equation}
D_\kappa=\overline D-\frac{\kappa}{2}\operatorname{Var}_p(D)
+O(\kappa^2).
\label{eq:moments}
\end{equation}
\begin{eqvars}
\item $\operatorname{Var}_p(D)=\sum_r p_r(D_r-\overline D)^2$ (\textbf{Regional dispersion}): Weighted regional signal variance.
\item $O(\kappa^2)$ (\textbf{Approximation remainder}): Terms of second order and above for fixed regional signals.
\end{eqvars}
\end{samepage}
It penalizes dispersion but is not generally equivalent to the soft minimum.
The approximation requires $\kappa$ times a typical regional signal difference to be small.
Data can compare the mean, variance penalty, and full bottleneck directly.

\subsection{Cumulative intensity and the collapse boundary}
The integrated regional coordinate enters through $x(t)=D_\kappa(t)/D_{\kappa,\mathrm{ref}}$.
At fixed serotonergic context and viability, greater coordinated dopamine input gives greater conscious intensity.
The cumulative response can saturate, while the viability gate permits a high-input collapse.
The compact notation suppresses controlled oxygen, temperature, and other declared physiological arguments of viability.

\begin{samepage}
\begin{align}
F_C(x,h)&=\sum_{i=1}^{n}\alpha_i(h)
\sig\left(\frac{x-\mu_i(h)}{\tau_i}\right),\label{eq:cumulative}\\
C(x,h,E,J)&=V(x,E,J)F_C(x,h).
\label{eq:multistate}
\end{align}
\begin{eqvars}
\item $x$ (\textbf{Normalized dopamine input}): Divides the bottleneck signal by its positive reference $D_{\kappa,\mathrm{ref}}$.
\item $h$ (\textbf{Serotonergic context}): Summarizes effective serotonergic control. Any effect on the response parameters needs a separately specified mapping.
\item $E,J$ (\textbf{Reserve and injury}): Energy reserve and accumulated injury.
\item $V(x,E,J)$ (\textbf{Viability factor}): Combines the low-signal boundary, upper-collapse gate, reserve, and injury factors. Its value lies between zero and one.
\item $F_C(x,h)$ (\textbf{Cumulative intensity response}): Sums nonnegative contributions as coordinated dopamine input increases.
\item $i,n$ (\textbf{Recruitment components}): Component index and number of components. They are not clinical-state probabilities.
\item $\alpha_i(h)$ (\textbf{Response amplitude}): Positive amplitude of component $i$, on the fixed reference scale.
\item $\mu_i(h)$ (\textbf{Recruitment location}): Midpoint of component $i$ on the dopamine-input axis.
\item $\tau_i$ (\textbf{Recruitment scale}): Positive width of the rising sigmoid.
\item $\sig(z)=1/(1+e^{-z})$ (\textbf{Activation function}): Smooth sigmoid function.
\end{eqvars}
\end{samepage}

Each contribution increases toward saturation and remains available as the input rises.
The reference normalization is fixed across the comparison, rather than recomputed for each intervention.
The numerical benchmark holds the amplitudes and locations constant across serotonergic contexts.
A coupled extension can specify $\alpha_i(h)$ and $\mu_i(h)$ independently, while preserving positive amplitudes and recruitment scales.

At fixed $h$, the cumulative response has a positive derivative:
\begin{samepage}
\begin{equation}
\frac{\partial F_C}{\partial x}
=\sum_i\frac{\alpha_i(h)}{\tau_i}s_i(1-s_i)>0,
\qquad s_i=\sig\left(\frac{x-\mu_i(h)}{\tau_i}\right).
\label{eq:intensityslope}
\end{equation}
\begin{eqvars}
\item $s_i$ (\textbf{Recruitment fraction}): Sigmoid output for component $i$, strictly between zero and one at finite input.
\item $\partial F_C/\partial x$ (\textbf{Dopamine response slope}): Positive slope at fixed serotonergic context and fixed response parameters.
\end{eqvars}
\end{samepage}

Regional influence is also nonnegative in Equation~\ref{eq:softweight}.
Thus, increasing every regional signal increases the bottleneck and the cumulative intensity response.
An increased arithmetic mean alone does not ensure this result when a limiting region loses signal.

At fixed reserve and injury, $\partial_x C=V\partial_xF_C+F_C\partial_xV$.
For $V>0$, a falling response requires $\partial_x\log V<-\partial_x\log F_C$.
The low-signal factor increases with input, while the upper gate declines near its collapse boundary (Equation~\ref{eq:collapse}).
Thus, the falling branch reflects declining viability, rather than disappearing recruitment contributions.
The smooth gate starts acting below $x_c$, so the increasing domain ends within the approach to collapse.
Changes in reserve, injury, or serotonergic context can also change the output over time.

The high-input psychosis label describes a possible presentation with broadly elevated coordinated input.
It does not assign every psychosis, schizophrenia presentation, or stimulant exposure to that range.
Impaired reality testing is distinct from reduced subjective intensity and needs its own observation measure.
Sleep and wake labels likewise need the observation functions in Equation~\ref{eq:measurements}.

A reference simulation uses the following locations:
\begin{samepage}
\begin{equation}
(\mu_1,\mu_2,\mu_3,\mu_4)=(0.20,0.50,0.80,2.50),
\qquad x_c=3.20.
\label{eq:anchors}
\end{equation}
\begin{eqvars}
\item $x_c$ (\textbf{Collapse location}): Illustrative collapse location, not a fifth component location.
\end{eqvars}
\end{samepage}
The recruitment locations correspond to 20, 50, 80, and 250\% of the waking reference.
The collapse location is 320\%.
A recruitment location is a sigmoid midpoint, not a clinical state boundary.
Appendix~\ref{app:simulation} fixes the remaining parameters and normalizes the numerical example to $C(1)=1$.
The first component location, $\mu_1=0.20$, is not a calibrated threshold for loss of consciousness.
Figure~\ref{fig:continuum} plots this numerical example on the same waking-reference scale as the sleep trajectory.

The example peaks at $x=2.958$, where $C=1.649$.
This input is 7.6\% below $x_c$, measured relative to $x_c=3.20$.
That separation describes the chosen curve, not a measured physiological safety margin.

\figinput{\begin{tikzpicture}
\begin{axis}[
    width=28cm,height=12cm,
    xlabel={Effective dopamine signal (\% of quiet waking reference)},
    ylabel={Subjective intensity $C$ (reference = 1)},
    xmin=0,xmax=400,ymin=-.12,ymax=1.9,
    xtick={0,50,100,150,200,250,320,400},ytick={0,.25,.5,.75,1,1.25,1.5,1.75},
    axis lines=left,axis line style={ultra thick},grid=both,
    minor grid style={gray!5},major grid style={gray!15},
    xlabel style={font=\Large\bfseries},ylabel style={font=\Large\bfseries},
    tick label style={font=\large},clip=false,
    title={\Large\bfseries The Dopamine-Consciousness Continuum},
    title style={at={(0.5,1.05)}}]
\fill[sleepcolor!20] (axis cs:0,-.12) rectangle (axis cs:50,1.9);
\fill[wakecolor!15] (axis cs:50,-.12) rectangle (axis cs:100,1.9);
\fill[consciouscolor!15] (axis cs:100,-.12) rectangle (axis cs:250,1.9);
\fill[psychosiscolor!20] (axis cs:250,-.12) rectangle (axis cs:320,1.9);
\fill[deathcolor!40] (axis cs:320,-.12) rectangle (axis cs:400,1.9);
\addplot[consciouscolor!40,line width=8pt,opacity=.3] table[x=percent,y=C] {figures/state_response.dat};
\addplot[consciouscolor,line width=3pt] table[x=percent,y=C] {figures/state_response.dat};
\addplot[only marks,mark=*,mark size=10pt,dopaminecolor!30] coordinates {(20,0.0862728359087) (50,0.412687765439) (100,1) (250,1.41198449805) (320,0)};
\addplot[only marks,mark=*,mark size=8pt,dopaminecolor!60] coordinates {(20,0.0862728359087) (50,0.412687765439) (100,1) (250,1.41198449805) (320,0)};
\addplot[only marks,mark=*,mark size=6pt,dopaminecolor] coordinates {(20,0.0862728359087) (50,0.412687765439) (100,1) (250,1.41198449805) (320,0)};
\addplot[only marks,mark=*,mark size=3pt,white] coordinates {(20,0.0862728359087) (50,0.412687765439) (100,1) (250,1.41198449805) (320,0)};

\draw[ultra thick,deathcolor,line width=5pt,dashed] (axis cs:320,-.12)--(axis cs:320,1.9);
\draw[gray!70,line width=.7pt] (axis cs:24,.49)--(axis cs:20,.13);
\node[rounded corners=8pt,font=\small\bfseries,inner sep=10pt,align=center]
 at (axis cs:23,.73) {\faIcon{bed}\\DEEP SLEEP\\\textcolor{gray}{Proposed domain}\\\textcolor{gray}{\scriptsize Reduced access}};
\node[font=\small\bfseries,inner sep=10pt,align=center] (remdesc)
 at (axis cs:76,.27) {\faIcon{cloud}\\REM / DREAMS\\\textcolor{gray}{Proposed domain}\\\textcolor{gray}{\scriptsize Dream experience}};
\draw[gray!70,line width=.7pt,shorten <=2pt,shorten >=7pt] (axis cs:70,.39)--(axis cs:50,0.412687765439);
\draw[gray!70,line width=.7pt] (axis cs:145,1.27)--(axis cs:105,1.01);
\node[font=\small\bfseries,inner sep=10pt,align=center]
 at (axis cs:165,1.37) {\faIcon{sun}\\NORMAL WAKE\\\textcolor{gray}{Quiet reference: 100\%}\\\textcolor{gray}{\scriptsize Coherent reality}};
\draw[gray!70,line width=.7pt] (axis cs:275,.55)--(axis cs:253,1.35);
\node[
  font=\small\bfseries,
  inner sep=10pt,
  align=center
]
at (axis cs:277,.37) {
  \faIcon{exclamation-triangle}\\
  HIGH-INPUT\\
  PSYCHOSIS\\
  \textcolor{gray}{Proposed domain}\\
  \textcolor{gray}{\scriptsize Altered reality testing}
};
\node[text=black!80,rounded corners=8pt,font=\small\bfseries,inner sep=10pt,align=center]
 at (axis cs:363,.33) {\faIcon{skull}\\STATE COLLAPSE\\\textcolor{black!80}{$x_c=3.20$ example}\\\textcolor{black!80}{\scriptsize Injury is a separate criterion}};
\node[text=black!80,font=\normalsize\bfseries,rotate=90,rounded corners=4pt,inner sep=6pt]
 at (axis cs:330,1.38) {ILLUSTRATIVE BOUNDARY};
\node[opacity=.1,font=\Huge] at (axis cs:25,1.75) {\faMoon};
\node[opacity=.1,font=\Huge] at (axis cs:172,1.75) {\faSun};
\node[opacity=.1,font=\Huge] at (axis cs:281,1.75) {\faBolt};
\end{axis}
\end{tikzpicture}}{Dopamine-state response from Equation~\ref{eq:multistate}, using the parameters in Appendix~\ref{app:simulation}. Quiet waking input is 100\%, with $C(1)=1$. Cumulative recruitment increases until the upper viability gate dominates near collapse. State names and colored domains are proposed associations, without defining diagnostic thresholds. The psychosis example requires broadly elevated coordinated input. The upper gate represents reversible state collapse, while death requires the separate injury mechanism.}{continuum}

\subsection{Serotonergic complexity and substrate capacity}
The complexity drive function below describes signaling through the \HT{} receptor:
\begin{samepage}
\begin{align}
\Psi_{\mathrm{drive}}(x,h,t)&=\Psi_0
\left[1+\lambda_c f_{\mathrm{const}}+\lambda_L O_H(t)\right]
e^{-\gamma(x/x_s)^2}R(t),\label{eq:complexity}\\
O_H(t)&=\frac{L_H(t)}{K_H+L_H(t)}.\label{eq:htoccupancy}
\end{align}
\begin{eqvars}
\item $\Psi_0$ (\textbf{Reference complexity}): Reference complexity scale.
\item $\lambda_c,\lambda_L$ (\textbf{Signaling gains}): Gains for constitutive and ligand-dependent \HT{} signaling.
\item $f_{\mathrm{const}}$ (\textbf{5-HT2A constitutive activity}): Fraction that represents ligand-independent \HT{} receptor activity.
\item $O_H$ (\textbf{5-HT2A occupancy}): Occupied fraction of the modeled \HT{} receptor pool.
\item $L_H,K_H$ (\textbf{5-HT2A ligand and occupancy constant}): $L_H$ is the local concentration of the modeled ligand at \HT{} receptors. $K_H$ is the equilibrium half-occupancy concentration for that ligand at \HT{} receptors.
\item $\gamma$ (\textbf{Dopamine attenuation}): Nonnegative strength of dopamine-related attenuation in this response function.
\item $x_s$ (\textbf{Attenuation scale}): Positive attenuation scale, separate from the collapse location.
\item $R(t)$ (\textbf{5-HT2A temporal response}): Time-dependent factor for the modeled \HT{} signaling response.
\end{eqvars}
\end{samepage}

The quantities $L_H$, $f_{\mathrm{const}}$, and the gains implement the serotonergic input represented by $h$.
At fixed dopamine input and temporal response, greater positive \HT{} drive increases predicted complexity until the substrate ceiling applies.
This relation concerns effective receptor signaling, which can differ from extracellular serotonin concentration.

Receptor constitutive activity and inverse agonism make the additional baseline term biologically plausible \cite{Weiner2001,Vanover2006}.
They do not establish its magnitude during REM or its contribution to experience.
The gains require a specified outcome scale.

For a discrete exposure, the onset and decay function is
\begin{samepage}
\begin{equation}
R(t)=(1-e^{-t/\tau_{\mathrm{on}}})e^{-t/\tau_{\mathrm{des}}},\quad t\geq0,
\qquad t^*=\tau_{\mathrm{on}}\log\left(1+\frac{\tau_{\mathrm{des}}}{\tau_{\mathrm{on}}}\right).
\label{eq:onset}
\end{equation}
\begin{eqvars}
\item $\tau_{\mathrm{on}}$ (\textbf{Onset time}): Positive onset time constant.
\item $\tau_{\mathrm{des}}$ (\textbf{Decay time}): Positive decay or desensitization time constant.
\item $t^*$ (\textbf{Response maximum}): Time of the maximum response after the exposure.
\end{eqvars}
\end{samepage}
Both time constants depend on the intervention and measurement.
Universal drug-specific values require data for the drug, intervention, and measurement.
For sustained endogenous activity, $R$ is a steady-state factor.
It is not a decay from an arbitrary time origin.

Structural capacity $\Phi$ places a ceiling on dynamic complexity:
\begin{samepage}
\begin{equation}
\Psi_{\mathrm{eff}}=\min\{\Psi_{\mathrm{drive}},\Psi_{\max}(\Phi)\},
\qquad C_{\mathrm{total}}=C(x,h,E,J).
\label{eq:ceiling}
\end{equation}
\begin{eqvars}
\item $\Phi$ (\textbf{Structural capacity}): Represents the proposed capacity of the substrate. It is not an established measure of integrated information.
\item $\Psi_{\max}(\Phi)$ (\textbf{Complexity ceiling}): Complexity ceiling permitted by that capacity.
\item $\Psi_{\mathrm{eff}}$ (\textbf{Effective complexity}): Complexity after application of the ceiling.
\item $C_{\mathrm{total}}$ (\textbf{State output}): Intensity output at the specified dopamine input, serotonergic context, reserve, and injury.
\end{eqvars}
\end{samepage}

The modeled state has two coordinates: $(C_{\mathrm{total}},\Psi_{\mathrm{eff}})$.
Intensity is not the sum or product of dopamine and serotonin signals.
The substrate ceiling acts on complexity.
Any additional effect on intensity needs an explicit coupling law.

This formulation implements the ceiling hypothesis.
It avoids multiplication of consciousness by a second, capacity factor.
The symbol $\Phi$ is not an established integrated-information measurement.
A differentiable alternative is $\Psi_{\max}\{1-e^{-\Psi_{\mathrm{drive}}/\Psi_{\max}}\}$ for positive capacity.
Its low-drive limit approaches $\Psi_{\mathrm{drive}}$.

\subsection{Coupling and measurement}
The two coordinates need not be independent:
\begin{samepage}
\begin{equation}
\begin{pmatrix}\dot x\\\dot h\end{pmatrix}
=\begin{pmatrix}-a_D&\kappa_{DH}\\\kappa_{HD}&-a_H\end{pmatrix}
\begin{pmatrix}x-x_0\\h-h_0\end{pmatrix}
+\begin{pmatrix}u_D\\u_H\end{pmatrix}.
\label{eq:coupling}
\end{equation}
\begin{eqvars}
\item $x_0,h_0$ (\textbf{Reference coordinates}): Reference coordinates.
\item $a_D,a_H$ (\textbf{Relaxation rates}): Local relaxation coefficients.
\item $\kappa_{DH}$ (\textbf{Serotonin-to-dopamine coupling}): Sets the influence of the serotonergic coordinate on the dopamine coordinate.
\item $\kappa_{HD}$ (\textbf{Dopamine-to-serotonin coupling}): Sets the influence of the dopamine coordinate on the serotonergic coordinate.
\item $u_D,u_H$ (\textbf{External inputs}): External inputs expressed as coordinate changes per unit time.
\item $\dot x,\dot h$ (\textbf{Coordinate rates}): Rates of coordinate change.
\end{eqvars}
\end{samepage}

The coupling coefficients can differ in magnitude and sign.
Their units depend on the specified coordinate scales.
For this linear form, local stability requires $a_D+a_H>0$ and $a_Da_H>\kappa_{DH}\kappa_{HD}$.
CSF correlations cannot identify these coefficients.

Measurements require their own functions:
\begin{samepage}
\begin{align}
Y_{\mathrm{sleep}}&=g(x,h,E,\text{circadian phase},\text{sleep pressure})+\epsilon_s,\\
Y_{\mathrm{report}}&=m(C,\Psi_{\mathrm{eff}},\text{memory},\text{report ability})+\epsilon_r,\\
Y_{\mathrm{CSF}}&=B\,z_{\mathrm{turnover}}+b_{\mathrm{clearance}}+\epsilon_c.
\label{eq:measurements}
\end{align}
\begin{eqvars}
\item $Y_{\mathrm{sleep}},Y_{\mathrm{report}},Y_{\mathrm{CSF}}$ (\textbf{Observed outcomes}): Observed sleep, experience-report, and metabolite outcomes.
\item $g,m$ (\textbf{Observation functions}): Observation functions for sleep and reported experience.
\item $B$ (\textbf{Metabolite mapping}): Mapping from transmitter turnover to measured metabolites.
\item $z_{\mathrm{turnover}}$ (\textbf{Transmitter turnover}): Vector of transmitter turnover quantities.
\item $b_{\mathrm{clearance}}$ (\textbf{Clearance contribution}): Contribution from clearance to measured metabolite values.
\item $\epsilon_s,\epsilon_r,\epsilon_c$ (\textbf{Observation errors}): Measurement and model errors.
\end{eqvars}
\end{samepage}
Shared turnover and clearance can generate metabolite covariance without identifying receptor signals.
PPMI sleep measurements can constrain $g$ under additional assumptions, but do not identify the full report function $m$.

\paragraph{Independent measurements of presence and content:}\label{sec:experiencemeasurement}
A prospective human test measures subjective presence, felt complexity, and felt significance separately.
Candidate rating instructions are:
\begin{itemize}
\item \textbf{Presence:} Rate how strongly any experience was present.
\item \textbf{Felt complexity:} Rate how complex and interconnected the contents felt.
\item \textbf{Felt significance:} Rate how meaningful or personally important the contents felt.
\end{itemize}

These instructions define proposed measurement targets.
Their reliability and separation require independent validation.
The presence rating targets $C$, while the two content ratings target $\Psi$.
Each rating uses fixed verbal anchors and a common reporting window across conditions.

Vividness, emotional intensity, confidence, recall, and response ability are recorded separately.
An absent response or absent recall is not automatically coded as absent experience.

For the primary content comparison, a transparent summary is
\begin{samepage}
\begin{equation}
Y_{\Psi}=\omega_{\Psi}Y_{\mathrm{cx}}+(1-\omega_{\Psi})Y_{\mathrm{sig}},
\qquad \omega_{\Psi}=\tfrac12.
\label{eq:contentscore}
\end{equation}
\begin{eqvars}
\item $Y_{\Psi}$ (\textbf{Content score}): Observed summary of the two content ratings, distinct from the latent coordinate $\Psi$.
\item $Y_{\mathrm{cx}},Y_{\mathrm{sig}}$ (\textbf{Content ratings}): Felt complexity and felt significance, each mapped to $[0,1]$ using fixed anchors.
\item $\omega_{\Psi}$ (\textbf{Observation weight}): Gives equal weight to both ratings in this proposed primary comparison. It is a measurement convention, not a receptor parameter.
\end{eqvars}
\end{samepage}
The two component ratings remain visible even when their summary is identical.
Flexibility is assessed separately, so rigidity does not force low felt significance or a low summary score.
Alternative weights must be declared before evaluation and cannot be selected for agreement with receptor measurements.

Independent calibration specifies how the presence rating and $Y_{\Psi}$ depend on their respective latent coordinates.
Those mappings and their error models then remain fixed across the evaluated interventions.
Calibration must test whether the same mapping applies across the selected states.
The normalization $C(1)=1$ sets a model unit, rather than a percentage scale of measured experience.
Ratio statements about subjective intensity require separate evidence for a ratio scale.
Tests can instead assess ordered changes and the regional sensitivity restrictions through a calibrated, locally nonzero observation slope.

\begingroup
\setlength{\intextsep}{8pt}
\setlength{\abovecaptionskip}{5pt}
\figinput{\begin{tikzpicture}[scale=1]
\tikzset{
    moderncard/.style={
        draw=none,
        fill=white,
        drop shadow={shadow xshift=0pt, shadow yshift=2pt, opacity=0.1},
        rounded corners=16pt,
        minimum width=7cm,
        minimum height=8cm
    },
    iconbg/.style={
        circle,
        minimum size=40pt,
        fill=#1!10,
        draw=#1!30,
        line width=1pt
    },
    modernheader/.style={
        font=\Large\bfseries\sffamily,
        text=#1!65!black
    },
    propertyrow/.style={
        font=\footnotesize\sffamily
    }
}
\node[moderncard] at (0,0) {};
\node[iconbg=dopaminecolor] at (0,4) {};
\node[dopaminecolor, font=\Large] at (0,4) {\faIcon{fire}};
\node[modernheader=dopaminecolor] at (0,2.9) {Dopamine System};
\draw[dopaminecolor!30, line width=1pt] (-3,2.45) -- (3,2.45);
\node[propertyrow, align=left] at (0,-0.7) {
    \begin{tabular}{@{}l@{\hspace{8pt}}l@{}}
    \textcolor{gray}{\faIcon{crosshairs}} & \textbf{Function}\\
    & Proposed intensity and access\\[6pt]
    \textcolor{gray}{\faIcon{network-wired}} & \textbf{Distribution}\\
    & Subcortical + Cortical networks\\[6pt]
    \textcolor{gray}{\faIcon{microchip}} & \textbf{Mechanism}\\
    & Receptor and circuit gain\\[6pt]
    \textcolor{gray}{\faIcon{clock}} & \textbf{Regulation}\\
    & Feedback and receptor adaptation\\[6pt]
    \textcolor{red!65!black}{\faIcon{exclamation-circle}} & \textbf{Risk}\\
    & \textcolor{red!65!black}{Context-dependent system failure}\\[6pt]
    \textcolor{orange!65!black}{\faIcon{vial}} & \textbf{Model Constraint}\\
    & \textcolor{orange!65!black}{Reserve-dependent boundary}
    \end{tabular}
};
\node[moderncard] at (8,0) {};
\node[iconbg=serotonincolor] at (8,4) {};
\node[serotonincolor, font=\Large] at (8,4) {\faIcon{brain}};
\node[modernheader=serotonincolor] at (8,2.9) {\emph{5-HT\textsubscript{2A}} System};
\draw[serotonincolor!30, line width=1pt] (5,2.45) -- (11,2.45);
\node[propertyrow, align=left] at (8.3,-0.7) {
    \begin{tabular}{@{}l@{\hspace{8pt}}l@{}}
    \textcolor{gray}{\faIcon{project-diagram}} & \textbf{Function}\\
    & Felt complexity and significance\\[6pt]
    \textcolor{gray}{\faIcon{brain}} & \textbf{Distribution}\\
    & Primarily cortical layers\\[6pt]
    \textcolor{gray}{\faIcon{code}} & \textbf{Mechanism}\\
    & Pathway-dependent modulation\\[6pt]
    \textcolor{gray}{\faIcon{rocket}} & \textbf{Regulation}\\
    & Onset, adaptation, and recovery\\[6pt]
    \textcolor{green!65!black}{\faIcon{shield-alt}} & \textbf{Risk}\\
    & \textcolor{green!65!black}{Bounded target response}\\[6pt]
    \textcolor{green!65!black}{\faIcon{infinity}} & \textbf{Model Constraint}\\
    & \textcolor{green!65!black}{Other hazards remain possible}
    \end{tabular}
};
\node[fill=gray!10, rounded corners=20pt, 
      minimum width=60pt, minimum height=30pt] at (4,0.5) {};
\draw[gray!40, line width=2pt, <->, >=stealth] (3.3,0.5) -- (4.7,0.5);
\node[font=\footnotesize\bfseries\sffamily, text=black!70] at (4,1.1) {COUPLED};
\begin{scope}[shift={(-1.5,-4.3)}]
    \fill[red!15] (0,0) rectangle (0.5,0.4);
    \fill[orange!40] (0.5,0) rectangle (1.5,0.4);
    \fill[green!20] (1.5,0) rectangle (3,0.4);
    \draw[gray!50, line width=0.5pt] (0,0) rectangle (3,0.4);
    \node[font=\tiny\sffamily, text=gray] at (0,-0.2) {Boundary};
    \node[font=\tiny\sffamily, text=gray] at (1.5,-0.2) {Reference};
    \node[font=\tiny\sffamily] at (1.5,-0.6) {Illustrative viable range};
\end{scope}
\begin{scope}[shift={(6.5,-4.3)}]
    \shade[left color=serotonincolor!10,right color=serotonincolor!60] (0,0) rectangle (3.0,0.4);
    \draw[gray!50, line width=0.5pt] (0,0) rectangle (3,0.4);
    \node[font=\tiny\sffamily, text=gray] at (0,-0.2) {0};
    \node[font=\tiny\sffamily, text=gray] at (3,-0.2) {High};
    \node[font=\tiny\sffamily] at (1.5,-0.6) {Target saturation only};
\end{scope}
\node[fill=dopaminecolor!8, draw=dopaminecolor!40, rounded corners=8pt,
      font=\tiny\sffamily\bfseries, inner sep=6pt, text=dopaminecolor!65!black] 
      at (0,-5.5) {Regional state control};
\node[fill=serotonincolor!8, draw=serotonincolor!40, rounded corners=8pt,
      font=\tiny\sffamily\bfseries, inner sep=6pt, text=serotonincolor!65!black] 
      at (8,-5.5) {Complexity modulation};
\node[opacity=0.03, font=\Huge] at (-2,-1) {\faIcon{microchip}};
\node[opacity=0.03, font=\Huge] at (10,-1) {\faIcon{code-branch}};
\node[font=\large\bfseries\sffamily, text=black!80] at (4,6) 
    {Neurotransmitter System Architecture};
\node[font=\footnotesize\sffamily, text=gray!75] at (4,5.5) 
    {Candidate roles in conscious state control};
\end{tikzpicture}}{Candidate roles of the two signaling systems. The cards contrast regional intensity control with complexity modulation. Their connection permits coupling. The lower strips illustrate a viable range and a saturating target response. Neither strip is a clinical safety scale.}{architecture}
\endgroup

\begin{landscape}
\section{The state space of consciousness}\label{sec:statespace}
\begin{figure}[H]\centering
\fitlandart{\definecolor{heldDAcolor}{HTML}{D9AD00}
\definecolor{heldHTcolor}{HTML}{16854B}
\begin{tikzpicture}[
    yscale=1,
    xscale=1.85,
    state/.style={
        rectangle,
        rounded corners=4pt,
        fill=none,
        draw=#1!70,
        line width=1.5pt,
        inner sep=4pt,
        font=\small\sffamily,
        align=center,
        minimum width=2cm,
        minimum height=0.8cm
    },
    keystate/.style={
        rectangle,
        rounded corners=6pt,
        fill=none,
        draw=#1!90,
        line width=2pt,
        inner sep=6pt,
        font=\footnotesize\sffamily\bfseries,
        align=center,
        minimum width=2.4cm,
        minimum height=1cm
    },
    danger/.style={
        rectangle,
        rounded corners=6pt,
        fill=none,
        draw=red!85,
        line width=2pt,
        inner sep=5pt,
        font=\footnotesize\sffamily\bfseries,
        align=center,
        text=red!90
    },
    quadrant/.style={
        font=\large\bfseries\sffamily,
        text=#1!65!black
    },
    axis/.style={
        font=\normalsize\bfseries\sffamily,
        fill=none,
        inner sep=3pt
    },
    path/.style={
        line width=2.5pt,
        opacity=0.5,
        ->,
        >=stealth,
        #1!70
    }
]
\fill[consciouscolor!6] (-7.5,0) rectangle (0,6.5);
\fill[flowcolor!6] (0,0) rectangle (7.5,6.5);
\fill[sleepcolor!6] (-7.5,-6.5) rectangle (0,0);
\fill[dopaminecolor!6] (0,-6.5) rectangle (7.5,0);
\fill[gray!18] (-7.5,-6.5) rectangle (-5.95,6.5);
\draw[gray!60,line width=1.5pt,dashed] (-5.95,-6.5)--(-5.95,6.5);
\node[font=\small\bfseries,text=gray!70!black,rotate=90] at (-6.72,2.8) {DEATH};
\fill[red!10] (5.9,-6.5) rectangle (7.9,6.5);
\draw[red!70,line width=2pt,dashed] (5.9,-6.5) -- (5.9,6.5);
\node[font=\small\bfseries,text=red!80,rotate=90] at (7,2.8)
 {COLLAPSE AND INJURY};
\foreach \y/\val in {-5/Low,5/High} {
 \draw[gray!45] (-.1,\y)--(.1,\y);
 \node[text=gray!70,font=\small,rotate=90] at (.35,\y) {\val};
}
\foreach \x/\val in {-5.1/Very low,-2/Low,3/High,5/Very high} {
 \draw[gray!45] (\x,-.1)--(\x,.1);
 \node[text=gray!70,font=\small] at (\x,-.5) {\val};
}
\node[quadrant=consciouscolor,font=\normalsize\bfseries\sffamily] at (-3.8,6.95)
 {DREAMING \& MEDITATIVE};
\node[quadrant=flowcolor,font=\normalsize\bfseries\sffamily] at (2.7,6.95)
 {FLEXIBLE \& ASSOCIATIVE};
\node[quadrant=sleepcolor,font=\normalsize\bfseries\sffamily] at (-3.8,-6.95)
 {LOW ACCESS \& WITHDRAWAL};
\node[quadrant=dopaminecolor,font=\normalsize\bfseries\sffamily] at (2.7,-6.95)
 {FOCUSED \& GOAL-DIRECTED};

\node[keystate=gray,fill=none] (baseline) at (0,0) {QUIET WAKE\\Reference};
\node[keystate=consciouscolor,text width=3cm] (rem) at (-3,3.1) {REM scenario\\Higher complexity};
\node[keystate=sleepcolor,text width=3cm] (remlo) at (-3,-1.8) {REM scenario\\Lower complexity};
\node[keystate=sleepcolor,text width=3cm] (sleep) at (-4.7,-4.5) {Low-input scenario\\Reduced state intensity};
\node[keystate=flowcolor,text width=3cm] (high) at (3.15,3.1) {Stronger intensity\\Higher complexity};
\node[keystate=dopaminecolor,text width=3cm] (focus) at (3.15,-3.1) {Stronger intensity\\Lower complexity};
\node[danger,text width=2.2cm] (terminal) at (6.9,-4.1) {Irreversible\\Injury\\$J\geq J_{\mathrm{irr}}$};
\draw[line width=2pt,gray!55] (-5.95,0) -- (baseline.west);
\draw[->,line width=2pt,gray!55] (baseline.east) -- (7.8,0)
 node[axis,right,align=center,fill=none,text=black!65,xshift=6pt] {Effective DA\\{\scriptsize Intensity axis}};
\draw[line width=2pt,gray!55] (0,-6.8) -- (baseline.south);
\draw[->,line width=2pt,gray!55] (baseline.north) -- (0,6.8)
 node[axis,above,align=center,fill=none,text=black!65] {Effective 5-HT\textsubscript{2A}\\{\scriptsize Complexity axis}};
\draw[draw=sleepcolor!85!black,line width=1.7pt,-{Stealth[length=6pt,width=4pt]},shorten >=3pt,shorten <=3pt]
 (baseline.south west) .. controls (-1.05,-3.5) and (-2.65,-4.5) .. (sleep.east);
\draw[draw=sleepcolor!85!black,line width=1.7pt,-{Stealth[length=6pt,width=4pt]},shorten >=3pt,shorten <=3pt]
 (sleep.north) .. controls (-4.7,-2.9) and (-4.4,-1.8) .. (remlo.west);
\draw[draw=sleepcolor!85!black,line width=1.7pt,-{Stealth[length=6pt,width=4pt]},shorten >=3pt,shorten <=3pt]
 (baseline.north west) .. controls (-1.5,.8) and (-1.55,2.0) .. (rem.south east);
\draw[draw=heldDAcolor,line width=3pt,{Stealth[length=7pt,width=5pt]}-{Stealth[length=7pt,width=5pt]},shorten >=3pt,shorten <=3pt]
 (remlo.north) -- (rem.south);
\node[fill=none,inner sep=2pt,font=\footnotesize\bfseries,rotate=90,text=heldDAcolor!65!black] at (-3.28,1.3) {Same DA input};
\draw[draw=heldHTcolor,line width=3pt,{Stealth[length=7pt,width=5pt]}-{Stealth[length=7pt,width=5pt]},shorten >=3pt,shorten <=3pt]
 (rem.east) -- (high.west);
\node[fill=none,inner sep=2pt,font=\footnotesize\bfseries,text=heldHTcolor!80!black] at (1.10,3.60) {Same 5-HT\textsubscript{2A} input};
\node[keystate=gray,text width=2.1cm,minimum width=0cm] (death) at (-6.72,-3.9)
 {Death\\Absent\\Experience};
\node[font=\tiny\sffamily,text=gray!80,align=center,text width=2.15cm] at (-6.72,-5.45)
 {Absorbing outcome.\\Not a DA threshold.};
\node[font=\tiny\sffamily,text=red!75,align=center,text width=2.6cm] at (6.9,-5.25) {Separate criterion.\\Outside the map.};

\node[fill=none,draw=gray!50,rounded corners=6pt,inner sep=8pt,
 font=\small\sffamily,align=center,text width=16cm] at (0,-8.4) {
 \textbf{Model directions within a viable rising branch}\\
 The same REM-entry input permits different complexity drives.\\
 Coordinates are ordinal. The wider experiential catalogue appears in Appendix~\ref{app:phenomenology}.
};
\end{tikzpicture}}
\caption{Core model directions on ordinal axes. Within the viable rising branch, intensity follows proposed dopamine signaling and complexity follows proposed \HT{} signaling. The yellow comparison fixes dopamine input. The green comparison fixes \HT{} input. Grey marks death as an absorbing outcome, not a dopamine threshold. Excess-related collapse uses separate state and injury mechanisms.}
\label{fig:statemap}
\end{figure}
\end{landscape}

Figure~\ref{fig:statemap} illustrates the theory through two aspects of experience.
The horizontal labels indicate very low, low, high, and very high proposed effective dopamine signaling.
The vertical ordering distinguishes sparse or selective content from richer experience.
The theory associates these directions with effective dopamine and \HT{} signaling, respectively.
The drawing uses this proposed correspondence rather than measured transmitter concentrations.

The main map shows comparisons implied by the proposed coordinates, including two REM scenarios at the same dopamine input.
Appendix~\ref{app:phenomenology} retains the detailed experiential catalogue as a hypothesis-generating illustration.
Its named experiences are not predicted drug rankings or fixed diagnostic positions.

These placements describe the viable rising branch of the state function.
Reserve, injury, history, and interactions define the wider quantitative model.
Excess-related collapse appears separately because greater input can eventually reduce intensity.
The grey death panel denotes the absorbing injury outcome outside the viable map.

Higher modeled intensity can coexist with poor reality testing or reduced task performance.
Likewise, rich experience can coexist with limited external responsiveness.
These distinctions require separate measures of content, access, memory, and motor response.

\paragraph{Regional patterns in psychosis:}\label{sec:psychosispatterns}
Psychosis does not imply globally elevated coordinated dopamine.
Howes and Kapur describe cortical dopamine deficiency with striatal excess as an earlier hypothesis \cite{HowesKapur2009}.
Their updated account emphasizes increased presynaptic striatal function and altered salience.
Neither account supplies calibrated values for the regional receptor signals in this model.

W0 makes a conditional prediction for such regional imbalance.
With prefrontal signal fixed at $0.60$, $x$ stays below $0.853$, even if all other regional signals increase without bound.
The pattern $(0.6,1,3,1,1)$ gives $x=0.797$ and $C=0.824$ under the reference conditions in Appendix~\ref{app:simulation}.
These inputs define a numerical scenario, rather than measured patient values.

This regional pattern cannot reach the high-input band in Figure~\ref{fig:continuum}.
That band describes a possible psychotic presentation with broadly elevated coordinated input.
Neither psychosis nor stimulant exposure alone establishes that regional pattern.
Lower intensity does not itself establish REM, which also requires the specified starting state and transition history.
Complexity follows the separate receptor-drive and substrate terms in Equation~\ref{eq:complexity}.
A lower-intensity, high-complexity presentation is therefore possible when those terms support it.

\paragraph{Regional psychosis and the two coordinates:}
The regional-deficit route gives a specific interpretation of the two-coordinate state.
A prefrontal deficit lowers the dopamine bottleneck and therefore lowers $C$, the proposed subjective-intensity coordinate.
Preserved or increased effective \HT{} signaling can raise $\Psi$, the proposed complexity coordinate.
The resulting state can combine reduced subjective presence with highly associative, salient, or meaning-laden content.

Here, $\Psi$ concerns how complex and significant the contents feel to the person (Section~\ref{sec:definitions}).
A fixed delusion can remain highly significant despite low flexibility.
Its measured profile can differ from a diffuse state with many personally meaningful associations.

The regional-deficit route differs from Kapur's dopamine-centred aberrant-salience account, which assigns the primary disturbance to striatal dopamine \cite{Kapur2003}.
Pimavanserin, a \HT{} inverse agonist \cite{Muneta2020}, gives a clinical precedent for a serotonergic contribution to psychotic content.
In a phase 3 trial in Parkinson's disease psychosis, it reduced hallucinations and delusions without blocking dopamine receptors or worsening motor function \cite{Cummings2014}.
This therapeutic effect supports receptor involvement, without identifying the patients' bottleneck signals or their $(C,\Psi)$ coordinates.
Dopaminergic and serotonergic contributions can coexist in the same presentation.

\paragraph{A prospective dissociation test:}
The fixed-response benchmark holds $\alpha_i$ and $\mu_i$ independent of serotonergic drive.
At matched dopamine bottleneck, reserve, injury, and substrate capacity, reduced effective \HT{} drive then lowers $\Psi$ below its ceiling while preserving $C$.
A blinded controlled comparison tests this prediction through the separate measurements in Section~\ref{sec:experiencemeasurement}.
It measures felt complexity and significance alongside hallucinations, delusions, reality testing, and subjective presence.
Reduced abnormal significance can occur without a comparable reduction in presence or a loss of flexibility.

The study declares a content-change target and a presence-equivalence margin before evaluation.
Preserved presence requires an interval within that margin.
A nonsignificant presence difference is insufficient.
Verified receptor-drive reduction without the predicted content change below the ceiling would challenge the proposed route.
A presence change beyond the equivalence margin would challenge the fixed-response separation under the matched conditions.
Coupled response functions must instead supply their own quantitative presence prediction before evaluation.

\paragraph{Anesthesia and upstream control:}
An intervention can change the proposed coordinates without binding directly to dopamine or \HT{} receptors.
Upstream circuits can alter regional release, occupancy, or signaling efficacy in Equations~\ref{eq:numerator}--\ref{eq:deff}.

GABA (gamma-aminobutyric acid) provides a concrete example.
In rat preparations, propofol increased inhibitory GABA input and reduced excitatory glutamate input to ventral periaqueductal gray dopamine neurons.
Their activity decreased, supporting an indirect route from anesthetic action to dopamine control \cite{Li2018Propofol}.
The proposed sequence places altered GABA transmission upstream of dopamine and reduced effective dopamine signaling upstream of conscious intensity loss.
Reduced release enters $q_r(t)$ in Equation~\ref{eq:occupancy}, which can lower occupancy and the regional bottleneck input.
The direction depends on the affected circuit, so GABA enhancement alone does not specify a universal dopamine decrease.

Dopamine-neuron stimulation restored righting and changed EEG activity during maintained isoflurane anesthesia in mice \cite{Taylor2016}.
This provides a concrete intervention for testing the proposed control route.

The stronger prediction concerns mediation.
The comparison fixes reserve, injury, substrate capacity, serotonergic drive, oxygen, temperature, and the other declared physiological inputs.
With matched state histories, anesthesia-related outcomes should follow the calibrated regional dopamine coordinate.
Restoring this coordinate should restore the predicted output on the same state branch.
Persistent suppression at matched modeled inputs would challenge the conditional sufficiency claim.

A useful experiment combines fixed anesthetic exposure with regional signaling measurements and selective dopamine stimulation.
Behavioral responses and neural state measures should be recorded together.
The comparison tests whether measured dopamine changes explain the state effect or accompany a separate recovery mechanism.
The same causal test applies to other receptor systems and ion channels.
Their effects must enter independently measured model inputs, with shared parameters across interventions.

\subsection{Regional heterogeneity and the prefrontal hypothesis}
Figure~\ref{fig:regions} links the anatomical hypothesis to the regional equations.
The VTA and substantia nigra represent candidate dopamine sources.
Cortical and striatal boxes represent candidate recipient circuits.
A source can change signals in other regions even when its own direct weight in the conscious-state readout is small.

The dots illustrate prior weights $p_r$, with a candidate prefrontal emphasis.
The receptor badges identify local measurements needed for $N_r$, $A_r$, and $\deff(r,t)$.
They do not specify measured receptor ratios.

Orange arrows identify selected dopamine projections.
Gray arrows identify selected corticostriatal projections.
SNc denotes the dopamine source in the substantia nigra pars compacta.
The striatal box combines recipient territories and does not imply that every projection reaches the same cells.
Animal tracing studies support distinct VTA and SNc projection systems \cite{Beier2015,Lerner2015}.

The prefrontal example concerns a mismatch between global activity and local functional capacity.
Other regions can have high signals while the integrated prefrontal signal is low.
The soft minimum then becomes more sensitive to the low region.
The weighted mean can remain high, so it can miss this regional limitation.

The limiting region depends on the selected function, regional scale, prior weights, and current signals.
These quantities must be specified independently of the outcome being predicted.
Source and recipient effects require care to avoid counting the same influence twice.
The regional readout summarizes available control signals rather than the integration of experienced contents.

A lower bottleneck coordinate implies lower conscious intensity only on a rising branch of the state function.
On a falling branch, the same change can have the opposite effect.
The regional mechanism determines the input coordinate, and Equation~\ref{eq:multistate} determines its state-dependent consequence.

\begin{prediction}[Regional rescue]
At matched global drive, the model predicts greater local sensitivity to restoration of a limiting region.
Recovery of conscious access also requires a positive calibrated observation slope on the tested response branch.
The prediction requires an independently identified limiting region and matched effects on motor response.
\end{prediction}

\figinput{\begin{tikzpicture}[scale=1.5]
\tikzset{
    regionbox/.style={
        draw=#1!80, 
        fill=#1!25, 
        rounded corners=6pt,
        inner sep=8pt,
        font=\small\bfseries,
        text width=2.4cm,
        align=center,
        drop shadow={shadow xshift=2pt, shadow yshift=-2pt, opacity=0.4}
    },
    glow/.style={
        fill=#1!20,
        circle,
        minimum size=25pt,
        opacity=0.3
    }
}
\shade[ball color=gray!10, opacity=0.5] 
    plot[smooth cycle, tension=0.8] coordinates {
    (-4.2,0) (-3.8,1.5) (-3.2,2.8) (-1.5,3.2) (0,3.0) (1.8,2.9) 
    (3.5,2.2) (4.0,1.0) (3.8,-0.5) (3.2,-2.0) (2.0,-2.8) 
    (0,-3.0) (-2.2,-2.7) (-3.5,-1.8) (-4.0,-0.8)
};
\draw[ultra thick, black!70, smooth] 
    plot[smooth cycle, tension=0.8] coordinates {
    (-4.2,0) (-3.8,1.5) (-3.2,2.8) (-1.5,3.2) (0,3.0) (1.8,2.9) 
    (3.5,2.2) (4.0,1.0) (3.8,-0.5) (3.2,-2.0) (2.0,-2.8) 
    (0,-3.0) (-2.2,-2.7) (-3.5,-1.8) (-4.0,-0.8)
};
\shade[left color=green!70, right color=green!50] 
    (-5.2,-2.5) rectangle (-4.6,-1.5);
\shade[left color=green!50, right color=yellow!50] 
    (-5.2,-1.5) rectangle (-4.6,-0.5);
\shade[left color=yellow!50, right color=orange!60] 
    (-5.2,-0.5) rectangle (-4.6,0.5);
\shade[left color=orange!60, right color=red!60] 
    (-5.2,0.5) rectangle (-4.6,1.5);
\shade[left color=red!60, right color=red!80] 
    (-5.2,1.5) rectangle (-4.6,2.5);
\draw[thick, black!60, line width=1.5pt] (-5.2,-2.5) rectangle (-4.6,2.5);
\foreach \y in {-2,-1,0,1,2} {
    \draw[black!60, line width=1pt] (-5.2,\y) -- (-5.0,\y);
}
\node[font=\footnotesize\bfseries, text=red!80] at (-4.9,2.8) {\faIcon{fire} HIGH};
\node[font=\footnotesize\bfseries, text=black!70, rotate=90] at (-5.5,0) {ILLUSTRATIVE WEIGHT};
\node[font=\footnotesize\bfseries, text=green!70] at (-4.9,-2.8) {\faIcon{shield-alt} LOW};
\node[glow=dopaminecolor] at (-2.8,1.6) {};
\node[regionbox=dopaminecolor] (pfc) at (-2.8,1.6) {
    \faIcon{brain}\\[3pt]
    PREFRONTAL\\CORTEX\\[3pt]
    \tikz{\fill[orange] circle (2pt);}\tikz{\fill[orange] circle (2pt);}\tikz{\fill[orange] circle (2pt);}\tikz{\fill[orange] circle (2pt);}
};
\node[glow=consciouscolor] at (2.8,1.6) {};
\node[regionbox=consciouscolor] (motor) at (2.8,1.6) {
    \faIcon{running}\\[3pt]
    MOTOR\\CORTEX\\[3pt]
    \tikz{\fill[orange] circle (2pt);}\tikz{\fill[orange] circle (2pt);}\tikz{\fill[orange] circle (2pt);}
};
\node[glow=serotonincolor] at (0,-0.15) {};
\node[regionbox=serotonincolor] (striatum) at (0,-0.15) {
    \faIcon{sitemap}\\[3pt]
    STRIATUM\\[3pt]
    \tikz{\fill[orange] circle (2pt);}\tikz{\fill[orange] circle (2pt);}
};
\node[glow=wakecolor] at (-2.8,-2.0) {};
\node[regionbox=wakecolor] (vta) at (-2.8,-2.0) {
    \faIcon{bolt}\\[3pt]
    VTA\\[3pt]
    \tikz{\fill[orange] circle (2pt);}
};
\node[glow=flowcolor] at (2.8,-2.0) {};
\node[regionbox=flowcolor] (sn) at (2.8,-2.0) {
    \faIcon{cog}\\[3pt]
    SUBSTANTIA\\NIGRA (SNc)\\[3pt]
    \tikz{\fill[orange] circle (2pt);}
};
\node[fill=white, draw=dopaminecolor!80, rounded corners=4pt, 
      font=\tiny\bfseries, inner sep=4pt, drop shadow] 
      at (-0.65,1.8) {$D_1,D_2$: measure};
\node[fill=white, draw=consciouscolor!80, rounded corners=4pt, 
      font=\tiny\bfseries, inner sep=4pt, drop shadow] 
      at (3.35,0.5) {$D_1,D_2$: measure};
\node[fill=white, draw=serotonincolor!80, rounded corners=4pt, 
      font=\tiny\bfseries, inner sep=4pt, drop shadow] 
      at (0,-1.1) {$D_1,D_2$: measure};
\tikzset{corticalpath/.style={gray!80,line width=2.8pt,-{Stealth[length=9pt,width=7pt]}},
         dapath/.style={dopaminecolor!85!black,line width=2.8pt,-{Stealth[length=9pt,width=7pt]}}}
\draw[corticalpath] (pfc.south east) to[out=-20,in=160] (striatum.north west);
\draw[corticalpath] (motor.south west) to[out=200,in=20] (striatum.north east);
\draw[dapath] (vta.north) to[out=110,in=250] (pfc.south);
\draw[dapath] (vta.north east) to[out=20,in=235] (striatum.south west);
\draw[dapath] (sn.north west) to[out=160,in=305] (striatum.south east);
\node[font=\Large\bfseries, text=black!80] at (0,4.2) 
      {Regional Bottleneck Hypothesis};
\node[font=\footnotesize, text=black!70] at (0,3.8) 
      {Candidate anatomy and model weights require independent measurements};
\node[draw=black!40, fill=white!95, rounded corners=6pt, 
      drop shadow={shadow xshift=2pt, shadow yshift=-2pt, opacity=0.3},
      inner sep=10pt, font=\footnotesize] at (5.8,1.8) {
    \begin{tabular}{@{}l@{}}
    \textbf{Assumed Weight:}\\[3pt]
    \tikz{\fill[orange] circle (2pt);}\tikz{\fill[orange] circle (2pt);}\tikz{\fill[orange] circle (2pt);}\tikz{\fill[orange] circle (2pt);} Very High\\[2pt]
    \tikz{\fill[orange] circle (2pt);}\tikz{\fill[orange] circle (2pt);}\tikz{\fill[orange] circle (2pt);} High\\[2pt]
    \tikz{\fill[orange] circle (2pt);}\tikz{\fill[orange] circle (2pt);} Medium\\[2pt]
    \tikz{\fill[orange] circle (2pt);} Low\\[4pt]
    \textbf{Icons:}\\[3pt]
    \faIcon{brain} Cortical processes\\[2pt]
    \faIcon{running} Motor control\\[2pt]
    \faIcon{sitemap} Circuit processes\\[2pt]
    \faIcon{bolt} Dopamine source\\[2pt]
    \faIcon{cog} Motor dopamine source\\[5pt]
    \textbf{Selected pathways:}\\[3pt]
    \textcolor{dopaminecolor!85!black}{$\longrightarrow$} Dopaminergic\\[2pt]
    \textcolor{gray!65}{$\longrightarrow$} Corticostriatal
    \end{tabular}
};
\definecolor{darkgreen}{RGB}{0,100,0}
\node[red!60, font=\tiny\bfseries, rotate=0] at (-2.8,2.6) {CANDIDATE LIMIT};
\node[darkgreen!80, font=\tiny\bfseries, rotate=0] at (2.8,-0.75) {REGIONAL INPUT};
\draw[red!50, dashed, line width=1pt, ->] (-4.6,2) to[out=0, in=180] (pfc.west);
\end{tikzpicture}}{Regional bottleneck hypothesis. Orange arrows show selected dopamine projections, and gray arrows show selected corticostriatal projections. Receptor badges identify measurements, while arrow widths do not encode connection strength. In W0, each regional dot counts as one unit, and the eleven units normalize to total weight one (Section~\ref{sec:fixedbenchmark}). Current influence also depends on integrated signals through Equation~\ref{eq:softweight}. Other projections and feedback paths are omitted.}{regions}

This prediction is stronger than an association between striatal biomarkers and fatigue.
A test needs regional signaling or a justified regional proxy, together with an experience-related outcome.
Training data must determine the weights, or the experimental plan must fix them before the intervention.
Choosing the critical region after each failed test would make the theory difficult to refute.

\subsection{Consciousness during a seizure}\label{sec:seizure}
The model permits both preserved and impaired consciousness during a seizure.
The outcome depends on which regional signals change, how long they change, and the operating branch of the state function.
Electrophysiological seizure activity and effective dopamine signaling are different variables.
Their relationship requires regional measurements.

When a seizure spares the limiting network, its bottleneck input can remain near the waking reference.
When it reduces signals across that network, Equation~\ref{eq:softmin} lowers the state input.
Equation~\ref{eq:multistate} then converts this change into a state output.
Loss occurs when that output crosses an independently specified experience-related boundary.
The rising branch permits loss from reduced input, while the upper falling branch permits loss from excessive coordinated input.

\paragraph{A numerical example:}
Use W0 weights and signals after the temporal filter in Equation~\ref{eq:memory}.
The reference vector gives $D_{\kappa,\mathrm{ref}}=1$, so $x=D_\kappa$.
Use the complete state function and reference parameters from Appendix~\ref{app:simulation}, with fixed serotonergic context, reference reserve, and no injury:
\begin{equation}
C_{\mathrm{ex}}(x)=C(x,h_{\mathrm{ref}},E_{\mathrm{ref}},0),
\qquad x=D_\kappa.
\label{eq:seizureexample}
\end{equation}
\begin{eqvars}
\item $C_{\mathrm{ex}}$ (\textbf{Illustrative state output}): Uses the same normalized response as Figure~\ref{fig:continuum}, with $C_{\mathrm{ex}}(1)=1$.
\item $C_{\mathrm{acc}}$ (\textbf{Experience-related boundary}): Separates preserved and impaired states under a specified observation model. The illustrative value $C_{\mathrm{acc}}=0.50$ is not a known physiological boundary.
\end{eqvars}
This parameter choice demonstrates possible model behavior, rather than a fitted seizure classifier.
Clinical evaluation must determine the state parameters and observation boundary before testing seizure outcomes.

\begin{table}[H]\centering\small
\caption{Preserved and impaired states using the shared state response. Signals follow the W0 regional order. The boundary is an illustrative observation criterion.}
\label{tab:seizureexample}
\begin{tabularx}{\textwidth}{@{}p{3.5cm}Y c c c@{}}\toprule
Input pattern & Signals (PFC, motor, striatum, VTA, SN) & $D_\kappa$ & $C_{\mathrm{ex}}$ & State \\ \midrule
Reference & $(1,1,1,1,1)$ & $1.000$ & $1.000$ & Preserved \\
Focal motor increase & $(1,2,1,1,1)$ & $1.078$ & $1.044$ & Preserved \\
Moderate mismatch & $(0.7,1.4,1,1,1)$ & $0.878$ & $0.904$ & Preserved \\
Distributed suppression & $(0.1,2,0.1,0.1,0.1)$ & $0.180$ & $0.072$ & Impaired \\
\bottomrule\end{tabularx}
\end{table}

A focal rise in the motor signal leaves the modeled state above the chosen boundary.
A moderate prefrontal reduction also preserves this state.
Distributed suppression lowers the output below the boundary even while the motor signal doubles.
Thus, high activity or signaling in one circuit can coexist with loss when other required circuits become limiting.
A short perturbation can have a smaller effect because the regional filter attenuates rapid changes.

Complete isolated prefrontal suppression has a W0 bottleneck floor of $0.245$ when the other signals remain at one.
That floor constrains the regional input, not the possible state output.
The shared state function gives $C_{\mathrm{ex}}(0.245)\approx0.119$, below its chosen observation boundary.
Section~\ref{sec:fixedbenchmark} derives this regional bound independently of the state function.
Figure~\ref{fig:seizure} illustrates the proposed local mismatch, while Table~\ref{tab:seizureexample} supplies complete regional inputs for the numerical example.

\paragraph{Graded impairment before state loss:}
The same equations permit graded changes before an observation boundary is crossed.
Consider W0 with the integrated striatal signal reduced from one to $0.50$, while all other signals remain at one.
The bottleneck becomes $0.807$, and Equation~\ref{eq:seizureexample} gives $C_{\mathrm{ex}}=0.835$.
The same reduction in the prefrontal signal gives a bottleneck of $0.700$ and an output of $0.716$.
Both outputs remain above the chosen $0.50$ boundary.

Thus, identical local reductions can have different state effects while both preserve the modeled state.
Motor impairment and conscious engagement need separate observation functions.
Larger deficits can make the affected region the limiting component.
Complete striatal suppression gives a bottleneck of $0.406$ and an output of $0.267$ in this example.
The proposed dissociation therefore concerns a finite perturbation range, not immunity to striatal failure.

Animal experiments show subcortical seizure activity with reduced cortical function across several regions \cite{Englot2009}.
Human recordings associate impaired responsiveness with bilateral frontoparietal slow activity \cite{Englot2010}.
These observations motivate distributed network disruption as a candidate route to impaired consciousness.
The dopamine hypothesis assigns a specific control mechanism to this broader pattern.

Excess ligand can also reduce effective local signaling when the receptor conditions in Appendix~\ref{app:receptorpeak} hold.
Increased demand can reduce reserve and viability through Sections~\ref{sec:regionalexcess} and \ref{sec:reservefeedback}.
These routes can combine with the regional bottleneck during prolonged seizures.
Selective measurements must distinguish them from synchronization, impaired communication, and other arousal mechanisms.

\figinput{\begin{tikzpicture}
\begin{axis}[
    width=16cm,
    height=9cm,
    xlabel={Time (illustrative units, onset at zero)},
    ylabel={Normalized signal or demand (\% reference)},
    xmin=-30, xmax=120,
    ymin=0, ymax=380,
    axis lines=left,
    grid=major,
    legend pos=north east,
    legend style={font=\footnotesize,fill=white,draw=none},
]
\fill[seizurecolor!20] (axis cs:0,0) rectangle (axis cs:60,380);
\node[seizurecolor, font=\small\bfseries] at (axis cs:30,360) {SIMULATED SEIZURE};
\addplot[ultra thick, dopaminecolor, smooth] coordinates {
    (-30,100) (-20,105) (-10,110) (-5,120) (0,140) (10,200) (20,260) 
    (30,300) (40,280) (50,240) (60,180) (80,140) (100,120) (120,105)
};
\addlegendentry{Subcortical signal (hypothesis)}
\addplot[ultra thick, consciouscolor, smooth] coordinates {
    (-30,100) (-20,102) (-10,105) (-5,108) (0,110) (5,95) (10,75) 
    (15,50) (20,30) (25,20) (30,15) (40,10) (50,8) (60,5) 
    (80,25) (100,60) (120,85)
};
\addlegendentry{PFC signal (hypothesis)}
\addplot[ultra thick, orange, dashed, smooth] coordinates {
    (-30,100) (-20,105) (-10,115) (0,130) (10,200) (20,280) 
    (30,320) (40,300) (50,260) (60,200) (80,150) (100,120) (120,110)
};
\addlegendentry{Local demand (hypothesis)}
\draw[red, ultra thick, dashed] (axis cs:0,0) -- (axis cs:0,380);
\node[red, font=\bfseries, rotate=90] at (axis cs:-5,230) {Seizure Onset};
\draw[red, ultra thick, dashed] (axis cs:60,0) -- (axis cs:60,380);
\node[red, font=\bfseries, rotate=90] at (axis cs:65,80) {Seizure End};
\end{axis}
\end{tikzpicture}}{Schematic seizure-related regional mismatch. The proposed subcortical signal and local demand rise while the prefrontal signal falls. Each trace has its own reference normalization. The drawing illustrates a candidate mechanism, with complete numerical input scenarios given separately in Table~\ref{tab:seizureexample}.}{seizure}

\section{Sleep architecture and REM mechanisms}
\subsection{Dopamine-dependent REM entry and awakening}\label{sec:rementrytheory}
The theory assigns REM entry to an increase in the global integrated effective dopamine coordinate.
Here, global means $D_\kappa$, the regional bottleneck after temporal integration, rather than the arithmetic mean of transmitter concentrations.
The starting state is NREM sleep within the viable domain.
The sole changing control input in this sleep-transition comparison is dopamine.
Effective \HT{} signaling specifies the complexity of the resulting dreams, rather than an additional REM gate.

A moderate sustained increase can support a dream state, while a larger sustained increase can support awakening.
On the viable rising branch, the larger coordinated input also produces greater modeled intensity at fixed context.
The proposed waking boundary concerns an observed sleep transition, not a universal threshold for having experience.
Dreaming can occur during both REM and NREM sleep \cite{Siclari2017,Solms2000}.

A finite entry time distinguishes residence in a REM range from passage through it toward wakefulness.
Let $t_0$ mark the start of an eligible NREM interval, with $x(t_0)<x_{\mathrm{REM}}$.
Define the two input ranges and their uninterrupted residence times:
\begin{samepage}
\begin{equation}
\begin{aligned}
\mathcal I_{\mathrm{REM}}&=[x_{\mathrm{REM}},x_{\mathrm W}),
&\mathcal I_{\mathrm W}&=[x_{\mathrm W},\infty),\\
T_j(t)&=\sup\{\ell\in[0,t-t_0]:
x(u)\in\mathcal I_j\ \text{for all }u\in(t-\ell,t]\},\\
t_j&=\inf\{t>t_0:T_j(t)\geq\Delta_j\},
&j&\in\{\mathrm{REM},\mathrm W\}.
\end{aligned}
\label{eq:rementryrule}
\end{equation}
\begin{eqvars}
\item $t_0$ (\textbf{NREM starting time}): Start of the viable NREM comparison, before either candidate exit.
\item $x=D_\kappa/D_{\kappa,\mathrm{ref}}$ (\textbf{Coordinated input}): Regional bottleneck on the waking-reference scale.
\item $x_{\mathrm{REM}},x_{\mathrm W}$ (\textbf{Sleep-transition boundaries}): Fixed boundaries with $0<x_{\mathrm{REM}}<x_{\mathrm W}$. Contemporaneous \HT{} drive does not shift them in this comparison.
\item $\mathcal I_j$ (\textbf{Entry range}): Input range for REM or waking entry, restricted to the declared viable rising domain.
\item $T_j(t)$ (\textbf{Residence time}): Uninterrupted time in range $j$ since $t_0$. Leaving that range resets its residence time to zero.
\item $\Delta_j$ (\textbf{Entry duration}): Positive duration required in range $j$, fixed before evaluation.
\item $t_{\mathrm{REM}},t_{\mathrm W}$ (\textbf{Candidate entry times}): First completed residence requirement for each exit. An unmet requirement gives an infinite candidate time.
\end{eqvars}
\end{samepage}
The first completed requirement selects the predicted exit from NREM.
A rapid passage through the REM range can therefore produce waking without a completed REM episode.
A plateau within that range instead permits REM entry.
The rule stops at the first exit.
It does not specify later maintenance, downward transitions, or collapse.

It uses the dopamine trajectory and fixed entry durations, without adding a serotonergic entry gate.
The boundaries, durations, reference state, and observation resolution must be fixed independently before evaluation.
The waking reference $x=1$ is not itself a fitted waking boundary.
Observed entry of the wrong type, or entry without its residence requirement, would contradict this rule within the declared domain.
An unmet predicted exit would also contradict it after accounting for the fixed observation resolution.

\paragraph{A worked distinction between REM and waking:}
Consider equal regional signals initially at $0.20$, with a constant receptor-input step to $d$.
The common regional filter gives $x(t)=d+(0.20-d)e^{-(t-t_0)/\tau_m}$.
For this example, set $x_{\mathrm{REM}}=0.45$, $x_{\mathrm W}=0.80$, $\Delta_{\mathrm{REM}}=1.5\tau_m$, and $\Delta_{\mathrm W}=0.25\tau_m$.

With $d=0.60$, REM entry occurs at $t-t_0=2.481\tau_m$, and the waking range is never reached.
With $d=1.00$, passage through the REM range lasts only $1.012\tau_m$, shorter than its entry duration.
Waking then occurs at $t-t_0=1.636\tau_m$, without a completed REM episode.
These values demonstrate the distinction with one filter and one fixed rule.
They are not physiological estimates.

In mice, a dopamine transient in the basolateral amygdala initiates REM through Drd2-expressing neurons \cite{Hasegawa2022}.
Local serotonin inhibition can release this dopamine increase \cite{Hasegawa2024}.
The theory treats this local event as a possible upstream controller of its global transition coordinate.
These studies establish a regional REM mechanism, while the global sufficiency claim requires wider regional measurements.
Neither study measured dream content.

VTA dopamine-neuron activity also rises before NREM-to-wake transitions, and optogenetic stimulation can initiate waking \cite{EbanRothschild2016}.
These findings motivate different destinations for dopamine increases, rather than identifying a measured global boundary.
Simultaneous regional measurements must test whether the destinations follow the proposed magnitude and residence requirements.

Solms linked dreaming to a probably dopaminergic forebrain mechanism that can operate independently of REM sleep \cite{Solms2000}.
His lesion and pharmacological evidence separates dream experience from REM measures.
This is a direct precedent for measuring sleep stage and experience separately.
The present global dopamine threshold for REM entry is a further hypothesis.
The proposed \HT{} relation to dream complexity is also a separate claim.

In the schematic global trajectory, all five integrated W0 signals share the same value, so their bottleneck equals that value.
The illustrated boundaries are $x_{\mathrm{REM}}=0.45$ and $x_{\mathrm W}=0.80$.
For the schematic time axis, use $\tau_m=1$ minute and the entry durations from the worked example.
The marked entry times are calculated from the drawn input trajectory.
Figure~\ref{fig:sleep} separates the proposed global dopamine rise from the subsequent range of serotonergic drive.
Both low-complexity and high-complexity REM states are possible under the same dopamine entry rule.

\figinput{\input{figures/06_sleep.tex}}{Schematic sleep transitions and alternative receptor-drive scenarios. Panel (a) shows coordinated dopamine, a separate BLA transient, and the REM and waking boundaries. Dots mark completed entry requirements, using 1.5 minutes for REM and 0.25 minutes for waking. Shading gives schematic stage intervals, with later maintenance and downward transitions left unspecified. Panel (b) compares lower and higher effective \HT{} drive at the same ligand input. Both receptor scenarios permit REM under Equation~\ref{eq:rementryrule}.}{sleep}

\subsection{Effective 5-HT2A signaling and dream complexity}\label{sec:remcomplexity}
The theory predicts greater felt complexity or significance of dream content when effective \HT{} signaling increases under otherwise matched conditions.
Lower effective signaling predicts a lower content score within the same state.
REM entry alone therefore does not specify whether dreams are vivid, strange, or sparse.
The comparison must separate complexity from perceived intensity, recall, and external responsiveness.

This distinction also gives a possible bridge between dreaming and psychosis.
In the regional-deficit example $(0.6,1,3,1,1)$, the model gives $x=0.797$ and $C=0.824$.
Preserved or increased effective \HT{} signaling can raise $\Psi$ at this lower-than-reference intensity.
The result is a waking, dream-like content profile with richer associations and greater felt significance.
This similarity concerns content, not REM generation or sleep-stage identity.

A diffuse early psychotic state may combine rich associations with strong felt significance.
A consolidated delusion may retain strong significance despite reduced diversity or flexibility.
Section~\ref{sec:experiencemeasurement} therefore retains both content ratings alongside their fixed summary.
Flexibility and reality testing remain separate outcomes.

Constitutive \HT{} activity provides a ligand-independent component of receptor signaling.
It has been measured in human cortical tissue, with pathway-dependent effects \cite{Muneta2020}.
Its relative share can increase when ligand concentration falls, even if its absolute contribution stays constant.
An increase in total drive requires a larger gain, reduced attenuation, or another change in the response factors.

An optional state-dependent gain gives the following REM complexity function:
\begin{samepage}
\begin{equation}
\Psi_{\mathrm{REM}}=\Psi_0
\{1+\lambda_cf_{\mathrm{const}}G(q)+\lambda_LO_H\}
e^{-\gamma(x/x_s)^2}R,\qquad G(q)>0.
\label{eq:remconstitutive}
\end{equation}
\begin{eqvars}
\item $\Psi_{\mathrm{REM}}$ (\textbf{REM complexity drive}): Proposed complexity drive during REM.
\item $q$ (\textbf{State-control process}): Independently measured state-control process.
\item $G(q)$ (\textbf{Constitutive gain}): Positive gain on the constitutive component, distinct from the temporal response factor $R$.
\end{eqvars}
\end{samepage}
At fixed $x$, $R$, and substrate capacity, higher positive ligand-dependent or constitutive drive increases predicted complexity.
If ligand occupancy falls by $|\Delta O_H|$, a gain increase can offset it when $\lambda_cf_{\mathrm{const}}\Delta G>\lambda_L|\Delta O_H|$.
This gives a concrete route to complex dreams during low ligand availability.
The gain increase is an optional, testable response, not a requirement for REM.
With unchanged gain, falling ligand input lowers this effective drive.

Figure~\ref{fig:sleep} uses $\lambda_cf_{\mathrm{const}}=0.40$, $\lambda_L=0.60$, and waking occupancy $O_H=0.80$.
Reference ligand concentration is four times $K_H$, with occupancy given by Equation~\ref{eq:htoccupancy}.
The receptor bracket is normalized by its waking value, $1+0.40+0.60(0.80)=1.88$.
One scenario keeps $G=1$, while another raises $G$ to three during REM.
The same ligand decline produces different effective signals because receptor gain differs.

For example, let occupancy fall from $0.80$ to $0.20$.
At $G=1$, normalized receptor drive becomes $1.52/1.88\approx0.809$.
At $G=3$, it becomes $2.32/1.88\approx1.234$.
The number of receptors need not increase in either case.
The difference concerns signaling per available receptor, which must be measured independently.

Dream complexity then follows this drive with the dopamine attenuation, temporal response, and substrate ceiling already specified.
A high receptor drive predicts greater complexity until that ceiling applies.
A low drive predicts simpler contents despite preserved REM entry.
Rat microdialysis supports lower ligand availability during REM, but does not measure this effective receptor drive or dream complexity \cite{Portas1998}.

\begin{prediction}[Constitutive contribution]
A selective inverse agonist reduces the proposed REM complexity effect more than a neutral antagonist at matched relevant receptor occupancy.
The comparison requires matched off-target effects and direct measurements of dream content or an independently validated neural outcome.
\end{prediction}

Pathway-specific assays must establish how the interventions affect constitutive and ligand-driven signaling.
A complementary test compares dream complexity at matched REM timing and intensity, using independently measured \HT{} drive.
The theory predicts greater complexity with greater effective drive until the substrate ceiling applies.
These comparisons distinguish REM generation from the proposed control of dream content.

\subsection{Transition dynamics and hysteresis}\label{sec:transition}
A dynamical extension can specify state transitions:
\begin{samepage}
\begin{equation}
\tau_z\dot z=u(x,h,p)-a z+bz^3-cz^5,
\qquad a,b,c,\tau_z>0.
\label{eq:transition}
\end{equation}
\begin{eqvars}
\item $z$ (\textbf{Transition state}): Latent transition state, not necessarily conscious intensity.
\item $u(x,h,p)$ (\textbf{Transition input}): Net input from the control coordinates and sleep pressure $p$.
\item $a,b,c$ (\textbf{Feedback coefficients}): Positive feedback coefficients, with units set by the chosen scale of $z$.
\item $\tau_z$ (\textbf{Transition time}): Positive transition time constant.
\end{eqvars}
\end{samepage}

Some parameter values permit multiple stable states and different transition points during rising and falling input.
Other parameters give one stable state and no persistent hysteresis.
The model must determine which regime fits the observations.

For the sleep-specific comparison, the input is restricted to $u=u_{\mathrm{sleep}}(x)$, with the other control inputs fixed.
A full implementation must specify NREM, REM, and waking observation sets in $z$ before evaluation.
It must predict whether REM completes before waking under each tested trajectory.

The residence rule in Equation~\ref{eq:rementryrule} provides a reduced alternative for these two exits.
It is not derived uniquely from the polynomial.
Each implementation has its own fixed transition times.
The two delays must not be added to the same predicted event.
Neither implementation adds contemporaneous serotonergic drive as a sleep-entry gate in this restricted comparison.

This extension complements the cumulative intensity response and its collapse gate.
Equation~\ref{eq:multistate} gives state intensity at given inputs.
Equation~\ref{eq:transition} gives the dynamics of transitions between states.
Empirical models can compare memory alone with memory plus multiple stable states.
The latter must predict new transition paths, rather than only fit a loop already observed.

\subsection{Formulation and the timing of REM}
The entry rule predicts REM or waking from the measured trajectory of $x(t)$ and its time in each range.
A formulation can shift that trajectory through absorption, clearance, autoreceptor feedback, and regional receptor effects.
Daily dose or release formulation alone does not determine which entry requirement completes first.

The observational analyses therefore compare formulation with REM entry, persistence, and conventional timing.
They identify treatment-associated timing patterns that a mechanistic exposure model must explain.
A direct test requires time-resolved global signaling and independently fixed transition boundaries.

\section{Exploratory observational tests}\label{sec:sleepresults}
\subsection{Sleep data and exposure}
PPMI provided wearable sleep records, medication logs, clinical assessments, and imaging records \cite{PPMI2011,PPMIUserGuide}.
The sleep records cover December 2018 to January 2021.
Initial processing retained 95,434 of 106,419 nights from 348 participants.
The analyses below use smaller samples with outcome-specific eligibility rules.
Both timing analyses require positive recorded total levodopa LEDD.

The wearable classifier has reported four-stage accuracy of 73\% in 29 healthy adults \cite{Verily2025}.
REM-timing accuracy in Parkinson's disease needs separate validation.
The classifier uses accelerometer and photoplethysmography signals \cite{Verily2025}.
The outcomes measure classified sleep, with possible medication and autonomic effects on classification.

If treatment changes movement or autonomic signals used by the classifier, staging error can differ by exposure.
REM sleep behavior disorder can further alter nocturnal movement in Parkinson's disease \cite{Zoetmulder2016}.
Validation must therefore compare treatment conditions against polysomnography, rather than assume that classification error is nondifferential.

Medication reconstruction removed exact duplicate intervals and flagged unresolved doses.
An absent medication log does not establish zero exposure.
Prescription dates have monthly resolution, without administration time or verified adherence.

The operational formulation category groups modified-release and other delivery labels, as specified in Appendix~\ref{app:sleepmethods}.
Release formulations can differ in absorption and bioavailability \cite{SinemetCR}.
No observed equal-total-dose switch occurred among 108 recorded regimen transitions with record gaps of at most 40 days.
The comparison is therefore an observational regression contrast.

Confounding by indication is a specific concern.
Nocturnal motor symptoms and wearing-off can influence treatment choice and also predict sleep outcomes.
Controlled-release levodopa has been studied for nocturnal akinesia, without consistent improvement in polysomnographic sleep structure \cite{Wailke2011}.
Participant effects remove stable differences, but do not remove changing nocturnal symptoms that influence prescriptions.
A randomized, exposure-verified comparison would address this pathway directly.

\subsection{Shared model and uncertainty}\label{sec:statisticalmodels}
For participant $i$ and night $t$, the formulation model is
\begin{samepage}
\begin{equation}
Y_{it}=a_i+\beta_T\frac{L_{it}}{100}
+\beta_E\frac{E_{it}}{100}+\boldsymbol\gamma^\mathsf{T}Z_{it}+e_{it}.
\label{eq:formulation}
\end{equation}
\begin{eqvars}
\item $i,t$ (\textbf{Observation indices}): Participant and night indices in this statistical model.
\item $Y_{it}$ (\textbf{Sleep outcome}): Selected nightly sleep outcome.
\item $a_i$ (\textbf{Individual baseline}): Participant-specific intercept.
\item $L_{it}$ (\textbf{Total dose}): Total levodopa LEDD, in mg/day.
\item $E_{it}$ (\textbf{Formulation dose}): Operational formulation component of that dose, in mg/day LEDD.
\item $\beta_T,\beta_E$ (\textbf{Dose coefficients}): Total-dose and formulation coefficients per 100 mg/day LEDD.
\item $Z_{it},\boldsymbol\gamma$ (\textbf{Covariates}): Covariate vector and its coefficients.
\item $e_{it}$ (\textbf{Residual}): Model residual.
\end{eqvars}
\end{samepage}

Here, $E_{it}$ denotes medication exposure, not theoretical energy reserve.
The coefficient $\beta_E$ compares 100 mg/day LEDD assigned to the formulation component at fixed total levodopa LEDD.
Covariates include onset sine and cosine, interval duration, stage coverage, agonist dose, calendar month, and weekday.
A second specification adds a separate linear time trend for each participant.

Pointwise 95\% intervals use participant-cluster standard errors and a $t$ reference with $G-1$ degrees of freedom.
Null-imposed wild cluster tests use 9,999 draws.
Holm adjustment covers the six-test entry family and the eight-test conventional timing family.
Both include the two temporal specifications.
Outcome definitions and model specifications were selected during exploratory analysis of this database.
The adjustments control multiplicity within these families, not across the wider analytical search.

\subsection{Sustained REM entry and persistence}
Formulation-dependent exposure can alter REM timing and the duration of episodes after entry.

Reconstruction starts from raw 30-second epochs and preserves unknown stages and gaps.
Sleep onset requires ten consecutive sleep epochs with exact timestamp continuity.
Eligible intervals last 4 to 12 hours, with at least 80\% scored coverage.
The entry analysis additionally requires 95\% coverage during the first 120 minutes.

Sustained entry requires six consecutive REM epochs.
Waiting ends at completion of these three minutes and is capped at 120 minutes from sleep onset.
Nights without completed entry by that limit receive 120 minutes.
The sample contains 10,537 nights from 137 participants, including 19 nights without REM.
On 3,077 nights, sustained entry was not completed before the limit.
Deep-sleep waiting uses the same six-epoch requirement and 120-minute limit.

Persistence measures the additional duration of the first qualifying episode, capped at 20 minutes.
An observed stage exit ends the interval.
A gap or recording end causes censoring.
The analysis excludes 21 early-censored intervals and 70 nights without a qualifying episode.
Its sample contains 10,446 nights from 137 participants.
Conditioning on successful entry can introduce selection bias.

The waiting-time contrast is $+1.220$ minutes per 100 mg/day formulation substitution (95\% interval $0.425$ to $2.014$).
Its wild-bootstrap $p$ value is 0.0091.
Individual trends reduce the estimate to $+0.905$ minutes ($-0.037$ to $1.847$), with $p=0.124$.
The six-test family comprises REM waiting, REM persistence, and deep-sleep waiting under both temporal specifications.

The minimum Holm-adjusted $p$ value is 0.0546.
Table~\ref{tab:mechanismresults} reports all six tests and the two secondary stage contrasts.
Figure~\ref{fig:rementry} compares their magnitudes and uncertainty.

\figinput{\begin{tikzpicture}
\begin{axis}[
 width=11.8cm,height=7.1cm,
 axis lines=left,axis line style={black!45},
 xmin=-1.3,xmax=2.7,ymin=-.48,ymax=3.48,
 xtick={-1,0,1,2},ytick={0,1,2,3},
 yticklabels={REM minus deep waiting,Deep-sleep waiting,REM persistence,Sustained REM waiting},
 tick label style={font=\small},yticklabel style={align=right},
 xlabel={Contrast (minutes per 100 mg/day substitution)},
 label style={font=\small},xmajorgrids,grid style={black!8},
 legend style={at={(.5,1.06)},anchor=south,draw=none,fill=none,
   font=\small,legend columns=2,column sep=10pt},
 clip=false]
\draw[black!50,dashed] (axis cs:0,-.48)--(axis cs:0,3.48);
\draw[black!20,densely dotted] (axis cs:-1.3,.5)--(axis cs:2.7,.5);

\addplot[only marks,mark=*,mark size=2.7pt,sleepcolor,line width=1pt,error bars/.cd,x dir=both,x explicit] coordinates {
(1.219675140874,3.12) += (0.794802119353,0) -= (0.794802119353,0)
(-0.184662633174,2.12) += (0.286157066826,0) -= (0.286157066826,0)
(0.142559982054,1.12) += (1.196639905540,0) -= (1.196639905540,0)
(1.077115158820,0.12) += (1.384813253174,0) -= (1.384813253174,0)
};
\addlegendentry{Calendar controls}
\addplot[only marks,mark=diamond*,mark size=2.7pt,wakecolor,line width=1pt,error bars/.cd,x dir=both,x explicit] coordinates {
(0.905152980870,2.88) += (0.942020719287,0) -= (0.942020719287,0)
(-0.290067599890,1.88) += (0.396530739694,0) -= (0.396530739694,0)
(0.389960620529,0.88) += (1.032332235711,0) -= (1.032332235711,0)
(0.515192360341,-0.12) += (0.973163004731,0) -= (0.973163004731,0)
};
\addlegendentry{Individual trends}
\end{axis}
\end{tikzpicture}}{Sustained REM entry, persistence, and comparison outcomes (137 participants). Symbols show formulation contrasts with pointwise 95\% participant-cluster intervals. Positive values indicate later entry or longer persistence. Waiting outcomes use 10,537 nights, and persistence uses 10,446. The dotted line separates the secondary stage difference from the six-test family. The minimum Holm-adjusted $p$ value is 0.0546.}{rementry}

\begin{table}[H]\centering\small
\caption{Restricted waiting and persistence contrasts. Effects are minutes per 100 mg/day LEDD assigned to the formulation component at fixed total levodopa LEDD. All rows have 137 participants. Intervals are pointwise participant-cluster intervals. The six-test Holm family uses null-imposed wild cluster tests with 9,999 draws. The direct stage difference is a secondary contrast.}
\label{tab:mechanismresults}
\begin{tabular}{@{}llrrll@{}}\toprule
Model & Outcome & Nights & Effect & 95\% interval & $p_{\rm Holm}$ \\\midrule
Calendar & REM waiting & 10,537 & $1.220$ & $0.425$ to $2.014$ & 0.0546 \\
Calendar & REM persistence & 10,446 & $-0.185$ & $-0.471$ to $0.101$ & 1.0000 \\
Calendar & Deep-sleep waiting & 10,537 & $0.143$ & $-1.054$ to $1.339$ & 1.0000 \\
Calendar & REM minus deep & 10,537 & $1.077$ & $-0.308$ to $2.462$ & Not in family \\
Trends & REM waiting & 10,537 & $0.905$ & $-0.037$ to $1.847$ & 0.6210 \\
Trends & REM persistence & 10,446 & $-0.290$ & $-0.687$ to $0.106$ & 1.0000 \\
Trends & Deep-sleep waiting & 10,537 & $0.390$ & $-0.642$ to $1.422$ & 1.0000 \\
Trends & REM minus deep & 10,537 & $0.515$ & $-0.458$ to $1.488$ & Not in family \\
\bottomrule\end{tabular}\end{table}

Persistence and direct stage contrasts have intervals spanning zero.
They therefore permit changes in maintenance and timing effects shared with deep sleep.
The secondary stage contrasts are outside the six-test family.

The formulation component varies in 36 participants.
Information concentration indices are 10.46 and 8.76 participants under the two temporal specifications.
These indices describe concentrated exposure information, not effective degrees of freedom.
The largest participant contributes 21.1\% and 24.5\% of residualized exposure information, respectively.
Medication exclusions retain the same general pattern.

\paragraph{Participant-deletion robustness:}
Only 36 participants have within-participant formulation variation.
All 137 single-participant deletion coefficients remain positive in each specification, but most deletions remove no participant with that variation.
The calendar estimates range from $+0.992$ to $+1.363$ minutes.
The individual-trend estimates range from $+0.482$ to $+1.174$ minutes.

Removing the largest information contributor gives $+0.992$ ($0.123$ to $1.862$) with calendar adjustment.
With trends, it gives $+0.482$ ($-0.308$ to $1.272$), compared with $+0.905$ in the full sample.
Joint deletion of the two largest contributors gives $+1.067$ and $+0.728$ minutes, respectively.
Joint deletion of the three largest gives $+1.203$ and $+0.878$ minutes.
All four joint-deletion intervals include zero (Table~\ref{tab:lopo}).

The rankings use full-sample residualized exposure information and remain fixed for these nested deletions.
They are specified separately for each temporal model.
The top three contributors supply 45.2\% and 49.7\% of original exposure information, respectively.
The coefficients remain positive, while the intervals show limited precision and the single-deletion result shows substantial magnitude sensitivity.
These influence checks do not provide independent replications or resolve confounding by indication.

\begin{table}[H]\centering\small
\caption{Sustained-entry influence of the largest exposure-information contributors. Effects are minutes per 100 mg/day LEDD substitution, with pointwise 95\% participant-cluster intervals.}
\label{tab:lopo}
\begin{tabularx}{\textwidth}{@{}Y Y Y@{}}
\toprule
Participants removed & Calendar adjustment & Individual trends \\
\midrule
None & $+1.220$ ($0.425$ to $2.014$) & $+0.905$ ($-0.037$ to $1.847$) \\
Largest contributor & $+0.992$ ($0.123$ to $1.862$) & $+0.482$ ($-0.308$ to $1.272$) \\
Two largest & $+1.067$ ($-0.002$ to $2.135$) & $+0.728$ ($-0.194$ to $1.651$) \\
Three largest & $+1.203$ ($-0.072$ to $2.478$) & $+0.878$ ($-0.213$ to $1.969$) \\
\bottomrule
\end{tabularx}
\par\smallskip\noindent\footnotesize Rankings are fixed from the full sample, separately for each model. All refits retain the original eligibility, outcome, and covariate rules. These are influence diagnostics, not additional independent tests.
\end{table}

\subsection{Conventional timing sensitivity}
A conventional timing analysis uses 10,582 REM-containing nights from 137 participants.
The calendar contrasts are $+0.402$ percentage points for REM phase and $+2.10$ minutes for first-epoch latency.
Their 95\% intervals are $0.112$ to $0.691$ and $0.71$ to $3.49$.
Individual trends give $+0.331$ points ($-0.021$ to $0.683$) and $+1.41$ minutes ($-0.28$ to $3.10$).

The minimum eight-test Holm-adjusted $p$ value is 0.130.
These overlapping records provide a consistency check, rather than independent replication.
Appendix~\ref{app:sleepmethods} gives the complete family and its sensitivities.
Figure~\ref{fig:formulation} compares the observed timing contrasts under both temporal specifications.

\figinput{\begin{tikzpicture}
\begin{groupplot}[group style={group size=2 by 1,horizontal sep=1.9cm},width=7.1cm,height=5cm,axis lines=left,xmajorgrids,grid style={black!8},tick label style={font=\small},label style={font=\small},title style={font=\small},ymin=-.5,ymax=1.5,ytick={0,1}]
\nextgroupplot[xmin=-.10,xmax=.80,title={A. REM phase},xlabel={Percentage points},yticklabels={Individual trends,Calendar}]
\draw[black!45,dashed] (axis cs:0,-.5)--(axis cs:0,1.5);
\addplot[only marks,mark=*,mark size=2.6pt,sleepcolor,line width=1pt,error bars/.cd,x dir=both,x explicit] coordinates {(0.401849876540,1) += (0.289570286562,0) -= (0.289570286562,0)};
\addplot[only marks,mark=*,mark size=2.6pt,wakecolor,line width=1pt,error bars/.cd,x dir=both,x explicit] coordinates {(0.331143267853,0) += (0.351889952085,0) -= (0.351889952085,0)};
\nextgroupplot[xmin=-.45,xmax=3.8,title={B. REM latency},xlabel={Minutes},yticklabels={,}]
\draw[black!45,dashed] (axis cs:0,-.5)--(axis cs:0,1.5);
\addplot[only marks,mark=*,mark size=2.6pt,sleepcolor,line width=1pt,error bars/.cd,x dir=both,x explicit] coordinates {(2.103163617254,1) += (1.389857652745,0) -= (1.389857652745,0)};
\addplot[only marks,mark=*,mark size=2.6pt,wakecolor,line width=1pt,error bars/.cd,x dir=both,x explicit] coordinates {(1.409834635720,0) += (1.693965706643,0) -= (1.693965706643,0)};
\end{groupplot}
\end{tikzpicture}}{Recorded formulation category and REM timing. Effects are per 100 mg/day LEDD substitution at fixed total levodopa LEDD, using 10,582 nights from 137 participants. Lines show pointwise 95\% participant-cluster intervals. The minimum eight-test Holm-adjusted $p$ value is 0.130. The observations do not measure the global dopamine crossing in Equation~\ref{eq:rementryrule}.}{formulation}

The sustained-entry contrast is the strongest treatment-associated timing pattern in these data.
Both temporal specifications give positive conventional phase and latency coefficients.
These findings support a prospective study with controlled exposure and direct entry and exit measurements.

The dopamine entry rule alone does not predict the sign of the formulation contrast.
A prospective test must fix an exposure-to-signal model before observing REM outcomes.
That model must specify administration times, absorption, clearance, receptor effects, and the starting state.
Its predicted trajectory must then determine which residence requirement completes first under Equation~\ref{eq:rementryrule}.

The present records are compatible with later sustained entry, but do not identify that directional dopamine mechanism.
The future comparison must also control treatment indication and validate sleep staging.
This separates a pharmacological prediction from an interpretation chosen after observing the outcome.

\subsection{Regional motor observations and competing response models}\label{sec:regionalresults}\label{sec:regionalmotor}
The regional analysis uses 640 paired OFF/ON visits from 429 participants with Parkinson's disease.
Each body-side score sums seven rigidity and bradykinesia items, with a range from 0 to 28.
The items cover arm and leg rigidity, finger tapping, hand movement, pronation/supination, toe tapping, and leg agility.
The putamen of each hemisphere maps to the opposite body side.
The observation model links transporter imaging to motor performance.

Images use analyzed 123I-DaTscan records and the Xing whole-putamen SBR pipeline with its common reference region.
Both putamen values must be positive.
The nearest eligible image is in the examination month or the preceding three months.
Pair selection requires valid items, recorded medication use, and no known prior or current DBS or HIFU.
Monthly dates and clock order do not establish a common examination day.
The medication-state labels are clinical observations, not randomized assignments.

Regional DAT can predict disease burden that the OFF examination measures with error.
The marker can then improve ON prediction without changing the latent response to medication.
A common latent response can also produce unequal observed score changes because the items are ordinal.
The joint comparison includes these explanations in its common-response model.

The first eligible earlier-OFF/current-OFF/current-ON triplet from each of 254 participants supplies 10,668 ordinal item values.
One triplet per person avoids repeated use of the same measurement in the joint likelihood.
The earlier examination requires an explicit OFF state and recorded medication use.
The median separation is seven months, with medication changes between many examinations.
The model includes progression over the examination gap and current dose-to-examination intervals.
It does not reconstruct treatment changes between examinations.

Both models use a cumulative probit link with four ordered thresholds for each item.
Thresholds are shared across body sides and examination states, with values fitted separately in each model.
The item loading and probit residual variance are fixed at one, with no baseline intercept.
These choices fix the latent scale.

Both models let mean and signed regional SBR predict latent baseline burden.
They include common and side-specific random burden, common progression, regional SBR-dependent progression differences, and a random common medication response.
The common response can depend on mean SBR, common burden, and medication intervals.
Progression innovation variance is proportional to elapsed years.

The two ON states are
\begin{samepage}
\begin{equation}
\eta_{s,\mathrm{ON}}^{(0)}=B_s-U,
\qquad
\eta_{s,\mathrm{ON}}^{(1)}=B_s-U-\gamma s\widetilde d_r.
\label{eq:ordinalresponse}
\end{equation}
\begin{eqvars}
\item $s$ (\textbf{Body-side index}): Body side, with $+1$ for left and $-1$ for right.
\item $B_s$ (\textbf{Latent baseline burden}): Current latent OFF burden on side $s$.
\item $U$ (\textbf{Common drug response}): Common latent medication response.
\item $\eta_{s,\mathrm{ON}}^{(0)},\eta_{s,\mathrm{ON}}^{(1)}$ (\textbf{Predicted latent states}): Latent ON states under the common and regional models.
\item $\widetilde d_r$ (\textbf{Scaled imaging contrast}): Signed SBR contrast divided by its training root mean square.
\item $\gamma$ (\textbf{Regional response coefficient}): Added regional response coefficient, with either sign permitted.
\end{eqvars}
\end{samepage}

Only $\gamma$ is added in the regional model.
It is a statistical coefficient, separate from the attenuation parameter in Equation~\ref{eq:complexity}.
The signed contrast is half the right-minus-left putamen SBR difference.
Both models use the same fixed weak regularization.

Five participant folds reserve each person once for evaluation.
All scaling and parameter fitting use the training participants.
The primary score is the log probability of the 14 ON items, conditional on both OFF examinations and the predictors.
Reserved ON values enter only the evaluation score.
Intervals use 10,000 bootstrap samples of saved participant losses and omit full model-refitting uncertainty.

\subsection{Regional results and their scope}\label{sec:regionalmean}
The regional-minus-common conditional gain is $-7.38\times10^{-5}$ log units per triplet, evaluated with 15 quadrature nodes.
Its 95\% participant-bootstrap interval is $[-3.72,3.68]\times10^{-3}$.
The interval spans improvements and reductions in conditional prediction.
The near-zero gain also changes sign across integration orders.
Appendix~\ref{app:regionalmethods} gives the numerical checks and secondary comparisons.

The score interval concerns predictive value, with no equivalence margin for biological response.
Interpretation depends on shared item thresholds, the progression model, and uncertain examination dates.
State-dependent regional threshold drift could be indistinguishable from a latent regional response.
A standardized drug challenge with repeated baselines would address these competing explanations.

\section{The upper boundary: collapse and death}
\subsection{A finite viable range}
The theory has an upper boundary on viable conscious states.
The collapse function defines a reversible gate:
\begin{samepage}
\begin{equation}
V_0(x)=\left[1-\exp\left\{-\left(\frac{x_c-x}{s_c}\right)^4\right\}\right]
\mathbf 1(x<x_c),\qquad x_c,s_c>0.
\label{eq:collapse}
\end{equation}
\begin{eqvars}
\item $V_0$ (\textbf{Viability gate}): Reversible viability gate between zero and one.
\item $x_c$ (\textbf{Collapse location}): Positive collapse location on the normalized signal axis.
\item $s_c$ (\textbf{Collapse width}): Positive transition scale.
\item $\mathbf 1(x<x_c)$ (\textbf{Boundary indicator}): One below the collapse location and zero otherwise.
\end{eqvars}
\end{samepage}

For the illustrative values $x_c=3.20$ and $s_c=0.15$, $V_0(3.00)=0.958$ and $V_0(3.10)=0.179$.
The gate reaches zero at $x=3.20$.
The numbers and fourth-power shape specify an illustrative model curve that requires biological calibration.

The state response also needs a lower boundary.
Without an additional factor, the cumulative sigmoid sum remains positive at $x=0$.
A minimal boundary factor is
\begin{samepage}
\begin{equation}
G_0(x)=1-\exp[-(x/x_\ell)^2],\qquad x\geq0,\quad x_\ell>0.
\label{eq:lowgate}
\end{equation}
\begin{eqvars}
\item $G_0(x)$ (\textbf{Low-signal factor}): Sets state intensity to zero at zero coordinated signal and approaches one as the input increases.
\item $x_\ell$ (\textbf{Low-signal scale}): Sets the range of this proposed boundary, independently of the component locations.
\end{eqvars}
\end{samepage}
The illustrative value is $x_\ell=0.10$.
This boundary condition changes the numerical state curve near zero, while preserving its increasing recruitment response.
Together with $V_0$, it permits state loss at either end of the coordinated signal range.

The distinction between collapse and death requires an additional state variable.
If $x$ decreases, Equation~\ref{eq:collapse} recovers automatically.
Irreversible injury therefore needs a separate accumulation law:
\begin{samepage}
\begin{align}
\dot E&=P(O_2,G_{\mathrm{glc}},T;\mathbf m_{\mathrm{met}})-Q(x,h,E),\\
\dot J&=k_J[Q-P]_+ + k_T[T-T_c]_+,
\qquad [z]_+=\max(z,0),\\
V(x,E,J)&=G_0(x)V_0\{x;x_c(E,O_2,T),s_c\}\,v_E(E)\,v_J(J).
\label{eq:injury}
\end{align}
\begin{eqvars}
\item $E,J$ (\textbf{Reserve and injury}): Energy reserve and accumulated injury.
\item $P$ (\textbf{Energy supply}): Usable metabolic energy supplied per unit time. ATP production provides its operational energy scale.
\item $O_2$ (\textbf{Local oxygen}): Oxygen availability in the modeled tissue, rather than arterial oxygen alone.
\item $G_{\mathrm{glc}}$ (\textbf{Local glucose}): Glucose availability in the modeled tissue, rather than food intake or blood glucose alone.
\item $T$ (\textbf{Tissue temperature}): Temperature that affects metabolic rates and injury.
\item $\mathbf m_{\mathrm{met}}$ (\textbf{Metabolic context}): Includes alternative fuels, perfusion, transport, and mitochondrial capacity. These factors require measurement or experimental control.
\item $Q(x,h,E)$ (\textbf{Energy demand}): Energy demand per unit time.
\item $k_J,k_T$ (\textbf{Injury accumulation}): Finite nonnegative coefficients that convert excess demand and temperature into injury rates.
\item $T_c$ (\textbf{Temperature threshold}): Temperature threshold in the injury model.
\item $[z]_+=\max(z,0)$ (\textbf{Positive part}): Positive part of a quantity.
\item $v_E(E),v_J(J)$ (\textbf{Reserve and injury factors}): Reserve and injury factors between zero and one.
\end{eqvars}
\end{samepage}

\paragraph{Fuel, oxygen, and temperature:}
Glucose supplies fuel, whereas oxygen supports oxidative ATP production.
Glycolysis can produce ATP without oxygen, but does not by itself establish adequate sustained supply for the brain \cite{Allen2005}.
Human studies show cerebral use of lactate and ketone bodies \cite{VanHall2009,Owen1967}.
The metabolic context includes these alternative fuels.
Tissue supply depends on local availability, perfusion, and transport, rather than food intake or a blood concentration alone.

Temperature is a condition for metabolism, not a fuel.
Human cooling measurements show changes in cerebral oxygen consumption \cite{McCullough1999}.
Temperature can affect supply and demand together.

The demand functions below suppress this temperature argument in their notation.
X0 and X1 compare demand at fixed temperature.
Section~\ref{sec:serotoninsafety} specifies the complementary thermal injury route.
A temperature comparison must calibrate both functions and keep the other declared conditions fixed.

The supply mapping and metabolic context require independent calibration and fixed specifications before outcome evaluation.
Glucose and alternative fuels affect the state through supply, reserve, and injury.

The collapse location can additionally depend on reserve, oxygen availability, and temperature.
An absorbing failure boundary can set $v_J=0$ after a specified injury threshold.
This injury law is an acute approximation with fixed susceptibility and negligible repair.
Section~\ref{sec:reservefeedback} specifies reserve feedback, and Section~\ref{sec:injuryhistory} extends the injury history.

The threshold depends on physiological reserve and context.
The theory predicts that the same regional signal can have different effects under different oxygen, fuel, or temperature conditions.
A universal human death concentration does not follow.
Figure~\ref{fig:collapse} separates reversible and irreversible states in the proposed cascade.

\figinput{\begin{tikzpicture}[
    yscale=1.28,
    node distance=2.8cm,
    processgreen/.style={
        rectangle, 
        draw=consciouscolor!80,
        fill=consciouscolor!20,
        line width=1.5pt,
        minimum width=5cm, 
        minimum height=2cm,
        rounded corners=8pt, 
        drop shadow,
        font=\small\bfseries,
        inner sep=10pt,
        align=center
    },
    processyellow/.style={
        rectangle, 
        draw=wakecolor!80,
        fill=wakecolor!20,
        line width=1.5pt,
        minimum width=5cm, 
        minimum height=2cm,
        rounded corners=8pt, 
        drop shadow,
        font=\small\bfseries,
        inner sep=10pt,
        align=center
    },
    processred/.style={
        rectangle, 
        draw=psychosiscolor!80,
        fill=psychosiscolor!20,
        line width=1.5pt,
        minimum width=5cm, 
        minimum height=2cm,
        rounded corners=8pt, 
        drop shadow,
        font=\small\bfseries,
        inner sep=10pt,
        align=center
    },
    death/.style={
        rectangle, 
        draw=deathcolor!90, 
        ultra thick,
        line width=3pt,
        fill=deathcolor!80,
        text=white, 
        minimum width=5cm, 
        minimum height=2.5cm,
        rounded corners=10pt, 
        drop shadow,
        font=\small\bfseries,
        inner sep=12pt,
        align=center
    },
    arrow/.style={
        ultra thick, 
        ->, 
        >=stealth,
        line width=2.5pt
    },
    sidebox/.style={
        rectangle,
        draw=gray!50,
        fill=gray!10,
        rounded corners=6pt,
        align=left, 
        font=\small,
        inner sep=8pt,
        text width=4.5cm
    }
]
\node[font=\LARGE\bfseries] at (0,2) {Candidate Collapse and Injury Cascade};
\node[processgreen] (normal) at (0,0) {
    \begin{tabular}{c}
    \faIcon{brain} Reference drive\\
    \textcolor{gray}{Preserved reserve}\\
    \textcolor{green!65!black}{$\bullet$ Stable \& Sustainable}
    \end{tabular}
};
\node[processred] (high) at (0,-3) {
    \begin{tabular}{c}
    \faIcon{fire} High neural demand\\
    \textcolor{gray}{Reserve under stress}\\
    \textcolor{orange!65!black}{$\bullet$ Unsustainable State}
    \end{tabular}
};
\node[processyellow] (toxic) at (0,-6) {
    \begin{tabular}{c}
    \faIcon{exclamation-triangle} Critical reserve failure\\
    \textcolor{gray}{State function can fail}\\
    \textcolor{red!65!black}{$\bullet$ Critical Danger}
    \end{tabular}
};
\node[death] (death) at (0,-9) {
    \begin{tabular}{c}
    \faIcon{skull} IRREVERSIBLE INJURY\\
    $J$ crosses an injury boundary\\
    \textcolor{white}{$\bullet$ Terminal failure}
    \end{tabular}
};
\node[sidebox] (mech1) at (5.5,0) {
    \textbf{\faIcon{chart-line} Candidate processes:}\\
    $\bullet$ Regulated signaling\\
    $\bullet$ Circuit activity\\
    $\bullet$ Metabolic upregulation\\
    $\bullet$ Energy supply meets demand
};
\node[sidebox] (mech2) at (5.5,-3) {
    \textbf{\faIcon{bolt} Candidate processes:}\\
    $\bullet$ Ca$^{2+}$ channel opening\\
    $\bullet$ Mitochondrial stress\\
    $\bullet$ ATP demand surge\\
    $\bullet$ Receptor overstimulation
};
\node[sidebox] (mech3) at (5.5,-6) {
    \textbf{\faIcon{bomb} Candidate processes:}\\
    $\bullet$ Excitotoxic cascade\\
    $\bullet$ Membrane depolarization\\
    $\bullet$ Injury accumulation\\
    $\bullet$ Cellular energy collapse
};
\node[sidebox] (symp1) at (-5.5,0) {
    \textbf{\faIcon{smile} Possible outcomes:}\\
    $\bullet$ Flow states\\
    $\bullet$ Enhanced performance\\
    $\bullet$ Increased confidence\\
    $\bullet$ Clear thinking
};
\node[sidebox] (symp2) at (-5.5,-3) {
    \textbf{\faIcon{meh} Possible outcomes:}\\
    $\bullet$ Racing thoughts\\
    $\bullet$ Grandiosity\\
    $\bullet$ Hyperthermia\\
    $\bullet$ Sleep disruption
};
\node[sidebox] (symp3) at (-5.5,-6) {
    \textbf{\faIcon{dizzy} Possible outcomes:}\\
    $\bullet$ Delusions\\
    $\bullet$ Convulsions\\
    $\bullet$ Coma\\
    $\bullet$ Multi-organ failure
};
\draw[arrow, consciouscolor!80] (normal) -- (high);
\draw[arrow, wakecolor!80] (high) -- (toxic);
\draw[arrow, deathcolor, line width=4pt] (toxic) -- (death);
\draw[gray!60, ->, line width=1.5pt] (normal.east) -- (mech1.west);
\draw[gray!60, ->, line width=1.5pt] (high.east) -- (mech2.west);
\draw[gray!60, ->, line width=1.5pt] (toxic.east) -- (mech3.west);
\draw[gray!60, <-, line width=1.5pt] (normal.west) -- (symp1.east);
\draw[gray!60, <-, line width=1.5pt] (high.west) -- (symp2.east);
\draw[gray!60, <-, line width=1.5pt] (toxic.west) -- (symp3.east);
\node[fill=green!20, rounded corners=4pt, font=\footnotesize\bfseries, 
      inner sep=6pt] at (-5.5,1.4) {Preserved reserve};
\node[fill=yellow!30, rounded corners=4pt, font=\footnotesize\bfseries, 
      inner sep=6pt] at (-5.5,-1.5) {Reserve decreases};
\node[fill=red!30, rounded corners=4pt, font=\footnotesize\bfseries, 
      inner sep=6pt] at (-5.5,-4.5) {Acute failure};
\draw[thick] (8.75,1) rectangle (9.25,-7);
\fill[green!60] (8.75,1) rectangle (9.25,-0.5);
\fill[yellow!60] (8.75,-0.5) rectangle (9.25,-2.5);
\fill[orange!70] (8.75,-2.5) rectangle (9.25,-4.5);
\fill[red!80] (8.75,-4.5) rectangle (9.25,-6);
\fill[deathcolor] (8.75,-6) rectangle (9.25,-7);
\node[font=\footnotesize\bfseries, rotate=90] at (9.75,-2.8) {MODEL VIABILITY};
\node[font=\tiny] at (9.6,0.9) {High};
\node[font=\tiny] at (8.45,-2.5) {Mid};
\node[font=\tiny] at (9.6,-6) {Low};
\node[font=\tiny] at (8.45,-7) {Zero};
\draw[deathcolor, ultra thick, dashed, line width=3pt] 
    (-8,-7.5) -- (10,-7.5);
\node[deathcolor, font=\footnotesize\bfseries, 
      inner sep=4pt, rounded corners=3pt] 
    at (-5.2,-8.05) {INJURY BOUNDARY};
\node[red!60, font=\small] at (-2,-6.45) {\faIcon{radiation}};
\node[red!60, font=\small] at (2,-6.45) {\faIcon{radiation}};
\end{tikzpicture}}{Candidate collapse and injury cascade. The central boxes separate preserved reserve, high demand, critical reserve failure, and irreversible injury. Side boxes give candidate processes and outcomes for measurement. Their sequence and time scales depend on physiological context.}{collapse}

\subsection{Regional excess and additive demand}\label{sec:regionalexcess}
The bottleneck is asymmetric by construction.
A low regional signal can limit coordinated availability, while a high signal in one region cannot replace the others.
For a set $\mathcal H$ raised without bound, with every remaining signal at $d$, its upper limit is
\begin{samepage}
\begin{equation}
\lim_{D_r\to\infty,\ r\in\mathcal H}D_\kappa
=d-\frac{\log(1-p_{\mathcal H})}{\kappa},
\qquad p_{\mathcal H}=\sum_{r\in\mathcal H}p_r<1.
\label{eq:excessceiling}
\end{equation}
\begin{eqvars}
\item $\mathcal H$ (\textbf{Raised-signal set}): Selects regions whose integrated signals increase while the other signals stay fixed.
\item $p_{\mathcal H}$ (\textbf{Raised weight}): Sums their fixed bottleneck weights.
\item $d$ (\textbf{Remaining reference signal}): Gives the common fixed signal outside the raised set.
\end{eqvars}
\end{samepage}

For W0 with $d=1$, raising only the PFC gives a ceiling of $1.113$.
Adding motor cortex gives $1.253$, adding striatum gives $1.426$, and leaving only SN at one gives $1.599$.
The coordinate $1.50$ can thus be reached without every region increasing.
Coordinates $2.50$ and $3.20$ require every included region above reference, but do not require equal regional signals.
These mathematical limits characterize the coordinate.
Clinical state boundaries require calibration over a finite signal range.

Systemic administration does not imply uniform regional signaling.
For example, low systemic methylphenidate doses preferentially increased prefrontal catecholamines in rats \cite{Berridge2006}.
High prefrontal D1 stimulation can suppress persistent neuronal firing and impair spatial tuning \cite{Vijayraghavan2007}.
This supports local functional impairment from excessive stimulation.
Energy demand and conscious intensity require separate measurements.

\paragraph{A receptor-mediated deficit route:}
Excess ligand and excess effective signal are different perturbations.
The optional receptor peak in Appendix~\ref{app:receptorpeak} can make increasing ligand reduce the effective regional signal.
A decline below reference requires both the peak condition and a suitable reference concentration.
Equation~\ref{eq:receptorbaseline} states the exact condition.
The positive high-ligand plateau limits the size of that decline.

This route changes the bottleneck input without prior reserve depletion.
Its speed depends on ligand kinetics, receptor response, and the regional memory time.
The D1 recording result motivates functional measurements alongside independent tests of the distinct D2 autoreceptor mechanism.

An isolated prefrontal decline remains subject to the W0 floor, while a distributed decline can lower the coordinate further.
A circuit-competence variable would require an independent assay if receptor signaling fails to explain local function.

A complementary route can represent regional demand while preserving the bottleneck for coordinated availability.
Energy demands add across regions, whereas functional availability need not add.
A quadratic law specifies the candidate relation between dopamine signaling and local demand:
\begin{samepage}
\begin{align}
\mathcal L_Q(\mathbf D)&=\sum_r w_r^Q D_r^2,
\qquad w_r^Q\geq0,\quad\sum_r w_r^Q=1,\\
Q_{\mathrm{reg}}(\mathbf D,h,E)&=Q_b(h,E)+c_Q(h,E)\mathcal L_Q(\mathbf D),
\qquad Q_b\geq0,\quad c_Q>0.
\label{eq:regionaldemand}
\end{align}
\begin{eqvars}
\item $\mathcal L_Q$ (\textbf{Regional load index}): Combines squared integrated signals on their declared reference scales.
\item $w_r^Q$ (\textbf{Demand weight}): Gives the region's share of the reference dopamine-sensitive cost, distinct from its bottleneck weight.
\item $Q_{\mathrm{reg}}$ (\textbf{Total regional demand}): Gives total energy demand per unit time in this candidate.
\item $Q_b(h,E)$ (\textbf{Background demand}): Gives the nonnegative demand independent of the regional dopamine inputs.
\item $c_Q(h,E)$ (\textbf{Demand scale}): Converts the dimensionless load index to energy per unit time at fixed context.
\end{eqvars}
\end{samepage}

Candidate X0 fixes $w_r^Q=p_r$ to the W0 weights and fixes the exponent at two.
Equal demand and bottleneck weights constitute an additional hypothesis.
The total energy scale is absorbed in $Q_b$ and $c_Q$.
Both functions require independent calibration and must stay fixed during evaluation.
The quadratic cost requires direct energy-demand measurements beyond receptor occupancy and the available PPMI outcomes.

The regional demand replaces $Q$ in the reserve and injury dynamics:
\begin{samepage}
\begin{align}
\dot E&=P(O_2,G_{\mathrm{glc}},T;\mathbf m_{\mathrm{met}})-Q_{\mathrm{reg}}(\mathbf D,h,E),\\
\dot J&=k_J[Q_{\mathrm{reg}}-P]_++k_T[T-T_c]_+.
\label{eq:regionalinjury}
\end{align}
\begin{eqvars}
\item $E,J$ (\textbf{Reserve and injury}): Give energy reserve and accumulated injury, with the units defined in Equation~\ref{eq:injury}.
\item $P,k_J,k_T,T_c$ (\textbf{Supply and injury terms}): Specify supply, injury coefficients, and the thermal threshold from Equation~\ref{eq:injury}.
\end{eqvars}
\end{samepage}
The state function remains $C(x,h,E,J)$, with $x$ determined by the bottleneck.
Excess effective signal can change reserve even when the bottleneck changes little.
The load uses the same temporally integrated signals as the bottleneck, so fast unfiltered peaks lie outside this candidate.
This demand route predicts effects through reserve and injury.
If excess ligand instead lowers the effective signal, the quadratic demand term also decreases.

For homogeneous signals $D_r=d$, $\mathcal L_Q=d^2$ and $D_\kappa=d$.
X0 then recovers the scalar demand choice $Q=Q_b+c_Qd^2$ within Equation~\ref{eq:injury}.
Subdivision also preserves demand when child signals match and their demand weights sum to the parent weight.
Regional injury localization requires region-specific reserve measurements.
The quadratic cost applies within its declared finite signal range.
Receptor saturation prevents an infinite ligand dose from implying an infinite effective signal.

X0 provides restrictions beyond a general claim about overload.
At fixed context, its demand satisfies
\begin{samepage}
\begin{align}
\mathcal L_Q&=\overline D^{\,2}+\operatorname{Var}_p(\mathbf D),\\
\frac{\partial Q_{\mathrm{reg}}/\partial D_r}
{\partial Q_{\mathrm{reg}}/\partial D_s}
&=\frac{p_rD_r}{p_sD_s},\qquad p_r,p_s,D_r,D_s>0.
\label{eq:demandrestrictions}
\end{align}
\begin{eqvars}
\item $\operatorname{Var}_p(\mathbf D)$ (\textbf{Weighted signal variance}): Equals $\sum_r p_r(D_r-\overline D)^2$ under the fixed W0 weights.
\item $r,s$ (\textbf{Demand comparison pair}): Selects two regions with positive weights and signals for small perturbations.
\end{eqvars}
\end{samepage}
At fixed weighted mean, demand changes by $c_Q$ times the change in regional variance.
Matched context and load predict matched demand even when regional maxima differ.
This provides a comparison with a demand model driven by the largest regional signal.
At fixed reserve and injury, the bottleneck sensitivity follows Equation~\ref{eq:rescueratio}.
During prolonged input, the additional reserve pathway changes the total outcome response.

A matched-bottleneck comparison isolates the demand route.
The vector $(4,1,1,1,1)$ gives $D_\kappa=1.113$ and $\mathcal L_Q=6.455$.
Uniform signals at that same bottleneck give $\mathcal L_Q=1.239$.
At matched initial reserve, supply, and serotonergic context, X0 predicts a reserve derivative difference of $-5.216c_Q$.
This prediction concerns reserve change.
The state and viability functions determine its consequences for conscious intensity.

A test must measure regional signaling, energy demand, supply, and a separately defined functional outcome.
The same calibrated cost scale must predict redistribution and small regional perturbations.
Failure of those demand restrictions would reject X0 without rejecting the spatial bottleneck.
Reproducible intensity differences at matched bottleneck inputs, histories, and physiological context would challenge the conditional state claim.

\subsection{A declared reserve feedback law}\label{sec:reservefeedback}
The sign of $\partial Q/\partial E$ determines the local reserve feedback.
At fixed inputs and supply, a positive demand derivative gives linear asymptotic stability at a scalar interior equilibrium.
A zero derivative requires further analysis.
An interior equilibrium must also exist within the viable domain.
A negative sign gives amplifying feedback around an interior equilibrium.

X0 fixes demand independent of reserve over its acute comparison interval.
Candidate X1 adds a specified reserve response:
\begin{samepage}
\begin{align}
e&=E/E_{\mathrm{ref}},\qquad
M_E(e)=\frac{e+\alpha_E/(e+\epsilon_E)}
{1+\alpha_E/(1+\epsilon_E)},\\
Q_{\mathrm{reg}}(\mathbf D,h,E)&=Q_{\mathrm{ref}}(\mathbf D,h)M_E(e).
\label{eq:reservefeedback}
\end{align}
\begin{eqvars}
\item $e$ (\textbf{Relative reserve}): Gives energy reserve relative to a positive reference $E_{\mathrm{ref}}$.
\item $M_E(e)$ (\textbf{Reserve demand factor}): Combines reduced activity demand with increased cost near reserve failure.
\item $\alpha_E,\epsilon_E$ (\textbf{Feedback parameters}): Set the strength and regularization of the low-reserve cost term.
\item $Q_{\mathrm{ref}}(\mathbf D,h)$ (\textbf{Reference-reserve demand}): Equals $Q_b(h,E_{\mathrm{ref}})+c_Q(h,E_{\mathrm{ref}})\mathcal L_Q(\mathbf D)$.
\end{eqvars}
\end{samepage}
X1 keeps the X0 regional weights and exponent, and fixes $\alpha_E=1/4$ and $\epsilon_E=1/20$.
These values define a prospective candidate for evaluation.
The function is positive and finite for nonnegative reserve, and $M_E(1)=1$.
The law specifies required demand, which can exceed actual ATP use when cellular processes fail.

The demand derivative is
\begin{samepage}
\begin{equation}
\frac{\partial Q_{\mathrm{reg}}}{\partial E}
=\frac{Q_{\mathrm{ref}}}
{E_{\mathrm{ref}}\{1+\alpha_E/(1+\epsilon_E)\}}
\left[1-\frac{\alpha_E}{(e+\epsilon_E)^2}\right].
\label{eq:reservefeedbacksign}
\end{equation}
\end{samepage}
For X1 with positive $Q_{\mathrm{ref}}$, feedback is protective above $e=0.45$ and amplifying below it.
These labels specify the direction of local reserve feedback.
With supply equal to $Q_{\mathrm{ref}}$, the reserve-only model has a stable equilibrium at $e=1$.
It also has an unstable equilibrium at $e=0.188$.
Both equilibria follow from the candidate parameters.

The stability statement fixes supply, regional inputs, and all other states.
If supply depends on reserve, the condition for linear asymptotic stability is $P_E-Q_E<0$ in that scalar comparison.
Coupled injury and repair require analysis of the full dynamics.
The same X1 parameters must predict both reserve perturbations and regional-load contrasts.
A precise failure of the declared derivative reversal would reject X1.
The function and its parameters must be fixed before evaluation.

\subsection{Injury history, adaptation, and recovery}\label{sec:injuryhistory}
Equation~\ref{eq:injury} describes acute injury at fixed susceptibility.
At matched initial injury and thermal exposure, equal integrated supply deficits produce equal final injury.
Repair and protective adaptation require additional state dynamics.
Brief ischemic pretreatment can protect gerbil hippocampal neurons from a later ischemic insult \cite{Kitagawa1990}.
This motivates a history-dependent injury law, independently of the proposed dopamine pathway.

Before an irreversible boundary, consider
\begin{samepage}
\begin{align}
r_E(E)&=\min(1,\max(0,E/E_{\mathrm{ref}})),\\
Q_{\mathrm{tot}}&=Q_{\mathrm{reg}}+c_R\rho_J r_E(E)J,\qquad
\sigma=\frac{[Q_{\mathrm{tot}}-P]_+}{Q_s},\\
\tau_{\mathrm{ad}}\dot a_{\mathrm{ad}}
&=r_E(E)\frac{\sigma}{1+\sigma}-a_{\mathrm{ad}},\\
\dot J&=\frac{k_J[Q_{\mathrm{tot}}-P]_+}{1+\eta_{\mathrm{ad}}a_{\mathrm{ad}}}
+k_T[T-T_c]_+-\rho_Jr_E(E)J.
\label{eq:injuryhistory}
\end{align}
\begin{eqvars}
\item $r_E(E)$ (\textbf{Recovery capacity}): Limits repair and adaptation when reserve falls, with $r_E(0)=0$.
\item $Q_{\mathrm{tot}}$ (\textbf{Total required demand}): Adds the modeled energy cost of repair to regional demand.
\item $c_R$ (\textbf{Repair energy cost}): Converts repaired injury units to energy units, with $c_R\geq0$.
\item $\rho_J$ (\textbf{Repair rate}): Sets the nonnegative repair rate per unit time at reference reserve.
\item $\sigma$ (\textbf{Normalized supply deficit}): Divides unmet demand by the fixed positive power scale $Q_s$.
\item $a_{\mathrm{ad}}$ (\textbf{Protective adaptation}): Represents an independently measurable adaptation state, distinct from autoreceptor feedback $A_r$.
\item $\tau_{\mathrm{ad}}$ (\textbf{Adaptation time}): Sets the positive acquisition and decay time in this candidate.
\item $\eta_{\mathrm{ad}}$ (\textbf{Protection strength}): Sets the nonnegative reduction in deficit-related injury susceptibility.
\end{eqvars}
\end{samepage}
For $a_{\mathrm{ad}}(0)\in[0,1]$, the adaptation stays in that interval.
The protection factor never eliminates injury from a sustained deficit.
The repair cost enters the energy balance, and repair stops at zero reserve.
The acute law is recovered when $\eta_{\mathrm{ad}}=\rho_J=0$.
The adaptation and repair parameters require independent calibration before a history comparison.

Reserve cannot become negative:
\begin{samepage}
\begin{equation}
\dot E=
\begin{cases}
P-Q_{\mathrm{tot}}, & E>0,\\
[P-Q_{\mathrm{tot}}]_+, & E=0.
\end{cases}
\label{eq:reserveboundary}
\end{equation}
\end{samepage}
Unmet demand can still accumulate injury when reserve reaches zero.
The state factor satisfies $v_E(0)=0$.
Before an irreversible threshold, reserve recovery and repair can restore the modeled state.

If $J$ reaches a declared threshold $J_{\mathrm{irr}}>0$, an irreversible-failure event occurs.
The injury gate then remains zero, even if a later unconstrained repair calculation would reduce $J$.
This absorbing event represents irreversible loss of the modeled substrate.
Its relation to organism death requires evidence about essential functions.

For insults with equal totals, order and spacing change adaptation, repair, reserve, and final injury.
Intervals can also dissipate protection, so repeated brief insults are not universally safer.
A test must compare matched challenges after specified pretreatment and recovery intervals.
The same parameters must predict both protection and its decay.

\paragraph{Distinct routes from signal extremes to state loss and death:}
The state equation reaches zero at zero coordinated signal and at the modeled upper viability boundary.
Excess ligand can also cause a receptor-mediated deficit under Equation~\ref{eq:receptorbaseline}.
These routes concern state loss, which can be reversible.

The model does not imply that every dopamine extreme causes death.
At zero regional signal, X0 demand reduces to its background term.
If supply meets that demand, low dopamine alone does not deplete reserve in this model.
A fatal low-drive pathway would require an additional demonstrated failure of essential support, such as sustained loss of oxygen or usable fuel supply.
Severe deficits or excessive demand can then accumulate irreversible injury through the same injury law.
Neither Parkinson's disease nor impaired task performance establishes such a fatal pathway.

\subsection{Optional solute and oxidative injury routes}\label{sec:wasteoxidative}
Solute accumulation and oxidative stress require production and removal balances alongside energy accounting.
Equation~\ref{eq:injuryhistory} includes demand and repair costs but has no explicit chemical burden state.
Two measured burdens extend the injury law while the regional bottleneck and state function specify functional output.

Mouse experiments found faster clearance of injected amyloid-beta and inulin during sleep \cite{Xie2013}.
Other mouse experiments found reduced clearance of a small fluorescent dye during sleep and anesthesia \cite{Miao2024}.
These results use different tracers and methods, and their interpretation remains disputed.
Recent experiments in naturally sleeping mice link noradrenergic vascular oscillations to fluid transport and clearance \cite{Hauglund2025}.
Tracer entry, redistribution, and removal must be distinguished.
The candidate therefore imposes no universal direction for a sleep effect on clearance.

Oxidative injury also needs a separate production and removal balance.
Experiments linked ischemic succinate accumulation and its subsequent oxidation to reperfusion injury in rodent heart and brain models \cite{Chouchani2014}.
This supports a route that can remain active after blood supply returns.
The mechanism concerns a specific metabolic intermediate and requires separate assessment of dopamine dependence and oxygen effects.

For a prespecified tissue compartment and chemical assay, define
\begin{samepage}
\begin{align}
\dot W&=u_W(t)-\lambda_W(t)W,\\
\dot Z&=u_Z(t)-\lambda_Z(t)Z.
\label{eq:burdenbalance}
\end{align}
\begin{eqvars}
\item $W$ (\textbf{Selected solute}): Nonnegative concentration of one specified, potentially harmful nonoxidant solute.
\item $Z$ (\textbf{Selected oxidant}): Nonnegative concentration of one specified reactive species, distinct from accumulated tissue injury.
\item $u_W,u_Z$ (\textbf{Formation rates}): Nonnegative inputs with concentration per time units, measured or independently calibrated before evaluation.
\item $\lambda_W,\lambda_Z$ (\textbf{Removal rates}): Nonnegative first-order clearance and detoxification rates, with inverse time units.
\end{eqvars}
\end{samepage}
All rate inputs must be finite and locally bounded over the evaluation interval.
Their measurements must separate production from disappearance.
Rate measurements and concentration ranges must be specified independently before outcome evaluation.
The first-order reduction applies only within its independently declared concentration range.

Before the irreversible boundary in Section~\ref{sec:injuryhistory}, extend Equation~\ref{eq:injuryhistory} as
\begin{samepage}
\begin{align}
\dot J={}&\frac{k_J[Q_{\mathrm{tot}}-P]_+}{1+\eta_{\mathrm{ad}}a_{\mathrm{ad}}}
+k_T[T-T_c]_+\\
&+k_W[W-W_c]_++k_Z[Z-Z_c]_+
-\rho_Jr_E(E)J.
\label{eq:extendedinjury}
\end{align}
\begin{eqvars}
\item $W_c,Z_c$ (\textbf{Burden boundaries}): Nonnegative concentrations that delimit the proposed injury response, calibrated independently of evaluation outcomes.
\item $k_W,k_Z$ (\textbf{Additional injury coefficients}): Finite nonnegative coefficients, with injury per concentration per time units.
\item $E,J,P,Q_{\mathrm{tot}}$ (\textbf{Reserve, injury, supply, and demand}): Use the definitions and units in Equations~\ref{eq:injury} and~\ref{eq:injuryhistory}.
\end{eqvars}
\end{samepage}
The sign restrictions in Equations~\ref{eq:injuryhistory} and~\ref{eq:burdenbalance} also apply.
Setting $k_W=k_Z=0$ exactly recovers the preceding injury history for identical inputs and coefficients.
Ordinary reference concentrations must lie within the declared noninjurious domain.
At $J=0$, the derivative remains nonnegative.
Equation~\ref{eq:reserveboundary} enforces nonnegative reserve, and Section~\ref{sec:injuryhistory} specifies the absorbing irreversible event.

The additional terms must not count the same injury twice.
Within this extension, the deficit and thermal terms represent residual routes after the explicit chemical contributions.
Coefficients fitted to total injury cannot be reused as independent direct contributions.
If the selected solute acts entirely through oxidants, set $k_W=0$ and separately establish its contribution to $u_Z$.
Removing a solute or oxidant is distinct from repairing the damage already represented by $J$.

Transport and detoxification costs must enter regional demand once.
Total demand includes the repair cost.
If measured regional demand already includes these costs, do not add them again.
Comparisons must hold total demand fixed or measure its change.
Zero reserve does not force both removal rates to zero.
Passive transport or maintained perfusion can remain, while energy-dependent removal requires its own measured relation to reserve.

For fixed formation rates and positive removal rates, the balances have stable equilibria $W_*=u_W/\lambda_W$ and $Z_*=u_Z/\lambda_Z$.
With zero formation, $W(t)=W(0)e^{-\lambda_Wt}$ and $Z(t)=Z(0)e^{-\lambda_Zt}$.
Nonnegative initial concentrations remain nonnegative.
Without removal, positive formation instead accumulates burden.

The primary discriminator concerns injury after supply recovery.
At $P\geq Q_{\mathrm{tot}}$ and $T\leq T_c$, the preceding law permits no positive injury derivative.
The extension predicts further injury when the added burden terms exceed repair.
After matched preceding insults, independently changing removal must predict the injury trajectory with the same calibrated coefficients.
This comparison requires measured burdens, supply, demand, temperature, reserve, and initial injury.
Evaluation therefore requires a dedicated chemical and metabolic dataset.

\subsection{A discriminating test}
A preclinical test can compare neural function during matched receptor signaling under different metabolic conditions.
The reserve extension predicts different collapse locations when supply changes.
A fixed-threshold model predicts the same location.
Simultaneous cardiovascular and neural measurements are needed to locate the primary failure.
PPMI contains no appropriate collapse or mortality measurements for this test.

\section{Serotonergic saturation and the safety hypothesis}
\label{sec:serotoninsafety}
The safety hypothesis concerns a specific receptor response.
Can a saturating receptor response limit one component of a drug-induced state change?
For Equation~\ref{eq:complexity}, fixed nonnegative parameters give
\begin{equation}
\lim_{L_H\to\infty}\Psi_{\mathrm{drive}}
=\Psi_0(1+\lambda_cf_{\mathrm{const}}+\lambda_L)
e^{-\gamma(x/x_s)^2}R.
\label{eq:saturation}
\end{equation}
This is a finite mathematical limit on the proposed complexity drive.
Other receptors, cardiovascular responses, temperature, behavior, and drug interactions contribute separately to adverse effects.

The safety hypothesis has a restricted form:
\begin{hypothesis}[Bounded target response]
At fixed physiological context, receptor saturation and adaptation limit a specified \HT{}-mediated state response.
They do not guarantee a decreasing total hazard or universal safety at higher exposure.
\end{hypothesis}

\paragraph{A serotonergic route to thermal injury:}
In rats, Ma et al. found hypothermia with mild serotonin toxicity and hyperthermia with severe toxicity \cite{Ma2008}.
Their receptor interventions implicate \HT{} and NMDA mechanisms in the hyperthermic response.
The temperature response thus depends on receptor recruitment and toxicity severity.

When it raises tissue temperature above $T_c$, Equation~\ref{eq:injury} adds injury through $k_T[T-T_c]_+$.
Serotonergic heating and regional dopamine demand can therefore reach the same injury variable through different terms.
These conditional routes are not exclusive to either transmitter.

Temperature remains an input.
A predictive serotonin-to-temperature law requires separate measurements and calibration of $k_T$ and $T_c$.
Thermal injury can increase after the complexity response saturates, alongside injury from other mechanisms.

Human psilocybin trials also report adverse effects \cite{Goodwin2022}.
A test of the bounded target response must measure the target-specific response and whole-system hazard at matched exposure.

One candidate hazard decomposition is
\begin{samepage}
\begin{equation}
\mathcal H(t)=\mathcal H_{2A}\{O_H(t),R(t)\}
+\mathcal H_{\mathrm{other}}\{L_H(t),T(t),M(t)\},
\qquad \mathcal H_{2A},\mathcal H_{\mathrm{other}}\geq0.
\label{eq:hazard}
\end{equation}
\begin{eqvars}
\item $\mathcal H(t)$ (\textbf{Total hazard}): Total modeled hazard rate.
\item $\mathcal H_{2A}$ (\textbf{Target-related hazard}): Nonnegative hazard component linked to the specified receptor response.
\item $\mathcal H_{\mathrm{other}}$ (\textbf{Other hazard}): Nonnegative hazard component from other processes.
\item $M(t)$ (\textbf{Exposure context}): Concurrent exposures and physiological context.
\end{eqvars}
\end{samepage}
Even if the target-related hazard saturates, the other hazard is unconstrained.

\clearpage
\section{Substrate, evolution, and comparative predictions}
\subsection{Critical dynamics as a capacity hypothesis}
Critical dynamics are a candidate property of a capable substrate.
Neuronal avalanche experiments establish relevant collective behavior in cortical preparations \cite{Beggs2003}.
The stronger claim that criticality is necessary for a nonzero conscious capacity is a hypothesis.
Its evaluation needs independent measures of experience alongside critical dynamics.

A candidate substrate score can combine recurrence, integration, diversity, and energy reserve:
\begin{samepage}
\begin{equation}
\Phi=F(\mathcal R,\mathcal I,\mathcal D,E),\qquad
\Psi_{\max}=M(\Phi),\quad M'(\Phi)\geq0.
\label{eq:substrate}
\end{equation}
\begin{eqvars}
\item $\mathcal R,\mathcal I,\mathcal D$ (\textbf{Substrate measures}): Represent candidate measures of recurrence, network integration, and diversity. Network integration differs from regional signal aggregation.
\item $F$ (\textbf{Capacity mapping}): Maps these substrate measures and energy reserve to structural capacity $\Phi$. Its biological form is not yet specified.
\item $M(\Phi)$ (\textbf{Complexity ceiling}): Nondecreasing capacity-to-complexity-ceiling function.
\end{eqvars}
\end{samepage}

These integration measures concern network organization, not the soft-minimum aggregation of regional dopamine signals.
The function $F$ requires operational definitions and validation.
Power-law fits need comparison with alternative distributions, finite-size behavior, and measurement effects.
The model predicts a capacity constraint across neuromodulatory conditions, rather than a neuron-count threshold alone.
Figure~\ref{fig:programme} gives the cross-scale validation programme.

\landfiginput{\begin{tikzpicture}[
    xscale = 1.3,  
    yscale = 1.05,
    node distance=1.5cm and 2.5cm,
    moderndata/.style={
        rectangle, 
        draw=blue!40,
        fill=white,
        line width=1.5pt,
        minimum width=3.8cm, 
        minimum height=2.8cm,
        rounded corners=10pt, 
        align=center, 
        font=\footnotesize\bfseries\sffamily,
        drop shadow={shadow xshift=0pt, shadow yshift=2pt, opacity=0.15},
        inner sep=6pt
    },
    modernprocess/.style={
        rectangle, 
        draw=green!50,
        fill=white,
        line width=1.5pt,
        minimum width=3.8cm, 
        minimum height=2.8cm,
        rounded corners=10pt, 
        align=center, 
        font=\footnotesize\bfseries\sffamily,
        drop shadow={shadow xshift=0pt, shadow yshift=2pt, opacity=0.15},
        inner sep=6pt
    },
    modernanalysis/.style={
        rectangle, 
        draw=orange!50,
        fill=white,
        line width=1.5pt,
        minimum width=3.8cm, 
        minimum height=2.8cm,
        rounded corners=10pt, 
        align=center, 
        font=\footnotesize\bfseries\sffamily,
        drop shadow={shadow xshift=0pt, shadow yshift=2pt, opacity=0.15},
        inner sep=6pt
    },
    modernresult/.style={
        rectangle, 
        draw=red!50,
        fill=white,
        line width=2pt,
        minimum width=8cm, 
        minimum height=2.5cm,
        rounded corners=12pt, 
        align=center, 
        font=\small\bfseries\sffamily,
        drop shadow={shadow xshift=0pt, shadow yshift=3pt, opacity=0.2},
        inner sep=12pt
    },
    modernarrow/.style={
        ->,
        >=stealth,
        line width=1.5pt,
        color=gray!40
    },
    layerlabel/.style={
        font=\large\bfseries\sffamily,
        text=black!70
    },
    iconcirc/.style={
        circle,
        fill=#1!15,
        minimum size=25pt,
        inner sep=0pt
    }
]
\fill[blue!5] (-2,1) rectangle (16,-2);
\fill[green!5] (-2,-2.5) rectangle (16,-5.5);
\fill[orange!5] (-2,-6) rectangle (16,-9);
\fill[red!5] (2,-10) rectangle (12,-13);
\node[layerlabel] at (-1.65,-0.5) {\rotatebox{90}{\parbox{3cm}{\centering DATA\\COLLECTION}}};
\node[layerlabel] at (-1.65,-4) {\rotatebox{90}{PROCESSING}};
\node[layerlabel] at (-1.65,-7.5) {\rotatebox{90}{ANALYSIS}};
\node[layerlabel] at (2.5,-11.5) {\rotatebox{90}{INTEGRATION}};
\node[moderndata] (eeg) at (1,-0.5) {
    \begin{tabular}{@{}>{\centering\arraybackslash}p{3.8cm}@{}}
    \textcolor{blue!60}{\faIcon{brain}}\\
    Wearable sleep recording\\
    \textcolor{consciouscolor!65!black}{\scriptsize\bfseries AVAILABLE}\\
    \textcolor{gray}{\scriptsize 95\,434 nights, 348 people}
    \end{tabular}
};
\node[moderndata] (pet) at (5,-0.5) {
    \begin{tabular}{@{}>{\centering\arraybackslash}p{3.8cm}@{}}
    \textcolor{blue!60}{\faIcon{atom}}\\
    Clinical assessment\\
    \textcolor{consciouscolor!65!black}{\scriptsize\bfseries AVAILABLE}\\
    \textcolor{gray}{\scriptsize UPDRS, MoCA, DaTscan}
    \end{tabular}
};
\node[moderndata] (bio) at (9,-0.5) {
    \begin{tabular}{@{}>{\centering\arraybackslash}p{3.8cm}@{}}
    \textcolor{blue!60}{\faIcon{vial}}\\
    CSF neurochemistry\\
    \textcolor{consciouscolor!65!black}{\scriptsize\bfseries AVAILABLE}\\
    \textcolor{gray}{\scriptsize HVA and 5-HIAA, 151 people}
    \end{tabular}
};
\node[moderndata] (clinical) at (13,-0.5) {
    \begin{tabular}{@{}>{\centering\arraybackslash}p{3.8cm}@{}}
    \textcolor{blue!60}{\faIcon{atom}}\\
    EEG and $[^{11}$C$]$raclopride PET\\
    \textcolor{deathcolor}{\scriptsize\bfseries PROPOSED}\\
    \textcolor{gray}{\scriptsize aperiodic exponent, D$_2$ availability}
    \end{tabular}
};
\node[modernprocess] (signal) at (1,-4) {
    \begin{tabular}{@{}>{\centering\arraybackslash}p{3.8cm}@{}}
    \textcolor{green!65!black}{\faIcon{wave-square}}\\
    Sleep staging\\
    Epoch quality control\\
    \textcolor{gray}{\scriptsize Scored-epoch coverage}
    \end{tabular}
};
\node[modernprocess] (kinetic) at (5,-4) {
    \begin{tabular}{@{}>{\centering\arraybackslash}p{3.8cm}@{}}
    \textcolor{green!65!black}{\faIcon{chart-line}}\\
    Clinical scoring\\
    Standardised metrics\\
    \textcolor{gray}{\scriptsize ON/OFF labels are incomplete}
    \end{tabular}
};
\node[modernprocess] (assay) at (9,-4) {
    \begin{tabular}{@{}>{\centering\arraybackslash}p{3.8cm}@{}}
    \textcolor{green!65!black}{\faIcon{flask}}\\
    Biochemical Assays\\
    Separate metabolite records\\
    \textcolor{gray}{\scriptsize Concentration Profiles}
    \end{tabular}
};
\node[modernprocess] (scoring) at (13,-4) {
    \begin{tabular}{@{}>{\centering\arraybackslash}p{3.8cm}@{}}
    \textcolor{green!65!black}{\faIcon{wave-square}}\\
    Signal processing\\
    Kinetic modelling\\
    \textcolor{gray}{\scriptsize Proposed}
    \end{tabular}
};
\node[modernanalysis] (spectral) at (1,-7.5) {
    \begin{tabular}{@{}>{\centering\arraybackslash}p{3.8cm}@{}}
    \textcolor{orange!65!black}{\faIcon{moon}}\\
    Sleep architecture\\
    REM phase and episodes\\
    \textcolor{gray}{\scriptsize Within-participant models}
    \end{tabular}
};
\node[modernanalysis] (mapping) at (5,-7.5) {
    \begin{tabular}{@{}>{\centering\arraybackslash}p{3.8cm}@{}}
    \textcolor{orange!65!black}{\faIcon{map}}\\
    Statistical modelling\\
    Model comparisons\\
    \textcolor{gray}{\scriptsize Participant-level inference}
    \end{tabular}
};
\node[modernanalysis] (correlation) at (9,-7.5) {
    \begin{tabular}{@{}>{\centering\arraybackslash}p{3.8cm}@{}}
    \textcolor{orange!65!black}{\faIcon{project-diagram}}\\
    Metabolite covariance\\
    Partial correlation\\
    \textcolor{gray}{\scriptsize Log scale, participant bootstrap}
    \end{tabular}
};
\node[modernanalysis] (modeling) at (13,-7.5) {
    \begin{tabular}{@{}>{\centering\arraybackslash}p{3.8cm}@{}}
    \textcolor{orange!65!black}{\faIcon{chart-area}}\\
    Spectral slope\\
    D$_2$ availability\\
    \textcolor{gray}{\scriptsize Proposed}
    \end{tabular}
};
\node[modernresult] (integration) at (7,-11.5) {
    \begin{tabular}{c}
    \textcolor{red!65!black}{\Large\faIcon{microscope}}\\[4pt]
    \Large\textbf{Integrated Analysis}\\
    Constrained predictions and alternative models\\
    \textcolor{gray}{\small Agreement and disagreement across measurements}
    \end{tabular}
};
\foreach \source/\target in {eeg/signal, pet/kinetic, bio/assay, clinical/scoring} {
    \draw[modernarrow] (\source) -- (\target);
}
\foreach \source/\target in {signal/spectral, kinetic/mapping, assay/correlation, scoring/modeling} {
    \draw[modernarrow] (\source) -- (\target);
}
\draw[modernarrow, red!40] (spectral.south) -- ++(0,-0.5) -| ([xshift=10pt]integration.north west);
\draw[modernarrow, red!40] (mapping.south) -- ++(0,-0.7) -| ([xshift=-1.15cm]integration.north);
\draw[modernarrow, red!40] (correlation.south) -- ++(0,-0.7) -| ([xshift=1.15cm]integration.north);
\draw[modernarrow, red!40] (modeling.south) -- ++(0,-0.5) -| ([xshift=-10pt]integration.north east);
\node[font=\Large\bfseries\sffamily, text=black!80] at (7,2.5) 
    {Multiscale Test Programme};
\node[font=\footnotesize\sffamily, text=gray!75] at (7,2) 
    {Convergent evidence across neuroimaging, neurochemistry, and clinical assessment};
\node[rounded corners=6pt, draw=mygray70!45, fill=white, inner sep=5pt,
      font=\tiny\sffamily] at (7,1.25)
      {\textcolor{consciouscolor!65!black}{$\blacksquare$}~Available \quad
       \textcolor{deathcolor}{$\blacksquare$}~Proposed \quad
       \textcolor{mygray40}{Separate measurement windows}};
\node[gray!30, font=\Large] at (7,-2.25) {\faIcon{chevron-down}};
\node[gray!30, font=\Large] at (7,-5.75) {\faIcon{chevron-down}};
\node[gray!30, font=\Large] at (7,-9) {\faIcon{chevron-down}};
\end{tikzpicture}}{Multiscale test programme. The diagram separates available sleep, clinical, and CSF measurements from proposed EEG and receptor-imaging studies. Each comparison needs appropriate temporal alignment and an explicit observation model. The analysis layer compares observations with constrained predictions and alternative explanations.}{programme}

\subsection{Species comparisons and proposed rankings}\label{sec:species}
Species comparisons can provide tests of candidate capacity rankings.
Neuron counts provide anatomical measurements for comparison with the proposed functional capacity score.
The human cerebral cortex contains about 16 billion neurons in the cited study \cite{Azevedo2009}.
An elephant estimate gives about 5.6 billion cortical neurons, despite a much larger total neuron count \cite{HerculanoHouzel2014}.
A stereological study gives an estimate of about 37 billion neocortical neurons in the long-finned pilot whale \cite{Mortensen2014}.

\begin{table}[H]
\centering\small
\caption{Anatomical comparisons that challenge a single count-based capacity ranking. Values are approximate study estimates from different regions, samples, and counting methods. The final column identifies questions for the capacity model.}
\label{tab:species}
\begin{tabularx}{\textwidth}{@{}l r Y@{}}
\toprule
Species & Cortical neurons (billions) & Question for the capacity model \\
\midrule
Human & 16 & Does recurrence explain performance beyond count? \\
African elephant & 5.6 & How do cortical and cerebellar organization contribute? \\
Long-finned pilot whale & 37 & Does the proposed capacity measure transfer across cortical architectures? \\
\bottomrule
\end{tabularx}
\end{table}

\subsubsection{Anatomy, energy, and functional organization}\label{sec:capacitymetrics}
The proposed complexity ceiling depends on how tissue supports organized dynamics, rather than brain size alone.
Neuron number, packing density, body scaling, and energy allocation describe different constraints on this capacity.
Their contribution must be tested through the same substrate mapping in Equation~\ref{eq:substrate}.

Four anatomical and energetic descriptors make these comparisons explicit:
\begin{equation}
\begin{aligned}
\rho_N&=\frac{N_{\mathcal S}}{V_{\mathcal S}},
& b_{\mathrm{brain}}&=\frac{M_{\mathrm{brain}}}{M_{\mathrm{body}}},\\
\mathrm{EQ}&=\frac{M_{\mathrm{brain}}}{a_{\mathrm{all}}M_{\mathrm{body}}^{b_{\mathrm{all}}}},
& f_{\mathrm{met}}&=\frac{P_{\mathrm{brain}}}{P_{\mathrm{body}}}.
\end{aligned}
\label{eq:anatomicalmetrics}
\end{equation}
\begin{eqvars}
\item $\rho_N$ (\textbf{Neuronal density}): Number of neurons per unit tissue volume, distinct from receptor density.
\item $N_{\mathcal S},V_{\mathcal S}$ (\textbf{Matched count and volume}): Neuron count and volume within the same specified structure $\mathcal S$.
\item $b_{\mathrm{brain}}$ (\textbf{Brain mass fraction}): Fraction of body mass occupied by brain mass.
\item $M_{\mathrm{brain}},M_{\mathrm{body}}$ (\textbf{Brain and body masses}): Masses measured in compatible units and comparable physiological conditions.
\item $\mathrm{EQ}$ (\textbf{Encephalization quotient}): Observed brain mass divided by the brain mass expected from the declared body-size reference.
\item $a_{\mathrm{all}},b_{\mathrm{all}}$ (\textbf{Allometric reference}): Positive coefficient and exponent fitted in a specified comparison group. The coefficient carries the required units.
\item $f_{\mathrm{met}}$ (\textbf{Brain metabolic fraction}): Brain metabolic power divided by whole-body metabolic power under matched conditions.
\item $P_{\mathrm{brain}},P_{\mathrm{body}}$ (\textbf{Metabolic powers}): Energy use per unit time in brain and whole body, measured with compatible methods.
\end{eqvars}

Density distinguishes many neurons packed into a small volume from the same count spread through a larger structure.
A study of 28 avian species found high neuronal packing in songbird and parrot brains \cite{Olkowicz2016}.
Large parrots and corvids had forebrain neuron counts comparable to monkeys with much larger brains.
Such comparisons show why equal brain mass need not imply equal cellular resources.
Cross-species measurements must specify cortical, pallial, cerebellar, or whole-brain compartments before comparing counts and volumes.

The brain fraction and EQ answer different questions.
Doubling body mass at unchanged brain mass halves the brain fraction.
With an allometric exponent of $2/3$, the same change multiplies EQ by \mbox{$2^{-2/3}\approx0.630$}.
This example shows why an EQ requires a declared reference rather than one universal exponent.
Neither ratio specifies neuronal density or circuit organization.

Metabolic fraction measures allocation, while absolute brain power measures the available rate of energy expenditure.
A large fraction of a small whole-body budget can still represent little absolute brain power.
The reserve equations additionally distinguish supply, demand, and stored energy.
Comparisons should therefore measure absolute power and regional reserve alongside the fraction.
The theory predicts that energetic limits constrain sustained recurrent processing, rather than that a fixed metabolic percentage establishes consciousness.

Packing can also change geometric distances, but conduction depends on more than distance.
For a path of length $\ell$ and conduction speed $v_{\mathrm{cond}}$, the transmission time is $\ell/v_{\mathrm{cond}}$, before synaptic delays.
Shorter paths can support faster reciprocal interactions when other factors are fixed.
Connectivity, conduction speed, and metabolic limits must establish whether a denser preparation actually supports that advantage.
Count, volume, and density also obey an exact relation, so a statistical model must avoid treating all three as independent predictors.

\subsubsection{Signal diversity and organized integration}
An explicit diversity measure begins with a defined set of activity patterns:
\begin{equation}
H_{\mathrm{sig}}=-\sum_{k=1}^{K_{\mathrm{pat}}}\pi_k\log_2\pi_k,
\qquad
\mathcal D_H=\frac{H_{\mathrm{sig}}}{\log_2K_{\mathrm{pat}}},
\qquad 0\leq\mathcal D_H\leq1.
\label{eq:signalentropy}
\end{equation}
\begin{eqvars}
\item $H_{\mathrm{sig}}$ (\textbf{Pattern entropy}): Shannon entropy of the specified activity-pattern distribution, measured in bits.
\item $\pi_k$ (\textbf{Pattern probability}): Probability of pattern $k$. Probabilities sum to one, with $0\log_2 0=0$.
\item $K_{\mathrm{pat}}$ (\textbf{Pattern alphabet}): Number of possible patterns in a dictionary fixed before comparison, with $K_{\mathrm{pat}}>1$.
\item $\mathcal D_H$ (\textbf{Normalized pattern diversity}): Entropy divided by its maximum for that dictionary. It is one candidate measure of $\mathcal D$.
\end{eqvars}

The channel count, spatial scale, sampling interval, pattern definition, and recording duration must be comparable.
This entropy describes pattern diversity at the chosen scale.
Temporal unpredictability additionally depends on ordering, which requires conditional entropy or another declared temporal estimator.
Independent noise can have high values for both measures.
Neither measure alone establishes integrated experience.

Consider eight binary channels with the same possible alphabet of $2^8=256$ patterns.
Independent fair channels have $H_{\mathrm{sig}}=8$ bits and $\mathcal D_H=1$, despite having no interactions.
Eight perfectly synchronized fair channels visit only two patterns, giving $H_{\mathrm{sig}}=1$ bit and $\mathcal D_H=0.125$.
Their high correlation can arise without a large repertoire or reciprocal causal influence.
These exact examples separate diversity, synchrony, and integration.

The candidate integration measure $\mathcal I$ should quantify interactions between parts, rather than correlation alone.
A local perturbation can test the spatial spread and differentiation of responses under controlled input.
Recurrence $\mathcal R$ additionally concerns return influence through circuit loops.
In network preparations, selective interruption of return paths can distinguish recurrent responses from forward propagation or persistent local excitability.
Perturbation-based complexity measurements have distinguished human wakefulness, sleep, and anesthesia, providing a methodological precedent \cite{Casali2013}.
Their transfer across species requires separate calibration.

\subsubsection{Predictions that connect anatomy to the ceiling}
The proposed chain is anatomy and energy supply, then recurrent network dynamics, then the accessible complexity ceiling.
It predicts that density and brain allocation matter through the functional organization they support.
They cannot substitute for measurements of that organization.
For example, changing connectivity at fixed neuron count and receptor input can change the ceiling without changing neuronal density.

\begin{prediction}[Organized capacity]
At matched effective dopamine and pathway input, preparations with greater differentiated recurrent responses have a higher proposed complexity ceiling.
Interrupting the required recurrent connections lowers that ceiling even when spontaneous pattern entropy remains high.
\end{prediction}

Tests must define the response ceiling independently of the measures used to construct $\Phi$.
They should compare neuron count, density, body scaling, and metabolic allocation as competing predictors of that ceiling.
The functional score must improve predictions in species or preparations excluded from calibration.
Phylogenetic relatedness, sensory specialization, recording scale, and physiological state enter the comparison explicitly.
This prevents a score from succeeding only because closely related species share anatomy or because measurements favor one species.

The link to \HT{} signaling gives a further test.
At matched substrate capacity, receptor drive should change complexity below a common ceiling.
At matched receptor drive, different substrates can permit different ceilings.
The ceiling function must predict both comparisons with parameters fixed before evaluation.
This makes species differences and within-species neuromodulation parts of the same model.

A proposed rank based on $\Phi$ must predict a defined outcome in species outside the fitting sample.
Appropriate outcomes include response diversity, integration across modalities, and state-dependent changes in controlled tasks.
No single task can establish the full richness of another species' experience.
The model can still fail if its capacity ordering repeatedly predicts the wrong observable ordering.

\subsection{Evolutionary persistence and psychosis}\label{sec:evolution}
The evolutionary proposal links useful associative exploration with vulnerability to unstable control.
A broad serotonergic complexity range could expand the set of candidate interpretations, while dopamine-supported engagement makes those candidates behaviorally effective.
Within a viable range, this combination could help when familiar solutions fail.
Near a control boundary, the same variation could impair filtering, coordination, or reality testing.
The proposed advantage concerns a context-dependent trait, rather than the clinical disorder itself.

Polygenic scores for schizophrenia and bipolar disorder were associated with creative occupation and artistic society membership \cite{Power2015}.
This connects inherited liability with a measurable exploratory phenotype.
It does not identify a receptor pathway or establish increased reproductive fitness.
A separate study found no reproductive advantage for high schizophrenia or bipolar disorder polygenic scores \cite{Mullins2017}.
Genomic findings also support background selection as a contributor to the persistence of weakly harmful variants \cite{Pardinas2018}.

\paragraph{A stable mixture of strategies:}
One candidate mechanism is negative frequency-dependent selection.
Exploratory behavior could have greater individual returns when few others use the same strategy.
As exploration becomes common, duplicated searches and coordination costs could reduce those returns.
Selection can then preserve a mixture of exploratory and conservative traits.
This requires an advantage to individual fitness that changes with frequency.

A coarse selection model makes this claim explicit:
\begin{equation}
\frac{df_{\mathrm{ex}}}{dg}
=f_{\mathrm{ex}}(1-f_{\mathrm{ex}})
\bigl(b_{\mathrm{ex}}-c_{\mathrm{ex}}f_{\mathrm{ex}}\bigr).
\label{eq:evolutionfrequency}
\end{equation}
\begin{eqvars}
\item $f_{\mathrm{ex}}$ (\textbf{Exploratory trait frequency}): Population frequency of a specified exploratory trait class, between zero and one. It is not the frequency of a diagnosis.
\item $g$ (\textbf{Generational time}): Time measured in generations.
\item $b_{\mathrm{ex}}$ (\textbf{Rare-trait advantage}): Relative fitness advantage of the exploratory trait when it is rare, per generation.
\item $c_{\mathrm{ex}}$ (\textbf{Frequency-dependent cost}): Positive decline in that advantage as the trait becomes common, per generation.
\end{eqvars}
This reduced model assumes transmission of the trait classes and omits migration, new mutation, and drift.
When $0<b_{\mathrm{ex}}<c_{\mathrm{ex}}$, it has an internal equilibrium $f_{\mathrm{ex}}^*=b_{\mathrm{ex}}/c_{\mathrm{ex}}$.
The local slope is $-c_{\mathrm{ex}}f_{\mathrm{ex}}^*(1-f_{\mathrm{ex}}^*)<0$, so small frequency changes return toward that equilibrium.
Thus, a stable minority can persist without a universal preferred population percentage.

\paragraph{From trait variation to clinical outcomes:}
A population distribution of control parameters can retain a vulnerable tail near the model's instability boundaries.
Environmental stress, regional imbalance, and reserve loss can move different individuals across those boundaries.
A broadly stable trait distribution can therefore sustain recurrent clinical cases across generations.
The clinical fraction also depends on exposure, episode duration, ascertainment, and population structure.

This account predicts persistence, rather than a fixed percentage of psychotic illness in every population.
The EU-GEI study found eightfold variation in treated psychosis incidence across standardized catchment areas \cite{Jongsma2018}.
Incidence counts new cases over time, while prevalence counts existing cases.
Neither is identical to the frequency of exploratory traits or to the prevalence of schizotypal personality disorder.

\paragraph{Linked tests:}
The same declared control functions must predict useful exploration and sustained impairment in different contexts.
Greater effective complexity should help when tasks reward novel associations, but provide less benefit when stable filtering is essential.
The advantage should decrease when regional coordination deteriorates or physiological reserve becomes limiting.
For the frequency-dependent mechanism, reproductive returns must also decline as the exploratory trait becomes more common.

These tests require separate measures of receptor function, regional coordination, task performance, family structure, and reproductive outcomes.
A creativity association alone cannot establish the proposed selection mechanism.
Mutation, pleiotropy, drift, and background selection remain alternative explanations of persistence.

\section{Developmental modulation of the control surface}
Age can enter the model through receptor density, efficacy, clearance, connectivity, and metabolic reserve.
A piecewise receptor trajectory gives an illustrative developmental function:
\begin{samepage}
\begin{equation}
\rho_{\mathrm{demo}}(a)=
\begin{cases}
1.25+0.15(a-6)/9,&6\leq a<15,\\
1.40-0.40(a-15)/10,&15\leq a<25,\\
1.00-0.15(a-25)/25,&25\leq a<50,\\
0.85-0.15(a-50)/20,&50\leq a\leq70.
\end{cases}
\label{eq:development}
\end{equation}
\begin{eqvars}
\item $a$ (\textbf{Age}): Age in years, restricted here to ages 6 through 70.
\item $\rho_{\mathrm{demo}}(a)$ (\textbf{Density trajectory}): Illustrative receptor density relative to a common adult reference.
\end{eqvars}
\end{samepage}
The function is continuous at the stated boundaries.
Its values specify a developmental simulation for comparison with age-specific receptor measurements.
Figure~\ref{fig:development} shows the trajectory with an explicit reference choice.

\figinput{\begin{tikzpicture}[scale=1.1]
\tikzset{
  age_zone/.style  = {ellipse, line width=1.8pt, minimum width=3cm,
                      minimum height=2.4cm, fill=#1!12, draw=#1!45, drop shadow},
  age_label/.style = {rounded corners=5pt, inner sep=5pt, line width=1.2pt,
                      font=\footnotesize\sffamily\bfseries, fill=none, align=center,
                      draw=none, text=#1!90},
  phenom_box/.style= {rounded corners=3pt, inner sep=3pt, font=\tiny\sffamily,
                      align=center, fill=#1!10, draw=#1!45, text=#1!80},
  trajectory_point/.style={circle, minimum size=11pt,
                      font=\footnotesize\bfseries, fill=#1!70, text=white,
                      draw=white, line width=1.8pt, drop shadow},
  density_curve/.style={line width=2.3pt, #1!60, opacity=0.85},
  axis/.style={line width=2.3pt, mygray40, arrows={-Stealth[length=5pt]}},
  scale_tick/.style={mygray40, line width=1.5pt},
  vulnerability/.style={rounded corners=6pt, minimum width=3.4cm,
                        minimum height=2.6cm, draw=red!50, fill=red!7,
                        dashed, line width=2pt}
}
\draw[mygray70!08] (-7.5,-6) grid[step=2.0] (7.5,6.5);
\draw[axis] (-7.3,0)--(7.5,0);
\draw[axis] (0,-5.8)--(0,6.4);
\node[font=\small\sffamily\bfseries,fill=white,inner sep=2pt] at (6.7,-0.75) {Dopamine $\rightarrow$};
\node[font=\small\sffamily\bfseries,fill=white,inner sep=2pt,rotate=90] at (-0.4,6) {5-HT$_{2A}$ $\rightarrow$};
\foreach \x/\lab in {-6/Low, -3/, 0/Reference, 3/, 6/High}
  { \draw[scale_tick] (\x, 0.12)--(\x,-0.12);
    \node[font=\scriptsize\sffamily, text=mygray40] at (\x,-0.38) {\lab}; }
\foreach \y/\lab in {-5/Min, 3/High}
  { \draw[scale_tick] (0.12,\y)--(-0.12,\y);
    \node[font=\scriptsize\sffamily, anchor=east, text=mygray40] at (-0.18,\y) {\lab}; }
\node[age_zone=sleepcolor]   (child)  at (-4.4, 1.3) {};  
\node[age_zone=flowcolor]    (teen)   at ( 1.6, 4.1) {};  
\node[age_zone=dopaminecolor](adult)  at ( 3.1, 1.2) {};  
\node[age_zone=consciouscolor](mid)   at ( 0.8,-2.0) {};  
\node[age_zone=mygray70]      (old)   at (-3.0,-3.9) {};  
\node[age_label=sleepcolor]    at (child) {Childhood\\3-10 yr};

\node[age_label=flowcolor]     at (teen)  {Adolescence\\12-18 yr};
\node[phenom_box=flowcolor]    at ( 1.6, 3.0) {Proposed flexibility\\and vulnerability};
\node[age_label=dopaminecolor] at (adult) {Early Adult\\25-40 yr};
\node[phenom_box=dopaminecolor]at ( 2.3,0.4) {Proposed focus};
\node[age_label=consciouscolor]at (mid) {Mid-life\\45-60 yr};
\node[phenom_box=consciouscolor] at (0.8,-3.2) {Proposed consolidation};
\node[age_label=mygray70]      at (old) {Late life\\65+ yr};
\node[phenom_box=mygray70]     at (-3.0,-5.1) {Capacity variation};
\draw[line width=4pt, gray!55, rounded corners=9pt,
      postaction={draw,white,line width=2pt,shorten >=2pt,shorten <=2pt},
      ->, >=Stealth] (child) .. controls +(60:3) and +(180:2) .. (teen)
                    .. controls +(0:2)  and +(120:1.5) .. (adult)
                    .. controls +(-60:1.7) and +(40:1.5).. (mid)
                    .. controls +(220:2) and +(0:2)     .. (old);
\begin{scope}
\draw[fill=white, draw=mygray70!45, rounded corners=3pt, line width=.8pt]
      (2.85,-6.15) rectangle (7.35,-2.35);
\draw[mygray40, line width=1pt, -Stealth] (3.75,-5.45) -- (7.20,-5.45);
\draw[mygray40, line width=1pt, -Stealth] (3.75,-5.45) -- (3.75,-2.60);
\node[font=\tiny\sffamily, text=mygray40] at (5.5,-5.97) {Age (years)};
\node[font=\tiny\sffamily, text=mygray40, rotate=90] at (3.03,-4.0)
      {density / adult reference};
\foreach \a/\x in {0/3.75, 25/4.90, 50/6.05, 75/7.20}
  { \draw[mygray40] (\x,-5.45) -- (\x,-5.55);
    \node[font=\tiny\sffamily, text=mygray40] at (\x,-5.62) {\a}; }
\foreach \v/\y in {0.7/-4.9805, 1.0/-4.0415, 1.4/-2.7895}
  { \draw[mygray40] (3.75,\y) -- (3.68,\y);
    \node[font=\tiny\sffamily, text=mygray40, anchor=east] at (3.66,\y) {\v}; }
\draw[mygray70!45, dashed, line width=.5pt] (3.75,-4.0415) -- (7.20,-4.0415);
\draw[density_curve=dopaminecolor]
  plot coordinates{(4.026,-3.259)(4.440,-2.7895)(4.900,-4.0415)(6.050,-4.511)(6.970,-4.9805)};
\node[font=\tiny\sffamily, text=dopaminecolor!65!black, fill=none, inner sep=1pt]
      at (4.45,-2.6) {D$_1$/D$_2$};
\draw[density_curve=consciouscolor, dashed]
  plot coordinates{(4.026,-3.7285)(4.440,-2.946)(4.900,-4.0415)(6.050,-4.4484)(6.970,-4.824)};
\node[font=\tiny\sffamily, text=consciouscolor!65!black, fill=none, inner sep=1pt]
      at (6.6,-4.4) {5-HT$_{2A}$};
\node[font=\tiny\sffamily, text=mygray40, anchor=north east] at (7.30,-2.45)
      {illustrative density};
\end{scope}
\node[phenom_box=sleepcolor]   at (-4.4, 2.35) {Learning and\\exploration};
\node[rounded corners=5pt, fill=white, draw=mygray70!55, drop shadow,
      inner sep=5pt, font=\tiny\sffamily, align=left] at (-5,4.5)
      {\textbf{Legend} \vspace{0.2cm}\\
       \textcolor{sleepcolor}{$\bullet$} Childhood scenario\\
       \textcolor{flowcolor}{$\bullet$} Adolescent scenario\\
       \textcolor{dopaminecolor!65!black}{$\bullet$} Adult reference\\
       \textcolor{consciouscolor!65!black}{$\bullet$} Mid-life scenario\\
       \textcolor{mygray40}{$\bullet$} Late-life scenario};
\begin{pgfonlayer}{background}
  \node[vulnerability] (crit) at (1.6, 4.1) {};
  \node[font=\scriptsize\bfseries, align=center, text=red!60] at (1.9,5.65) {Candidate sensitivity window \\ Location requires measurement};
\end{pgfonlayer}
\end{tikzpicture}}{Conceptual developmental trajectory. The age regions and behavioral labels show proposed patterns. The inset has its own age axis. Its dopamine curve follows Equation~\ref{eq:development}, and the serotonin curve is a separate illustrative scenario. A common adult reference is required for density comparisons. Both curves use illustrative inputs.}{development}

Reference choice changes the prediction.
If both challenge and baseline signals contain the same multiplicative density, that density cancels in their ratio.
An age-dependent horizontal shift thus requires a common reference or another age-dependent mechanism.

For the simple response $S=A\rho L/(K+L)$, a target response $S^*$ requires
\begin{samepage}
\begin{equation}
L^*=\frac{K S^*}{A\rho-S^*},\qquad A\rho>S^*.
\label{eq:devdose}
\end{equation}
\begin{eqvars}
\item $S^*$ (\textbf{Target response}): Fixed target receptor response.
\item $L^*$ (\textbf{Required concentration}): Ligand concentration required to reach that target.
\item $A$ (\textbf{Efficacy}): Positive efficacy scale in this reduced developmental response.
\item $\rho$ (\textbf{Relative density}): Receptor density relative to the common reference.
\item $K$ (\textbf{Occupancy constant}): Positive half-occupancy concentration.
\end{eqvars}
\end{samepage}

Here, $A$ is an efficacy coefficient rather than the autoreceptor feedback term in Equation~\ref{eq:deff}.
The familiar inverse-density scaling is only a low-occupancy approximation.
The exact expression predicts loss of an attainable target when capacity falls below $S^*$.
This is a stronger, falsifiable developmental prediction.
It requires receptor and exposure measurements, not age alone.

\paragraph{A concrete density comparison:}
Consider the same target $S^*=0.50$, with $A=K=1$.
Relative densities of $1.4$, $1.0$, and $0.7$ require concentrations of $0.556$, $1.000$, and $2.500$, respectively.
These values describe model preparations, rather than measured age groups.
The required concentration rises steeply as receptor capacity approaches the target.
At $\rho\leq0.50$, no finite concentration reaches that target.

Receptor loss can therefore remove previously attainable responses while preserving others.
Development can also change the two axes differently if dopamine and \HT{} receptor densities follow different trajectories.
Their accessible state range can change in shape, rather than undergo one uniform shift.
Tests must compare matched ligand exposure and a common reference across ages.
Independent receptor measurements can then distinguish density changes from efficacy, clearance, and network development.

\section{Intracellular pathways and complexity}
The pathway extension separates receptor activation from its downstream effects:
\begin{samepage}
\begin{equation}
\Psi_{\mathrm{drive}}=\Psi_0
\left[1+\lambda_c f_{\mathrm{const}}+\sum_j\lambda_j f_j(L,t)\right]
e^{-\gamma(x/x_s)^2}R(t).
\label{eq:pathways}
\end{equation}
\begin{eqvars}
\item $j$ (\textbf{Pathway index}): Downstream pathway index.
\item $f_j(L,t)$ (\textbf{Pathway activity}): Pathway activity at ligand concentration $L$ and time $t$.
\item $\lambda_j$ (\textbf{Pathway weight}): Pathway contribution to the complexity drive.
\end{eqvars}
\end{samepage}
The pathway activities $f_j$ require assay-specific definitions.
Candidate branches include G-protein signaling, arrestin-related processes, and interactions with glutamatergic systems.
Their overlap requires care because a sum can count the same downstream effect twice.
A multivariate pathway model can give interaction estimates when sufficient interventions are available.

Experiments show ligand-dependent arrestin contributions to receptor signaling \cite{Schmid2008}.
Other experiments identify functional \HT{} and mGluR2 complexes in cortex \cite{GonzalezMaeso2008}.
These findings motivate separate pathway comparisons at matched receptor occupancy.

The proposed links to sensory richness, flexibility, and reality testing are mechanistic hypotheses.
Human cortical assays show pathway-dependent effects of pimavanserin \cite{Muneta2020}.
This finding strengthens the need for separate pathway measurements.

\figinput{\begin{tikzpicture}[scale=1.15]
\tikzset{
  mem/.style      = {double, double distance=3pt, line width=3pt, mygray40},
  bilip/.style    = {circle, fill=mygray70!30, minimum size=2.2pt},
  receptor/.style = {rounded corners=4pt, minimum width=28pt, minimum height=48pt,
                     line width=2.3pt, drop shadow},
  ligand/.style   = {star, star points=6, star point ratio=0.58,
                     minimum size=13pt, line width=1.4pt},
  protein/.style  = {ellipse, minimum width=25pt, minimum height=17pt,
                     font=\small\sffamily\bfseries, text=white,
                     line width=1.8pt, drop shadow},
  enzyme/.style   = {diamond, minimum width=18pt, minimum height=18pt,
                     inner sep=1.5pt, font=\footnotesize\sffamily, line width=1.8pt},
  sms/.style      = {circle, minimum size=9pt, font=\tiny\sffamily\bfseries,
                     line width=1.4pt, text=white},
  arrow/.style    = {-{Stealth[length=5pt,width=4pt]}, line width=1.6pt, shorten >=1.5pt, shorten <=1.5pt},
  arrow2/.style   = {arrow, dash pattern=on 3pt off 2pt},
  box/.style      = {rounded corners=12pt, dashed, line width=1.8pt,
                     minimum width=50pt, minimum height=32pt},
  effect/.style   = {ellipse, drop shadow, font=\small\sffamily\bfseries, align=center,
                     line width=1.8pt, minimum width=70pt, minimum height=30pt}
}
\fill[consciouscolor!8]  (-8,2.1) rectangle ( 9.5,5.1);
\node[font=\scriptsize\sffamily, text=consciouscolor!65!black] at (-6.8,4.65)
     {Extracellular};
\fill[mygray70!05] (-8,-5.1) rectangle (9.5,2.1);
\node[font=\scriptsize\sffamily, text=mygray40!75] at (-6.8,-4.65) {Cytoplasm};
\draw[mem] (-8,2.1).. controls (-4,2.35) and (4,1.85).. (9.5,2.1);
\foreach \x in {-7.5,-7,...,7.5} {
  \node[bilip] at (\x, 2.28) {};
  \node[bilip] at (\x, 1.92) {};
}
\node[receptor,fill=consciouscolor!38,draw=consciouscolor!70] (r5ht) at (-4,2.1) {};
\node[rotate=90, font=\footnotesize\bfseries, text=white] at (r5ht) {5-HT$_{2A}$};
\node[receptor,fill=sleepcolor!35,draw=sleepcolor!70,minimum width=24pt,
      minimum height=40pt] (mglur) at (-2.75,2.1) {};
\node[rotate=90, font=\footnotesize\bfseries, text=white] at (mglur) {mGluR2};
\foreach \j/\xx/\yy in {1/-5.6/3.7, 2/-4.6/4.1, 3/-3.6/3.7,
                        4/-2.6/4.1, 5/-1.6/3.7, 6/-0.6/4.1}
  \node[ligand,fill=consciouscolor!70,draw=consciouscolor!90] (L\j) at (\xx,\yy) {};
\node[font=\small\sffamily\bfseries, text=consciouscolor!65!black] at (-4.8,4.75) {5-HT};
\draw[arrow, consciouscolor] (L3) -- (r5ht.north);
\node[protein, fill=dopaminecolor!55, draw=dopaminecolor!75] (Gq) at (-3,0.55) {G$\alpha_q$};
\draw[arrow, dopaminecolor] (r5ht.south) -- (Gq);
\node[enzyme, fill=dopaminecolor!40, draw=dopaminecolor!75] (plc) at (-0.9,0.55) {PLC$\beta$};
\draw[arrow, dopaminecolor] (Gq) -- (plc);
\draw[dopaminecolor!30, line width=7pt] (-1.3,1.2)--(-0.5,1.2);
\node[font=\tiny\sffamily, text=mygray40] at (-0.87,1.4) {PIP$_2$};
\node[sms, fill=dopaminecolor!70, draw=dopaminecolor!90] (IP3) at (1.4,0.8) {IP$_3$};
\node[sms, fill=dopaminecolor!70, draw=dopaminecolor!90] (DAG) at (1.4,-0.4) {DAG};
\draw[arrow, dopaminecolor] (plc) -- (IP3);
\draw[arrow, dopaminecolor] (plc) -- (DAG);
\node[box, fill=dopaminecolor!9, draw=dopaminecolor!50] (er) at (3.4,0.6) {};
\node[font=\footnotesize\sffamily\bfseries, text=dopaminecolor!65!black] at (3.4,1.2) {ER};
\draw[arrow, dopaminecolor] (IP3) -- (er.west);
\foreach \k in {0,0.18,0.36}
  \node[sms,fill=dopaminecolor!85,draw=dopaminecolor!85,scale=0.55]
       at (3+2*\k,0.48+0.8*\k) {\tiny Ca$^{2+}$};
\node[enzyme, fill=dopaminecolor!40, draw=dopaminecolor!75] (PKC) at (3,-1.55) {PKC};
\draw[arrow, dopaminecolor] (DAG) -- (PKC);
\node[circle,fill=yellow!75,draw=orange!65,scale=0.6] at (3.32,-1.25) {P};
\node[enzyme, fill=mygray70!35, draw=mygray70!75, scale=0.8] (GRK) at (-5,0) {GRK};
\draw[arrow, mygray70] (GRK) -- (r5ht);
\foreach \yy in {2.15,1.9,1.65}
  \node[circle,fill=yellow!75,draw=orange!65,scale=0.55] at (-3.3,\yy) {P};
\node[protein, fill=flowcolor!55, draw=flowcolor!80] (BARR) at (-4,-1.55) {$\beta$-Arr2};
\draw[arrow, flowcolor] (r5ht.south) -- (BARR);
\node[enzyme, fill=flowcolor!45, draw=flowcolor!80] (Raf) at (0,-1.55) {Raf};
\node[enzyme, fill=flowcolor!45, draw=flowcolor!80] (MEK) at (0,-2.85) {MEK};
\node[enzyme, fill=flowcolor!45, draw=flowcolor!80] (ERK) at (0,-4.15) {ERK};
\draw[arrow, flowcolor] (BARR) -- (Raf);
\draw[arrow, flowcolor] (Raf) -- (MEK);
\draw[arrow, flowcolor] (MEK) -- (ERK);
\node[box, fill=flowcolor!9, draw=flowcolor!55, minimum width=70pt] (Nuc) at (2,-4.15) {};
\node[font=\footnotesize\bfseries\sffamily, text=flowcolor!65!black] at (2,-3.6) {Nucleus};
\draw[arrow2, flowcolor] (ERK) -- (Nuc.west);
\node[protein, fill=flowcolor!55, draw=flowcolor!80, scale=0.8] (CREB) at (2,-4.15) {CREB};
\node[circle,fill=yellow!75,draw=orange!65,scale=0.55] at (3.4,-4.7) {P};
\node[protein, fill=sleepcolor!55, draw=sleepcolor!80] (Gi) at (-1.5,-0.45) {G$\alpha_i$};
\draw[arrow, sleepcolor,preaction={draw=white,line width=3.6pt}] (mglur.south) -- (Gi);
\node[enzyme, fill=sleepcolor!45, draw=sleepcolor!80] (AC) at (1.2,-1.25) {AC};
\draw[-{Bar[width=5pt]},line width=1.4pt,sleepcolor] (Gi) -- (AC);
\node[sms, fill=sleepcolor!70, draw=sleepcolor!90] (cAMP) at (3.4,-2.8) {cAMP$\downarrow$};
\draw[arrow2, sleepcolor] (AC) -- (cAMP);
\node[effect, fill=dopaminecolor!12, draw=dopaminecolor!55,
      text=dopaminecolor!65!black] (percept) at (7.5,0.6) {Perceptual effects\\(hypothesis)};
\node[effect, fill=sleepcolor!12, draw=sleepcolor!55,
      text=sleepcolor!65!black] (inteffect) at (7.5,-2) {Integration\\(hypothesis)};
\node[effect, fill=flowcolor!12, draw=flowcolor!55,
      text=flowcolor!65!black] (plasticity) at (7.5,-4.1) {Plasticity and learning\\(hypothesis)};
\draw[arrow2,dopaminecolor!85!black] (PKC.east) -- (4.8,-1.55) |- (percept.west);
\draw[arrow2,sleepcolor!85!black]   (cAMP.east) -- (5,-2.8) |- (inteffect.west);
\draw[arrow2,flowcolor!85!black]    (Nuc.east) -- (plasticity.west);
\node[font=\scriptsize\sffamily,text=mygray40] at (-1,-5.0) {Pathway timing requires direct measurement};
\node[fill=white,rounded corners=5pt,draw=mygray70!55,drop shadow,
      inner sep=5pt,font=\tiny\sffamily,align=left] at (-5.5,-3.1)
{ \textbf{Pathway Key}\\
  \textcolor{dopaminecolor!65!black}{$\bullet$} G$_q$/PLC branch\\
  \textcolor{flowcolor}{$\bullet$} $\beta$-arrestin branch\\
  \textcolor{sleepcolor}{$\bullet$} G$_i$ branch\\[2pt]
  \textcolor{yellow!70}{$\circ$} Phosphorylation\\
  Bar: inhibition\\Dashed: indirect or proposed link };
\end{tikzpicture}}{Candidate intracellular routes. Molecular branches and proposed experience dimensions occupy separate layers. Dashed arrows mark indirect steps or proposed mappings. Biased ligands and pathway perturbations can distinguish these mappings at matched receptor occupancy. The molecular assignments remain candidate explanations of the indicated experience dimensions.}{pathways}

\paragraph{Pathway-specific predictions:}
Table~\ref{tab:pathwaypredictions} states which experience-related changes each branch is proposed to influence most strongly.
These assignments extend the scalar complexity sum in Equation~\ref{eq:pathways} to separately measured outcomes.
Their test requires matched receptor occupancy, exposure, dopamine state, and baseline performance.
Biochemical selectivity must be measured in the relevant pathway assays before a phenomenological contrast is interpreted.

\begin{table}[H]\centering\small
\caption{Proposed relative contributions of intracellular pathways. These are directional hypotheses for the routes in Figure~\ref{fig:pathways}, not established rankings of named compounds.}
\label{tab:pathwaypredictions}
\begin{tabularx}{\textwidth}{@{}p{3.0cm}Y Y@{}}\toprule
Pathway & Predicted emphasis & Discriminating outcome \\ \midrule
G$_q$/PLC/IP$_3$/DAG & Rapid sensory richness and perceptual alteration. & Greater sensory alteration under stronger measured G$_q$ drive, with arrestin recruitment and exposure matched. \\
$\beta$-arrestin-related signaling & Learning, persistent flexibility, and changes in meaning. Insight is a candidate experiential outcome. & Greater later learning or flexibility under stronger measured arrestin signaling, at matched acute sensory alteration. \\
mGluR2-associated G$_i$ signaling & Contextual integration and preservation of reality testing during altered experience. & Better discrimination of internal and external evidence under stronger measured pathway engagement, at matched sensory intensity. \\
\bottomrule\end{tabularx}
\end{table}

A dissociation would support the proposed division of function.
For example, increased sensory alteration without increased learning would separate perceptual and plasticity-related predictions.
Equivalent outcomes despite verified pathway separation would challenge the proposed assignments.

\paragraph{Connecting the experimental levels:}
Molecular assays first estimate $f_j(L,t)$ across concentration and time, including a condition without added agonist.
Receptor controls separate constitutive signaling from other sources of baseline assay activity.
These measurements constrain the input functions before any experience-related outcome is fitted.

Network preparations then compare response diversity, recurrence, and recovery after perturbation under matched pathway input.
Changing connectivity at matched receptor input tests whether substrate organization changes the response ceiling.
Random activity with high diversity provides a comparison for the proposed role of organized integration.
Section~\ref{sec:capacitymetrics} specifies the anatomical and dynamical measurements needed for this comparison.

Human comparisons should measure sensory alteration, learning, and contextual interpretation separately, alongside specified neural readouts.
Pathway weights estimated from a calibration panel must predict interventions excluded from that panel.
A single weighted complexity score cannot identify different pathway profiles.
Constitutive and ligand-driven effects require separate tests before they receive the same pathway weights.

\section{Receptor priming and rebound}
\subsection{The RPR hypothesis}
Receptor priming and rebound (RPR) is a translational extension of the theory.
It proposes that receptor adaptation can change the response to a later dopaminergic input.
The relevant quantity is functional gain after adaptation, not receptor count alone.
Existing animal work shows that antipsychotic exposure patterns can alter subsequent dopamine-related behavior \cite{Bedard2011}.

\begin{hypothesis}[Conditional receptor priming]
A priming intervention increases later useful signal only if postsynaptic gain exceeds autoreceptor feedback, residual blockade, and loss of signaling efficacy.
The effect also requires preserved circuit function and physiological reserve.
\end{hypothesis}

The receptor dynamics can be written as
\begin{samepage}
\begin{equation}
\dot\rho_j=\frac{\rho_{j,\mathrm{eq}}(B_j)-\rho_j}{\tau_{\rho,j}},
\qquad \dot B_j=-k_{B,j}B_j+u_{B,j}(t).
\label{eq:rprdynamic}
\end{equation}
\begin{eqvars}
\item $\rho_j$ (\textbf{Relative receptor density}): Receptor density relative to its reference density.
\item $\rho_{j,\mathrm{eq}}(B_j)$ (\textbf{Adapted equilibrium}): Sets the candidate equilibrium density under the selected antagonist input.
\item $\tau_{\rho,j}$ (\textbf{Adaptation time}): Positive adaptation time constant.
\item $B_j$ (\textbf{Antagonist input}): Effective antagonist occupancy or a specified exposure proxy.
\item $k_{B,j}$ (\textbf{Antagonist clearance}): Clearance constant for that antagonist input.
\item $u_{B,j}(t)$ (\textbf{Input rate}): Input rate, with units consistent with $B_j$.
\end{eqvars}
\end{samepage}

The receptor-equilibrium function and adaptation time are unknown.
They can differ between autoreceptors and postsynaptic receptors.
Figure~\ref{fig:rprtime} shows the two-phase concept in dimensionless time.

\figinput{\begin{tikzpicture}[xscale=1.4,yscale=1.1]
\tikzset{
    phasebox/.style={
        rectangle,
        rounded corners=12pt,
        minimum width=4.5cm,
        minimum height=1.5cm,
        font=\small\bfseries\sffamily,
        align=center,
        drop shadow={shadow xshift=0pt, shadow yshift=2pt, opacity=0.15}
    },
    timeline/.style={
        draw=gray!50,
        line width=2pt,
        ->,
        >=stealth
    },
    annotation/.style={
        font=\footnotesize\sffamily,
        text=black!70,
        align=center
    }
}
\shade[left color=red!5, right color=green!5] (-2,-6.8) rectangle (11,4);
\draw[timeline] (0,0) -- (10,0);
\node[annotation, above] at (0,0.2) {Start};
\node[annotation, above] at (3,0.2) {Priming};
\node[annotation, above] at (6,0.2) {Washout};
\node[annotation, above] at (10,0.2) {Probe};
\node[phasebox, fill=red!20, draw=red!40] (phase1) at (1.5,2) {
    \textcolor{red!65!black}{\faIcon{ban}} \textbf{Phase 1}\\
    Antagonist Priming\\
    \textcolor{gray}{\scriptsize Controlled receptor perturbation}
};
\node[phasebox, fill=green!20, draw=green!40] (phase2) at (7.5,2) {
    \textcolor{green!65!black}{\faIcon{rocket}} \textbf{Phase 2}\\
    Agonist Rebound\\
    \textcolor{gray}{\scriptsize Matched ligand exposure}
};
\draw[ultra thick, orange!60, dashed] (4.5,-4) -- (4.5,3.5);
\node[orange!70, font=\small\bfseries\sffamily] at (4.5,3.8) {Measure Adaptation};
\begin{scope}[shift={(0,-2.5)}]
    \draw[gray!30, line width=1pt] (0,0) -- (10,0);
    \node[annotation, align=right, anchor=east] at (-0.25,0.7) {Receptor\\density};
    \draw[blue!50, line width=1pt, dashed] (0,0.3) -- (10,0.3);
    \node[annotation, text=blue!60] at (10.5,0.3) {Baseline};
    \draw[ultra thick, blue!70, smooth] plot coordinates {
        (0,0.3) (1,0.3) (2,0.5) (3,0.9) (4,1.4) (4.5,1.5) 
        (5,1.3) (6,1.0) (7,0.7) (8,0.5) (9,0.4) (10,0.35)
    };
    \node[blue!70, font=\footnotesize\bfseries, anchor=west] at (5.2,2.1) {Density peak (illustrative)};
    \draw[blue!70, ->] (5.2,2) -- (4.5,1.5);
\end{scope}
\begin{scope}[shift={(0,-4)}]
    \draw[gray!30, line width=1pt] (0,0) -- (10,0);
    \node[annotation, align=right, anchor=east] at (-0.25,0) {Experimental\\state};
    \node[fill=red!30, rounded corners=4pt, font=\tiny\bfseries\sffamily,
          inner sep=4pt] at (2.5,0.3) {Blocked signaling};
    \node[fill=orange!30, rounded corners=4pt, font=\tiny\bfseries\sffamily,
          inner sep=4pt] at (4.5,0.3) {Transition};
    \node[fill=green!30, rounded corners=4pt, font=\tiny\bfseries\sffamily,
          inner sep=4pt] at (7.5,0.3) {Response to measure};
\end{scope}
\node[red!60, font=\small] at (2.5,-0.5) {\faIcon{exclamation-triangle}};
\node[orange!60, font=\small] at (4.3,-0.5) {\faIcon{radiation}};
\node[green!60, font=\small] at (7.5,-1.05) {\faIcon{question-circle}};
\node[fill=white, draw=gray!40, rounded corners=8pt, 
      font=\tiny\sffamily, inner sep=6pt, align=left] at (5,-5.6) {
    \textbf{Measurements:}\\
    - Receptor density\\
    - Occupancy and efficacy\\
    - Autoreceptor feedback\\
    - Circuit function
};
\node[font=\Large\bfseries\sffamily] at (5,4.5) {Receptor Priming and Rebound Hypothesis};
\end{tikzpicture}}{Conceptual RPR sequence. The receptor-density curve illustrates possible adaptation and recovery. The time axis is dimensionless and the curve is illustrative. Both receptor compartments, residual occupancy, and circuit function require measurement before a net-gain claim is possible.}{rprtime}

\subsection{A gain condition}
At matched ligand input, the idealized gain is
\begin{samepage}
\begin{equation}
G_{\mathrm{RPR}}=u_N\frac{1+A}{1+u_AA},
\qquad G_{\mathrm{RPR}}>1
\ \Longleftrightarrow\ 
u_N>\frac{1+u_AA}{1+A}.
\label{eq:rprgain}
\end{equation}
\begin{eqvars}
\item $G_{\mathrm{RPR}}$ (\textbf{Response gain}): Ratio of effective signal after adaptation to its baseline value.
\item $u_N$ (\textbf{Postsynaptic change}): Multiplicative change in the effective postsynaptic numerator.
\item $u_A$ (\textbf{Feedback change}): Multiplicative change in autoreceptor feedback.
\item $A$ (\textbf{Baseline feedback}): Baseline nonnegative autoreceptor feedback.
\end{eqvars}
\end{samepage}
Residual antagonist occupancy and reduced efficacy can lower $u_N$.
The inequality gives a direct rejection criterion.
If autoreceptor adaptation dominates, RPR can reduce the subsequent response.
Figure~\ref{fig:rprgain} shows the receptor mechanism and its gain condition.

\figinput{\begin{tikzpicture}[font=\small\sffamily]
\tikzset{
 panel/.style={rounded corners=10pt,draw=#1!40,fill=#1!5,line width=1pt},
 rec/.style={rectangle,rounded corners=1pt,minimum width=5pt,minimum height=10pt,
             inner sep=0pt,draw=#1!80,fill=#1!45,line width=.7pt},
 tag/.style={rounded corners=4pt,draw=#1!45,fill=white,inner sep=6pt,
             text width=2.8cm,minimum height=1.35cm,align=center,font=\footnotesize\sffamily},
 arr/.style={-{Stealth[length=6pt]},line width=1.6pt},
 outcome/.style={rounded corners=6pt,draw=#1!60,fill=#1!7,minimum width=4.2cm,
                 minimum height=1.25cm,align=center,font=\small\sffamily}
}
\node[font=\Large\bfseries] at (7,5.7) {Candidate Receptor Mechanism of RPR};
\node[font=\footnotesize,text=mygray40] at (7,5.15)
 {Postsynaptic adaptation and autoreceptor feedback must be measured separately};
\draw[panel=sleepcolor] (-.4,-.9) rectangle (6.6,4.65);
\draw[panel=flowcolor] (7.4,-.9) rectangle (14.4,4.65);
\node[font=\small\bfseries,text=sleepcolor!80!black] at (3.1,4.25) {REFERENCE AND LOW-GAIN STATES};
\node[font=\small\bfseries,text=flowcolor!80!black] at (10.9,4.25) {PRIMING AND MATCHED PROBE};
\foreach \xx/\name/\col/\count in {1.3/Reference/sleepcolor/7,4.9/Low gain/mygray70/3,9.1/Prime/wakecolor/9,12.7/Probe/flowcolor/9} {
 \begin{scope}[shift={(\xx,2.35)}]
  \node[font=\footnotesize\bfseries] at (0,1.25) {\name};
  \fill[\col!12] (0,0) circle (.70);
  \draw[\col!75,line width=1.6pt] (0,0) circle (.70);
  \node[text=\col!80!black,font=\tiny,align=center] at (0,0) {Recipient\\circuit};
  \foreach \j in {1,...,\count} {
   \pgfmathsetmacro{\ang}{360*\j/\count}
   \node[rec=\col,rotate=\ang-90] at (\ang:.70) {};
  }
  \ifnum\count=9
   \ifdim\xx pt<10pt
    \foreach \ang in {40,120,200,280} {
     \fill[mygray40] (\ang:.96) ++(-.055,-.055) rectangle ++(.11,.11);
    }
   \else
    \foreach \ang in {25,100,175,250,325} \fill[dopaminecolor] (\ang:1.04) circle (2.3pt);
   \fi
  \else
   \foreach \ang in {25,100,175,250,325} \fill[dopaminecolor] (\ang:1.04) circle (2.3pt);
  \fi
 \end{scope}
}
\draw[arr,mygray70] (2.5,2.35) -- (3.7,2.35);
\draw[arr,wakecolor!85!black,rounded corners=8pt] (1.3,1.45) -- (1.3,1.12) -- (8.96,1.12) -- (8.96,1.5);
\node[font=\tiny\bfseries,text=wakecolor!75!black,fill=white,draw=wakecolor!60,
      rounded corners=3pt,inner sep=2.5pt] at (7,1.12) {Antagonist exposure};
\draw[arr,flowcolor] (10.3,2.35) -- (11.5,2.35);
\node[font=\tiny,align=center,text=mygray40] at (10.9,1.65) {Washout\\then probe};
\node[tag=sleepcolor] at (1.3,.05) {Reference signal\\$N/(1+A)$};
\node[tag=mygray70] at (4.9,.05) {Density or efficacy\\can limit the signal};
\node[tag=wakecolor] at (9.1,.05) {Blockade and adaptation\\must be measured};
\node[tag=flowcolor] at (12.7,.05) {Matched ligand\\Measure net response};
\node[font=\tiny,text=mygray40] at (7,-1.4)
 {\textcolor{dopaminecolor}{$\bullet$} Dopamine \quad
  \textcolor{mygray40}{$\blacksquare$} Antagonist \quad
  Membrane bars: receptors \quad Symbol counts are illustrative};
\node[outcome=sleepcolor] (post) at (2.2,-2.75)
 {\textbf{Postsynaptic factor}\\$N\longrightarrow u_N N$};
\node[outcome=flowcolor] (auto) at (11.8,-2.75)
 {\textbf{Feedback factor}\\$A\longrightarrow u_A A$};
\node[panel=flowcolor,inner sep=10pt,align=center,minimum width=4.8cm,
      minimum height=1.65cm] (gain) at (7,-2.75)
 {\textbf{Net signaling gain}\\[5pt]
  $\displaystyle G_{\mathrm{RPR}}=\frac{u_N(1+A)}{1+u_A A}$};
\draw[arr,sleepcolor] (post.east) -- (gain.west);
\draw[arr,flowcolor] (auto.west) -- (gain.east);
\node[font=\footnotesize,align=center] at (7,-4.65)
 {Gain exceeds one only if $u_N>(1+u_A A)/(1+A)$.\\[3pt]
 Example: $u_N=1.4$, $u_A=2$. Only baseline feedback $A$ changes below.};
\node[outcome=consciouscolor] at (2.2,-6.1)
 {$A=0.2$ \quad $G=1.200$\\\textbf{Signal increases}};
\node[outcome=wakecolor] at (7,-6.1)
 {$A=1$ \quad $G=0.933$\\\textbf{Signal decreases}};
\node[outcome=deathcolor] at (11.8,-6.1)
 {$A=3$ \quad $G=0.800$\\\textbf{Signal decreases}};
\node[font=\footnotesize,text=mygray40,align=center] at (7,-7.3)
 {A receptor increase does not guarantee a signal increase or a clinical benefit.\\
 Residual blockade and efficacy changes enter $u_N$.};
\end{tikzpicture}}{Candidate receptor mechanism of RPR. Receptor and ligand symbols are schematic and do not show measured counts. The two panels compare a loss of effective response with a possible adaptation sequence that starts from the reference state. The lower panels show three gains from Equation~\ref{eq:rprgain} at fixed multipliers and different baseline feedback.
Residual blockade and efficacy changes enter $u_N$. Receptor upregulation alone does not establish benefit.}{rprgain}

\paragraph{Regional leverage of receptor adaptation:}
The numerator multiplier $u_N$ depends on the postsynaptic D$_2$ share of the original regional output in Equation~\ref{eq:numerator}.
Consider positive D$_2$ contributions of $75\%$ and $25\%$ of that numerator, with the other contribution unchanged.
A $40\%$ increase in the D$_2$ contribution gives $u_N=1.30$ and $u_N=1.10$, respectively.
With $A=1$ and $u_A=1.40$, Equation~\ref{eq:rprgain} gives gains of $1.083$ and $0.917$.
Thus, identical receptor adaptation can increase output in one circuit and decrease it in another.

The receptor terms and measured coefficient $\theta_r$ determine this regional difference.
A negative $\theta_r$ can reverse the direct postsynaptic contribution, within the domain where the total numerator remains nonnegative.
A local signaling gain also need not produce a large global change.
Equation~\ref{eq:softweight} determines how strongly the adapted region changes the bottleneck coordinate.
RPR must therefore predict both local signaling gain and its transfer through the regional bottleneck and state function.

RPR is an experimental hypothesis, not a clinical recommendation.
Its first tests concern receptor density, occupancy, signaling efficacy, and circuit function in preclinical systems.
Motor impairment, maladaptive sensitization, and destabilized state control are relevant competing outcomes.
The model gives no validated human dose, timing, or expected benefit.

\subsection{A path to validation}
The first comparison measures both receptor compartments after a controlled perturbation.
The next comparison measures signal at matched subsequent ligand exposure.
A selective manipulation of autoreceptor adaptation permits a direct test of the gain inequality.
Only evidence of useful circuit-level effects can motivate later translational work.
Figure~\ref{fig:rprpath} gives the staged programme with explicit decision criteria.

\figinput{\begin{tikzpicture}[
    node distance=0.8cm,
    stagebox/.style={
        rectangle,
        draw=#1!50,
        fill=white,
        line width=1.5pt,
        rounded corners=10pt,
        text width=10.5cm,
        minimum height=2.4cm,
        align=left,
        font=\small\bfseries\sffamily,
        drop shadow={shadow xshift=0pt, shadow yshift=2pt, opacity=0.15},
        inner xsep=8pt,
        inner ysep=10pt
    },
    riskbox/.style={
        fill=red!10, rounded corners=6pt, font=\tiny\sffamily,
        text width=2.2cm, minimum height=1.05cm,
        inner sep=4pt, align=center
    },
    milestone/.style={
        diamond,
        draw=orange!60,
        fill=orange!20,
        minimum width=30pt,
        minimum height=30pt,
        font=\tiny\bfseries\sffamily,
        inner sep=2pt
    },
    flowarrow/.style={
        ->,
        >=stealth,
        line width=2pt,
        #1!50
    }
]
\node[font=\Large\bfseries\sffamily, text=black!80] at (0,5) {RPR: Staged Tests and Decision Criteria};
\node[stagebox=blue] (invitro) at (0,3) {
    \begin{tabular}{@{}l@{}}
    \textcolor{blue!70}{\faIcon{flask}} \textbf{Stage 1: In Vitro Validation}\\[3pt]
    - iPSC-derived neurons from PD patients\\
    - Measure both receptor compartments\\
    - Measure occupancy and signaling efficacy
    \end{tabular}
};
\node[milestone] at ($(invitro.west)+(-1.0,0)$) {Test 1};
\node[stagebox=green, below=of invitro] (animal) {
    \begin{tabular}{@{}l@{}}
    \textcolor{green!65!black}{\faIcon{mouse}} \textbf{Stage 2: Pre-clinical Animal Models}\\[3pt]
    - 6-OHDA and MPTP rodent/primate models\\
    - Compare matched ligand responses and adverse effects\\
    - PET imaging of receptor dynamics
    \end{tabular}
};
\node[milestone] at ($(animal.west)+(-1.0,0)$) {Test 2};
\node[stagebox=purple, below=of animal] (biomarker) {
    \begin{tabular}{@{}l@{}}
    \textcolor{purple!70}{\faIcon{atom}} \textbf{Stage 3: Biomarker Development}\\[3pt]
    - [$^{11}\text{C}$] raclopride PET protocols\\
    - CSF dopamine metabolite profiles\\
    - Predictions in independent samples
    \end{tabular}
};
\node[milestone] at ($(biomarker.west)+(-1.0,0)$) {Test 3};
\node[stagebox=red, below=of biomarker] (clinical) {
    \begin{tabular}{@{}l@{}}
    \textcolor{red!65!black}{\faIcon{user-md}} \textbf{Stage 4: Translational Assessment}\\[3pt]
    - Requires prior evidence of useful circuit function\\
    - Requires an independent benefit-risk assessment\\
    - No human regimen or sample size is established
    \end{tabular}
};
\node[milestone] at ($(clinical.west)+(-1.0,0)$) {Decision};
\draw[flowarrow=blue] (invitro) -- (animal);
\draw[flowarrow=green] (animal) -- (biomarker);
\draw[flowarrow=purple] (biomarker) -- (clinical);
\node[riskbox] at ($(invitro.east)+(1.7,0)$) {
    \textbf{Key Risk:}\\
    Species differences\\
    in receptor dynamics
};
\node[riskbox] at ($(animal.east)+(1.7,0)$) {
    \textbf{Key Risk:}\\
    Severe motor effects\\
    in primates
};
\node[riskbox] at ($(biomarker.east)+(1.7,0)$) {
    \textbf{Key Risk:}\\
    Individual variability\\
    in response
};
\node[riskbox] at ($(clinical.east)+(1.7,0)$) {
    \textbf{Key Risk:}\\
    Rebound psychosis\\
    Ethical concerns
};
\node[draw=red!50, fill=red!5, rounded corners=8pt,
      font=\footnotesize\sffamily, inner sep=8pt, text width=10.5cm] [below=0.6cm of clinical] {
    \textcolor{red!65!black}{\faIcon{exclamation-triangle}} \textbf{Research Status:} 
    This hypothesis requires 
    preclinical evidence and independent review before any human application.
};
\end{tikzpicture}}{RPR validation sequence. Each stage requires a measured result before the next stage becomes justified. Failure of net signaling gain rejects the simple priming mechanism. A signaling gain without useful circuit function does not establish therapeutic benefit. The sequence specifies research stages.}{rprpath}

\section{Sharper predictions and future tests}
\subsection{Quantitative neural predictions}
The EEG and phase-amplitude coupling equations are candidate response families:
\begin{samepage}
\begin{align}
\alpha_{\mathrm{EEG}}(x)&=a_0+a_1x+a_2x^2,\label{eq:eeg}\\
\mathrm{PAC}(x)&=p_0\sqrt{x}\exp\left[-\frac{(x-m)^2}{2s^2}\right],\qquad x\geq0.
\label{eq:pac}
\end{align}
\begin{eqvars}
\item $\alpha_{\mathrm{EEG}}$ (\textbf{Spectral exponent}): Candidate exponent of the nonperiodic EEG spectrum.
\item $a_0,a_1,a_2$ (\textbf{EEG coefficients}): Coefficients of the candidate EEG response.
\item $\mathrm{PAC}$ (\textbf{Phase-amplitude coupling}): Specified phase-amplitude coupling measure.
\item $p_0$ (\textbf{Coupling amplitude}): Nonnegative PAC amplitude coefficient.
\item $m,s$ (\textbf{Curve location and width}): PAC curve location parameter and positive width.
\item $x$ (\textbf{Normalized dopamine input}): Integrated effective dopamine coordinate relative to its reference.
\end{eqvars}
\end{samepage}
The EEG exponent requires a defined spectral range and separation of periodic peaks.
PAC requires a defined montage, frequency range, estimator, and artifact control.
Neither quantity is identical to signal diversity or reported experience.
These equations predict shape only within a specified calibrated domain.

Illustrative EEG coefficients are $(a_0,a_1,a_2)=(1.821,-0.628,0.107)$ under $x=\mathrm{DA}/100$.
Their minimum occurs at $x=2.935$.
Illustrative PAC parameters are $(p_0,m,s)=(0.12,1.8,\sqrt{0.4})$.
Its positive maximum occurs at
\begin{samepage}
\begin{equation}
x_{\mathrm{PAC}}^*=\frac{m+\sqrt{m^2+2s^2}}{2}=1.905.
\label{eq:pacpeak}
\end{equation}
\begin{eqvars}
\item $x_{\mathrm{PAC}}^*$ (\textbf{Peak input}): Normalized input at the positive maximum of the illustrative PAC curve.
\end{eqvars}
\end{samepage}
These are mathematical properties of the illustrative curves.
They are not measured locations or confidence limits.
Future studies can fit the coefficients.
Separate tests must assess curve shape in participants outside training.

\paragraph{A joint shape test:}
The two curves predict different responses to the same input change.
For the illustrative coefficients, increasing $x$ from $1$ to $2.5$ lowers the spectral exponent throughout that interval.
Coupling instead rises and then falls around its interior maximum at $x=1.905$.
This separates a monotonic spectral change from a nonmonotonic coupling response within the specified interval.

One candidate implementation measures theta-phase to gamma-amplitude coupling, with predefined bands of $4$--$8$ Hz and $30$--$80$ Hz.
The estimator, recording montage, and artifact procedure remain fixed across the comparison.
Effective dopamine must be measured independently of these EEG outcomes.
Calibration data set the curve parameters, while separate data test both predicted shapes.
Measurements must cover both sides of the coupling peak.
A monotonic segment alone cannot distinguish the proposed curve from a monotonic alternative.

\subsection{Shared-parameter predictions of the dopamine model}
The strongest tests require the same parameters to predict several different perturbations.
A flexible dose-to-signal map and a flexible sleep observation function can reproduce many associations.
The model thus needs independent calibration of its input and observation functions.
The following tests impose exact restrictions on spatial change, temporal change, and receptor adaptation.

\paragraph{Move the deficit while preserving the mean:}
Consider two equally weighted regions with signals $d+\delta$ and $d-\delta$, where $d\geq|\delta|$.
Equation~\ref{eq:softmin} gives
\begin{samepage}
\begin{equation}
D_\kappa=d-\frac{\log\cosh(\kappa\delta)}{\kappa},\qquad
D_\kappa(\mathbf D+a\mathbf 1)=D_\kappa(\mathbf D)+a.
\label{eq:redistribution}
\end{equation}
\begin{eqvars}
\item $d$ (\textbf{Common mean}): Fixed mean of the two regional signals.
\item $\delta$ (\textbf{Regional displacement}): Signed regional deviation from that mean.
\item $\mathbf D$ (\textbf{Signal vector}): Vector of regional signals.
\item $a$ (\textbf{Common shift}): Common additive signal shift in this identity.
\item $\mathbf 1$ (\textbf{All-ones vector}): Vector with every entry equal to one.
\end{eqvars}
\end{samepage}
Regional imbalance lowers this coordinate at fixed mean.
A common regional shift produces the same shift in the coordinate.
A mean-only model predicts no effect of the redistribution.
For $\kappa=4$ and $\delta=0.3$, the illustrative reduction is 0.148 normalized units.
This is an equation output, not an expected biological effect size.

The local sensitivity also obeys an exact ratio:
\begin{samepage}
\begin{equation}
\frac{\partial D_\kappa/\partial D_r}{\partial D_\kappa/\partial D_s}
=\frac{p_r}{p_s}\exp[-\kappa(D_r-D_s)].
\label{eq:rescueratio}
\end{equation}
\begin{eqvars}
\item $r,s$ (\textbf{Region pair}): The two regions being compared.
\item $p_r/p_s$ (\textbf{Prior-weight ratio}): Their prior-weight ratio.
\item $D_r-D_s$ (\textbf{Signal difference}): Their integrated signal difference.
\end{eqvars}
\end{samepage}
An experiment can move a controlled deficit between two regions and compare small restoration inputs.
The predicted sensitivity must follow the deficit with the same weights and $\kappa$.
A fixed anatomical gate instead favors the same region across these changes.
A nonspecific network model may also respond to imbalance, so selective signaling measurements are essential.

The direction of an access or report change requires a calibrated observation slope.
Before the collapse transition, lower coordinated input gives lower intensity at fixed context.
Within the collapse transition, reducing input can instead restore intensity.
The experiment must fix the response branch before the test.
At fixed context, a common nonzero observation slope cancels from the ratio of small regional effects.
This makes Equation~\ref{eq:rescueratio} more restrictive than one successful regional stimulation.

\paragraph{Predict amplitude and timing together:}
For small sinusoidal inputs, the first-order memory filter predicts
\begin{samepage}
\begin{equation}
G(\omega)=\frac{1}{\sqrt{1+(\omega\tau_m)^2}},\qquad
\varphi(\omega)=\arctan(\omega\tau_m),\qquad
G=\cos\varphi.
\label{eq:gainphase}
\end{equation}
\begin{eqvars}
\item $G(\omega)$ (\textbf{Amplitude gain}): Response amplitude relative to its value at zero frequency.
\item $\omega$ (\textbf{Input frequency}): Angular input frequency, in radians per unit time.
\item $\varphi(\omega)$ (\textbf{Phase lag}): Phase lag in radians.
\item $\tau_m$ (\textbf{Memory time}): The same positive time constant as the regional memory filter.
\end{eqvars}
\end{samepage}
One time constant must explain both quantities across input frequencies.
The actual effective input and output require measurement.
An additional observation filter can alter this relationship and must be calibrated independently.

A study can fit the time constant on one waveform and predict other frequencies, reverse ramps, and long plateaus.
Pure lag produces loops that shrink toward zero as a smooth ramp becomes slower.
Persistent separation of transition thresholds requires an additional state mechanism.
Clock time, sleep pressure, noise-driven switching, and measurement delay are competing explanations.

For the feedback model in Equation~\ref{eq:transition}, ordinary folds require
\begin{samepage}
\begin{equation}
5cz^4-3bz^2+a=0,\qquad 9b^2>20ac.
\label{eq:foldcondition}
\end{equation}
\begin{eqvars}
\item $z$ (\textbf{Transition state}): Transition state from Equation~\ref{eq:transition}.
\item $a,b,c$ (\textbf{Feedback coefficients}): The same feedback coefficients used in that transition equation.
\end{eqvars}
\end{samepage}
If $b^2>4ac$, the symmetric model has three stable equilibria at zero input, separated by unstable equilibria.
Thus, this polynomial does not automatically define a two-state switch.
Its fitted parameters must reproduce the number of stable states and their transitions in independent input sequences.

\paragraph{Predict response gain and reversal:}
Equation~\ref{eq:rprgain} predicts when the same postsynaptic adaptation has opposite effects in preparations with different autoreceptor feedback.
For illustrative changes $u_N=1.4$ and $u_A=2$, baseline feedback $A=0.2$, $1$, and $3$ gives gains 1.20, 0.933, and 0.800.
The receptor-count-only account predicts benefit too broadly in this example.
Independent measurements of both compartments must predict the later response before its measurement.
Residual occupancy, efficacy, ligand input, and circuit function need separate controls.

These tests preserve a common model across conditions.
Successful prediction of redistribution, rate dependence, and gain reversal would constrain alternatives more strongly than another dose correlation.
Failure with precise input and output measurements would locate a specific equation that needs modification.
The present PPMI files do not contain the simultaneous regional or receptor measurements required for these decisive tests.

\figinput{\input{figures/22_structural_predictions.tex}}{Exact predictions of the spatial and temporal reductions. Panel (a) holds the mean of two equally weighted regions at one. The soft minimum decreases with imbalance, while the mean stays constant. Panel (b) compares one first-order filter with two equal filters in series. Both gains are normalized at zero frequency. These equation-generated curves are experimental targets, not participant observations.}{structuralpredictions}

\begin{samepage}
\subsection{Fixed regional benchmark and prospective tests}\label{sec:fixedbenchmark}
Benchmark W0 gives a numerical version of the five-box hypothesis in Figure~\ref{fig:regions}.
It converts the dot counts into relative weights by a declared convention.
Neither the counts nor their numerical conversion are biological measurements.
The benchmark fixes the following values:
\begin{eqvars}
\item $p_{\mathrm{PFC}}=4/11$ (\textbf{Prefrontal weight}): Gives the prefrontal box the largest prior contribution.
\item $p_{\mathrm{motor}}=3/11$ (\textbf{Motor cortical weight}): Gives the motor cortical box the second-largest prior contribution.
\item $p_{\mathrm{striatum}}=2/11$ (\textbf{Striatal weight}): Gives the striatal box a smaller, positive prior contribution.
\item $p_{\mathrm{VTA}}=1/11$ (\textbf{VTA weight}): Specifies the direct contribution of the VTA box to this candidate readout.
\item $p_{\mathrm{SN}}=1/11$ (\textbf{Nigral weight}): Specifies the direct contribution of the substantia nigra box.
\item $\kappa=4$ (\textbf{Fixed bottleneck strength}): Uses the regional signal scale on which every reference signal equals one.
\end{eqvars}
\end{samepage}

The weights sum to one.
The value $\kappa=4$ follows the illustrative mathematical checks, not a fitted receptor parameter.
These values define one candidate for future evaluation.
They were not preregistered for the present PPMI analyses.
A precise contradiction would reject W0 within its declared domain.
Changing its weights after evaluation would define a new candidate that requires new evaluation data.

\paragraph{How far can a regional deficit lower the coordinate?}
An isolated deficit need not bring the bottleneck close to zero.
Consider a set $\mathcal F$ with integrated signals equal to $\ell$, while all remaining signals equal $d>\ell$.
For $\kappa>0$, Equation~\ref{eq:softmin} gives the exact two-level expression
\begin{samepage}
\begin{equation}
D_\kappa=\ell-\frac{1}{\kappa}\log\left[p_{\mathcal F}
+(1-p_{\mathcal F})e^{-\kappa(d-\ell)}\right],
\qquad p_{\mathcal F}=\sum_{r\in\mathcal F}p_r.
\label{eq:deficitfloor}
\end{equation}
\begin{eqvars}
\item $\mathcal F$ (\textbf{Reduced-signal set}): Identifies regions with the same lower integrated signal.
\item $\ell$ (\textbf{Lower signal}): Gives their nonnegative integrated signal after temporal filtering.
\item $d$ (\textbf{Remaining signal}): Gives the common integrated signal in the other regions, with $d>\ell$.
\item $p_{\mathcal F}$ (\textbf{Affected weight}): Sums the fixed prior weights of the reduced-signal regions.
\end{eqvars}
\end{samepage}
For a target $\ell\leq D_\star<d$, the condition $D_\kappa\leq D_\star$ is equivalent to
\begin{samepage}
\begin{equation}
p_{\mathcal F}\geq
\frac{e^{-\kappa(D_\star-\ell)}-e^{-\kappa(d-\ell)}}
{1-e^{-\kappa(d-\ell)}}.
\label{eq:reachability}
\end{equation}
\begin{eqvars}
\item $D_\star$ (\textbf{Target coordinate}): Specifies a target on the bottleneck scale, without assigning a clinical state.
\end{eqvars}
\end{samepage}
The condition concerns affected weight, rather than the number of anatomical boxes.
It is unchanged when a region is subdivided with its signal and total weight preserved.
It applies exactly to this two-level pattern, not to every possible regional configuration.

In W0, complete prefrontal suppression with all other signals at one gives a lower bound of $0.245$.
The target $0.20$ would require approximately a prefrontal signal of $-0.046$, which violates the nonnegative-signal assumption.
At $\ell=0$, $d=1$, and $\kappa=4$, the required affected weight is approximately $0.439$.
The prefrontal weight $4/11$ does not satisfy this condition.
Several reduced signals can pass this target without changing W0 (Table~\ref{tab:reachability}).

\begin{table}[H]
\centering\small
\caption{Reachability under fixed W0. Signals follow the declared regional order and are already temporally integrated. Each row is a mathematical input scenario. No row is a measured seizure or a clinical state classification.}
\label{tab:reachability}
\begin{tabularx}{\textwidth}{@{}Y c c@{}}
\toprule
Integrated signals (PFC, motor, striatum, VTA, SN) & $D_\kappa$ & $D_\kappa\leq0.20$ \\
\midrule
$(1,1,1,1,1)$ & $1.000$ & No \\
$(0,1,1,1,1)$ & $0.245$ & No \\
$(0,0.3,1,1,1)$ & $0.198$ & Yes \\
$(0.1,1,0.1,0.1,0.1)$ & $0.177$ & Yes \\
$(0,0,1,1,1)$ & $0.110$ & Yes \\
\bottomrule
\end{tabularx}
\end{table}

Changing $\kappa$ has different limits for different input vectors.
For $(0,1,1,1,1)$, the target $0.20$ becomes reachable at $\kappa\approx4.999$.
For $(0.7,1.4,1,1,1)$, the bottleneck instead approaches $0.70$ as $\kappa$ increases without bound.
Neither change is part of W0.
An alternative weight or $\kappa$ must define a separate candidate before evaluation.

Reaching $0.20$ does not establish loss of consciousness.
The value is a component location $\mu_1$ in Equation~\ref{eq:anchors}, not a calibrated loss threshold.
For W0, the reference bottleneck is one, so the normalized state input $x$ equals $D_\kappa$.

Use the normalized state curve in Appendix~\ref{app:simulation}, with reserve and injury factors fixed at one.
Then $C(1)=1.000$, $C(0.245)=0.119$, and $C(0.20)=0.086$.
The waking reference lies on a rising branch, and isolated prefrontal suppression lowers the output in this example.
Section~\ref{sec:seizure} uses the same response with an illustrative access threshold.
That observation criterion does not establish a universal clinical boundary.

Distributed signal reduction provides a candidate route to lower bottleneck coordinates.
The observation boundary and state function determine whether that reduction produces impaired experience.
A complete test requires all regional inputs, their temporal integration, and independently calibrated state and observation functions.
Calibration must establish the response branch and the operational criterion for impaired access before evaluation.

\paragraph{A concrete redistribution:}
Unequal weights require preservation of the weighted mean, rather than the arithmetic mean.
Start with every regional signal equal to $d$.
Change two signals to $D_r=d+\delta$ and $D_s=d-(p_r/p_s)\delta$, with all signals nonnegative.
Keep every other signal equal to $d$.
Then $\overline D=d$, while Equation~\ref{eq:softmin} gives
\begin{samepage}
\begin{equation}
D_\kappa=d-\frac{1}{\kappa}\log\left[
1-p_r-p_s+p_r e^{-\kappa\delta}
+p_s e^{\kappa(p_r/p_s)\delta}\right]\leq d.
\label{eq:weightedspread}
\end{equation}
\begin{eqvars}
\item $d$ (\textbf{Common starting signal}): Sets the baseline in all five regions and the preserved weighted mean.
\item $r,s$ (\textbf{Perturbed regions}): Select two regions with positive weights.
\item $\delta$ (\textbf{Redistribution size}): Increases region $r$ while region $s$ changes by the compensating weighted amount.
\end{eqvars}
\end{samepage}

For W0, the signal vector $(0.7,1.4,1,1,1)$ follows the region order listed above.
Its weighted mean stays at one, but its bottleneck signal is $0.878$.
The prefrontal-to-motor local sensitivity ratio changes from $4/3$ to $21.926$.
These numbers are exact-model predictions rounded for presentation, not measured effects on conscious intensity.
An intensity prediction also needs the fixed state function and its response branch.
Equation~\ref{eq:redistribution} remains the special case of two equally weighted regions.

\paragraph{A restriction that cancels the unknown weights:}
An additional comparison can assess fixed regional weights without knowing their values.
Both selected regions must have positive weights.
Let $b_r^{(A)}$ and $b_s^{(A)}$ be local outcome slopes per measured regional signal change at baseline state $A$.
A common nonzero observation slope cancels from their ratio within that state.
Repeat the regional perturbations at baseline state $B$, with the same weights and $\kappa$.
Equation~\ref{eq:rescueratio} then implies
\begin{samepage}
\begin{equation}
\log\frac{b_r^{(B)}/b_s^{(B)}}{b_r^{(A)}/b_s^{(A)}}
=-\kappa\left[(D_r^{(B)}-D_s^{(B)})-(D_r^{(A)}-D_s^{(A)})\right].
\label{eq:ratiotransfer}
\end{equation}
\begin{eqvars}
\item $A,B$ (\textbf{Baseline states}): Identify two measurement conditions, separate from the autoreceptor term $A_r$.
\item $b_r^{(A)}$ (\textbf{Local outcome slope}): Gives the small outcome change per measured signal change in region $r$, at state $A$.
\item $D_r^{(A)}$ (\textbf{Baseline regional signal}): Gives the integrated signal before the small perturbation in state $A$.
\end{eqvars}
\end{samepage}

The prior-weight ratio cancels across states.
The observation slopes can differ between states, but must be common across the regional perturbations within each state.
Opposite regional response signs within one state would contradict this scalar readout under these conditions.
Near-zero slopes make the ratio unreliable.
Joint inference for the regional slopes must propagate this uncertainty without removing results because they are near zero.

A calibration contrast with a nonzero change in the regional signal difference can identify $\kappa$ under the model.
Its estimate and uncertainty must then predict different contrasts.
For W0, the prediction instead uses $\kappa=4$ without recalibration.
Neither approach permits refitting on the evaluation outcomes.

\paragraph{Requirements before evaluation:}
Numerical weights alone do not specify an executable experiment.
The study plan must fix anatomical boundaries, hemisphere treatment, reference normalization, signal assays, and the outcome.
It must also fix the state branch, physiological context, temporal filter, uncertainty model, and rejection tolerance.
Independent calibration data can determine the observation mapping before evaluation.
Anatomical subdivisions must preserve the weight assigned to each parent region.

The figure contains both dopamine sources and recipient circuits.
Source stimulation can change several regional signals, even when its direct readout weight is small.
Predictions must use the measured full regional change, rather than equating stimulation dose with one local signal change.
Selective perturbations or an identifiable perturbation matrix are needed to recover separate regional slopes.
The study must control direct effects on reporting, movement, serotonergic signaling, and physiological reserve.
These requirements are not met by the present PPMI measurements.

\subsection{Priority for empirical evaluation}
The shared-parameter predictions above require direct regional measurements and controlled perturbations.
Fixed regional weights and a common state readout must precede evaluation.
Independent data must assess the predicted redistribution, rescue, and sensitivity-transfer relationships.
These comparisons can distinguish the proposed regional law from a mean-signal account.

A prospective REM study should record administration time, exposure, and validated sleep stages.
It should specify entry and exit outcomes, zero-REM nights, and unknown-epoch handling before outcome inspection.
The competing NREM exits to REM and waking must be evaluated together with fixed boundaries and entry durations.
A dopamine rise that ends in waking must not be counted as evidence for REM entry.
A controlled formulation comparison also needs matched exposure and a direct stage-specific contrast.
The sustained-entry pattern motivates this prospective design.

A stronger regional response study needs verified same-day OFF and ON examinations, repeated baselines, and concurrent imaging.
Its observation model must allow baseline measurement error and ordinal score limits.
Direct experience measures are necessary before this motor comparison can assess conscious intensity.
These measurements would connect the motor comparison to the regional intervention tests specified above.

\section{Discussion}
The model maps coordinated regional dopamine signaling to an increasing recruitment response with an upper collapse gate.
Serotonergic pathways contribute to dynamic complexity within a substrate-dependent ceiling.
Here, complexity concerns the experienced richness and significance of contents, measured separately from subjective presence and cognitive flexibility.
Regional heterogeneity and temporal dynamics connect this architecture to transitions and failures of state control.
The later extensions ask how the same architecture changes across reserve, development, and receptor adaptation.

The architecture proposes laws of conscious state control.
It does not specify which physical processes constitute an experience.
This structural limit is separate from the measurement limits of PPMI.
The conditional sufficiency claim can still fail through different outcomes at matched modeled inputs and physiological context.

The mathematical constraints make these claims specific.
Normalized regional weights prevent arbitrary changes of scale.
The receptor reduction distinguishes monotonic attenuation from a peak under an explicit affinity condition.
The injury variable separates a reversible gate from death.
The RPR inequality states when receptor adaptation can help, fail, or reverse the intended effect.
These constraints yield different predictions for competing mechanisms.

Regional heterogeneity provides a common candidate mechanism for the contrasting state effects discussed above.
Causal attribution requires interventions that separate this mechanism from alternative explanations.

Both timing definitions point toward later REM entry as the formulation component increases at fixed total levodopa LEDD.
The sustained-entry coefficients remain positive after single deletions and the specified joint deletions.
Removing the largest contributor reduces the trend-adjusted estimate from $+0.905$ to $+0.482$ minutes.
Joint-deletion intervals include zero, consistent with concentrated information and limited precision.
Confounding by indication and exposure-dependent staging error remain distinct alternatives to a dopamine-mediated entry effect.
Prospective entry and exit measurements must distinguish selective timing effects from general changes in sleep.

The regional ordinal comparison permits imaging to explain latent baseline burden under a common medication response.
The interval for additional predictive value spans zero in 254 participants.
Repeated baselines and standardized drug challenges would better distinguish regional response from baseline disease burden.

The five commitments in Table~\ref{tab:commitments} distinguish fixed tests from extensions that need calibration.
Regional weights, observation functions, and state-specific parameters can fit many patterns if they are unrestricted.
Scientific progress thus requires a small number of fixed predictions that share parameters across conditions.
Failures must lead to a narrower domain or a revised mechanism, rather than a new unmeasured variable for every discrepancy.
The proposed tests identify several places where this discipline is possible.

Benchmark W0 fixes one numerical candidate, while the two-state sensitivity ratio removes the unknown prior weights from another prediction.
Both require direct regional measurements and independent evaluation.

W0 also limits how far an isolated deficit can lower the bottleneck.
Complete prefrontal suppression does not reach $0.20$ when the other signals remain at reference.
Distributed deficits can cross that coordinate, but a lower coordinate does not by itself establish lower intensity.
Independent calibration must link the numerical input to state intensity before a clinical prediction is evaluated.

X0 complements the bottleneck with an explicit regional cost model.
It separates coordinated signal availability from demand, with exact restrictions for matched-bottleneck and fixed-mean comparisons.
The quadratic cost and its demand weights remain unvalidated.
Neither local excess nor increased demand automatically establishes a high conscious state or loss of consciousness.

The optional receptor peak supplies a separate deficit route only when its limiting signal lies below the specified reference.
Its time scale requires kinetic measurements.
X1 declares protective and amplifying reserve regimes, while the injury extension adds repair and bounded adaptation.
The low-signal factor enforces zero intensity at zero coordinated input.
Fatal outcomes still require irreversible injury and failure of essential functions.

\section{Conclusion}
The theory proposes a specific regional law for conscious state control.
Its distinctive contribution is a set of transferable restrictions, especially the sensitivity ratio that cancels unknown regional weights.
These restrictions define what future measurements must satisfy, independently of the success of a medication association.
The REM entry pattern provides a specific target for prospective evaluation.
The regional control law requires direct evaluation beyond these observational measures.
The next decisive evidence needs regional measurements, controlled perturbations, and direct outcomes related to conscious experience.

\section*{Data, code, and declarations}
The analysis uses existing PPMI data \cite{PPMI2011}.
Access to participant records follows PPMI requirements.
The source package contains figure sources and a reproduction guide.
Its ancillary files include numerical-check scripts and aggregate results.
Participant records are not included in the manuscript package.

Appendix~\ref{app:empiricalmethods} reports the complete timing families, supporting methods, and numerical model checks.
No new participant intervention was performed for this analysis.
The proposed experiments require their own approvals before implementation.

\paragraph{Computational assistance:}
Generative AI tools were used to assist with mathematical exploration, the development of analysis code and literature searches.
Numerical results were verified using executable scripts.
The author takes full responsibility for all scientific claims, analyses, references, text and figures.

\paragraph{PPMI data and funding:}
\begingroup
\small
The analysis used the Parkinson's Progression Markers Initiative (PPMI) database (RRID:SCR\_006431).
The first data download was in July 2025, followed by later updates.
Data access follows the PPMI process at \url{https://www.ppmi-info.org/access-data-specimens/download-data}.
For current study information, see \url{https://www.ppmi-info.org}.

PPMI (a public-private partnership) is funded by the Michael J. Fox Foundation for Parkinson's Research and the following funding partners:

\begin{table}[H]
\centering
\begingroup
\footnotesize
\itshape
\renewcommand{\arraystretch}{1}

\begin{tabularx}{1\linewidth}{@{}>{\hsize=1.15\hsize}Y>{\hsize=1.05\hsize}Y>{\hsize=.8\hsize}Y@{}}
AbbVie & Critical Path Institute & Neumora \\
Alamar Biosciences & DaCapo Brainscience & Neuron23 \\
Aligning Science Across Parkinson's (ASAP) & Denali & Novartis \\
Arrowhead Pharma & Edmond J. Safra Foundation & Olink \\
Arvinas & Eli Lilly & Regeneron \\
AskBio & Gain Therapeutics & Roche \\
BIAL & GE Healthcare & Sanofi \\
BioArctic & Genentech & Tenvie \\
Biohaven & GSK & UCB \\
BlueRock Therapeutics & Insitro & Vanqua Bio \\
Bristol Myers Squibb & Johnson \& Johnson Innovative Medicine &
Voyager Therapeutics \\
Calico Labs & Lundbeck & The Weston Family Foundation \\
Capsida Biotherapeutics & Merck & \\
\end{tabularx}

\endgroup
\end{table}
\endgroup

The author declares no competing financial interests related to this work.

\bibliographystyle{unsrt}
\begingroup\small
\bibliography{references}

@article{Li2018Propofol,
  author = {Li, Jia and Yu, Tian and Shi, Fu and Zhang, Yu and Duan, Zikun and Fu, Bao and Zhang, Yi},
  title = {Involvement of Ventral Periaqueductal Gray Dopaminergic Neurons in Propofol Anesthesia},
  journal = {Neurochemical Research},
  year = {2018}, volume = {43}, number = {4}, pages = {838--847},
  doi = {10.1007/s11064-018-2486-y},
  note = {\href{https://doi.org/10.1007/s11064-018-2486-y}{DOI: 10.1007/s11064-018-2486-y}}
}

@article{Olkowicz2016,
  author = {Olkowicz, Seweryn and Kocourek, Martin and Lu{\v c}an, Radek K. and Porte{\v s}, Michal and Fitch, W. Tecumseh and Herculano-Houzel, Suzana and N{\v e}mec, Pavel},
  title = {Birds have primate-like numbers of neurons in the forebrain},
  journal = {Proceedings of the National Academy of Sciences},
  year = {2016}, volume = {113}, number = {26}, pages = {7255--7260},
  doi = {10.1073/pnas.1517131113},
  note = {\href{https://doi.org/10.1073/pnas.1517131113}{DOI: 10.1073/pnas.1517131113}}
}

@article{Casali2013,
  author = {Casali, Adenauer G. and Gosseries, Olivia and Rosanova, Mario and Boly, M{\'e}lanie and Sarasso, Simone and Casali, Karina R. and Casarotto, Silvia and Bruno, Marie-Aur{\'e}lie and Laureys, Steven and Tononi, Giulio and Massimini, Marcello},
  title = {A theoretically based index of consciousness independent of sensory processing and behavior},
  journal = {Science Translational Medicine},
  year = {2013}, volume = {5}, number = {198}, pages = {198ra105},
  doi = {10.1126/scitranslmed.3006294},
  note = {\href{https://doi.org/10.1126/scitranslmed.3006294}{DOI: 10.1126/scitranslmed.3006294}}
}

@article{Siclari2017,
  author = {Siclari, Francesca and Baird, Benjamin and Perogamvros, Lampros and Bernardi, Giulio and LaRocque, Joshua J. and Riedner, Brady and others},
  title = {The neural correlates of dreaming},
  journal = {Nature Neuroscience},
  year = {2017},
  volume = {20},
  number = {6},
  pages = {872--878},
  doi = {10.1038/nn.4545},
 note = {\href{https://doi.org/10.1038/nn.4545}{DOI: 10.1038/nn.4545}}
}

@article{Taylor2016,
  author = {Taylor, Norman E. and Van Dort, Christa J. and Kenny, Jonathan D. and Pei, JunZhu and Guidera, Jennifer A. and Vlasov, Ksenia Y. and others},
  title = {Optogenetic activation of dopamine neurons in the ventral tegmental area induces reanimation from general anesthesia},
  journal = {Proceedings of the National Academy of Sciences},
  year = {2016},
  volume = {113},
  number = {45},
  pages = {12826--12831},
  doi = {10.1073/pnas.1614340113},
 note = {\href{https://doi.org/10.1073/pnas.1614340113}{DOI: 10.1073/pnas.1614340113}}
}

@article{Hasegawa2022,
  author = {Hasegawa, Emi and Miyasaka, Ai and Sakurai, Katsuyasu and Cherasse, Yoan and Li, Yulong and Sakurai, Takeshi},
  title = {Rapid eye movement sleep is initiated by basolateral amygdala dopamine signaling in mice},
  journal = {Science},
  year = {2022},
  volume = {375},
  number = {6584},
  pages = {994--1000},
  doi = {10.1126/science.abl6618},
 note = {\href{https://doi.org/10.1126/science.abl6618}{DOI: 10.1126/science.abl6618}}
}

@article{Hasegawa2024,
  author = {Hasegawa, Emi and Li, Yulong and Sakurai, Takeshi},
  title = {Regulation of {REM} sleep in mice: The role of dopamine and serotonin function in the basolateral amygdala},
  journal = {Neuroscience Research},
  year = {2024},
  volume = {200},
  pages = {28--33},
  doi = {10.1016/j.neures.2023.09.003},
  note = {\href{https://doi.org/10.1016/j.neures.2023.09.003}{DOI: 10.1016/j.neures.2023.09.003}}
}

@article{Madsen2019,
  author = {Madsen, Martin K. and Fisher, Patrick M. and Burmester, Daniel and Dyssegaard, Agnete and Stenb{\ae}k, Dea S. and Kristiansen, Sara and others},
  title = {Psychedelic effects of psilocybin correlate with serotonin {2A} receptor occupancy and plasma psilocin levels},
  journal = {Neuropsychopharmacology},
  year = {2019},
  volume = {44},
  number = {7},
  pages = {1328--1334},
  doi = {10.1038/s41386-019-0324-9},
 note = {\href{https://doi.org/10.1038/s41386-019-0324-9}{DOI: 10.1038/s41386-019-0324-9}}
}

@article{Schartner2017,
  author = {Schartner, Michael M. and Carhart-Harris, Robin L. and Barrett, Adam B. and Seth, Anil K. and Muthukumaraswamy, Suresh D.},
  title = {Increased spontaneous {MEG} signal diversity for psychoactive doses of ketamine, {LSD} and psilocybin},
  journal = {Scientific Reports},
  year = {2017},
  volume = {7},
  pages = {46421},
  doi = {10.1038/srep46421},
 note = {\href{https://doi.org/10.1038/srep46421}{DOI: 10.1038/srep46421}}
}

@article{PPMI2011,
  author = {{Parkinson Progression Marker Initiative}},
  title = {The {Parkinson Progression Marker Initiative} ({PPMI})},
  journal = {Progress in Neurobiology},
  year = {2011},
  volume = {95},
  number = {4},
  pages = {629--635},
  doi = {10.1016/j.pneurobio.2011.09.005},
 note = {\href{https://doi.org/10.1016/j.pneurobio.2011.09.005}{DOI: 10.1016/j.pneurobio.2011.09.005}}
}

@techreport{Verily2025,
  author = {Rainaldi, Erin},
  title = {Derived Metrics for the {Verily Study Watch}},
  institution = {Verily Life Sciences},
  year = {2025},
  note = {PPMI data-user technical note, distributed as PPMI\_Verily\_Derived\_Data\_Rev\_20250327.pdf. The document has an inconsistent legacy header date. Not formally peer reviewed}
}

@techreport{PPMIUserGuide,
  author = {{Parkinson's Progression Markers Initiative}},
  title = {{PPMI} Data User Guide},
  institution = {Parkinson's Progression Markers Initiative},
  year = {2024},
  url = {https://www.ppmi-info.org/sites/default/files/docs/PPMI%20Data%20User%20Guide.pdf},
  note = {\href{https://www.ppmi-info.org/sites/default/files/docs/PPMI%20Data%20User%20Guide.pdf}{Source document}. Section 4.1 specifies SEX coding. Accessed September 5, 2026}
}

@misc{SinemetCR,
  author = {{DailyMed}},
  title = {{SINEMET CR} (carbidopa levodopa) Sustained-Release Tablets},
  howpublished = {U.S. National Library of Medicine, prescribing information},
  url = {https://dailymed.nlm.nih.gov/dailymed/downloadpdffile.cfm?setid=69e575b9-f8a5-494f-b736-2520ef505cb0},
  note = {\href{https://dailymed.nlm.nih.gov/dailymed/downloadpdffile.cfm?setid=69e575b9-f8a5-494f-b736-2520ef505cb0}{Source document}. Clinical Pharmacology and Dosage and Administration. Accessed September 5, 2026}
}

@article{Weiner2001,
  author  = {Weiner, David M. and Burstein, Ethan S. and Nash, Norman and
             Croston, Glenn E. and Currier, Erika A. and Vanover, Kimberly E.
             and Harvey, Susan C. and Donohue, Elaine and Hansen, Hans C.
             and Andersson, Carl M. and Spalding, Tracy A. and Gibson,
             Donald F. C. and Krebs-Thomson, Kirsten and Powell, Susan B.
             and Geyer, Mark A. and Hacksell, Uli and Brann, Mark R.},
  title   = {5-Hydroxytryptamine\textsubscript{2A} receptor inverse agonists
             as antipsychotics},
  journal = {Journal of Pharmacology and Experimental Therapeutics},
  volume  = {299}, number = {1}, pages = {268--276}, year = {2001},
 note = {\href{https://pubmed.ncbi.nlm.nih.gov/11561089/}{Source document}}
}

@article{Vanover2006,
  author  = {Vanover, Kimberly E. and Weiner, David M. and Makhay, Manuel
             and Veinbergs, Ingrid and Gardell, Luis R. and Lameh, Jelveh
             and Del Tredici, Andria L. and Piu, Fabrice and Schiffer,
             Hans H. and Ott, Thomas R. and Burstein, Ethan S. and Uldam,
             Anne K. and Thygesen, Mikkel B. and Schlienger, Nathalie and
             Andersson, Carl M. and Son, Tae Y. and Harvey, Susan C. and
             Powell, Susan B. and Geyer, Mark A. and Tolf, Bo-Ragnar and
             Brann, Mark R. and Davis, Robert E.},
  title   = {Pharmacological and behavioral profile of
             N-(4-fluorophenylmethyl)-N-(1-methylpiperidin-4-yl)-N'-
             (4-(2-methylpropyloxy)phenylmethyl) carbamide (ACP-103), a
             novel 5-hydroxytryptamine\textsubscript{2A} receptor inverse
             agonist},
  journal = {Journal of Pharmacology and Experimental Therapeutics},
  volume  = {317}, number = {2}, pages = {910--918}, year = {2006},
 note = {\href{https://doi.org/10.1124/jpet.105.097006}{Source document}}
}

@article{Azevedo2009,
  author  = {Azevedo, Frederico A. C. and Carvalho, Ludmila R. B. and Grinberg, Lea T. and Farfel, Jos{\'e} M. and Ferretti, Renata E. L. and Leite, Renata E. P. and others},
  title   = {Equal numbers of neuronal and nonneuronal cells make the human brain an isometrically scaled-up primate brain},
  journal = {Journal of Comparative Neurology},
  volume  = {513},
  number  = {5},
  pages   = {532--541},
  year    = {2009},
  doi     = {10.1002/cne.21974},
 note = {\href{https://doi.org/10.1002/cne.21974}{DOI: 10.1002/cne.21974}}
}

@article{Mortensen2014,
  author  = {Mortensen, Heidi S. and Pakkenberg, Bente and Dam, Maria and Dietz, Rune and Sonne, Christian and Mikkelsen, Bjarni and others},
  title   = {Quantitative relationships in delphinid neocortex},
  journal = {Frontiers in Neuroanatomy},
  volume  = {8},
  pages   = {132},
  year    = {2014},
  doi     = {10.3389/fnana.2014.00132},
 note = {\href{https://doi.org/10.3389/fnana.2014.00132}{DOI: 10.3389/fnana.2014.00132}}
}

@article{HerculanoHouzel2014,
  author  = {Herculano-Houzel, Suzana and Avelino-de-Souza, Kamilla and Neves, Kleber and Porfirio, Jairo and Messeder, Debora and Mattos Feij{\'o}, Larissa and others},
  title   = {The elephant brain in numbers},
  journal = {Frontiers in Neuroanatomy},
  volume  = {8},
  pages   = {46},
  year    = {2014},
  doi     = {10.3389/fnana.2014.00046},
 note = {\href{https://doi.org/10.3389/fnana.2014.00046}{DOI: 10.3389/fnana.2014.00046}}
}

@article{Beggs2003,
 author = {Beggs, John M. and Plenz, Dietmar},
 title = {Neuronal Avalanches in Neocortical Circuits},
 journal = {The Journal of Neuroscience},
 year = {2003},
 volume = {23},
 number = {35},
 pages = {11167-11177},
 doi = {10.1523/JNEUROSCI.23-35-11167.2003},
 note = {\href{https://doi.org/10.1523/JNEUROSCI.23-35-11167.2003}{DOI: 10.1523/JNEUROSCI.23-35-11167.2003}}
}

@article{Bedard2011,
 author = {Bédard, Anne-Marie and Maheux, Jérôme and Lévesque, Daniel and Samaha, Anne-Noël},
 title = {Continuous, but not Intermittent, Antipsychotic Drug Delivery Intensifies the Pursuit of Reward Cues},
 journal = {Neuropsychopharmacology},
 year = {2011},
 volume = {36},
 number = {6},
 pages = {1248-1259},
 doi = {10.1038/npp.2011.10},
 note = {\href{https://doi.org/10.1038/npp.2011.10}{DOI: 10.1038/npp.2011.10}}
}

@article{Ma2008,
 author = {Ma, Zhiyuan and Zhang, Gongliang and Jenney, Chris and Krishnamoorthy, Swapna and Tao, Rui},
 title = {Characterization of serotonin-toxicity syndrome (toxidrome) elicited by 5-hydroxy-l-tryptophan in clorgyline-pretreated rats},
 journal = {European Journal of Pharmacology},
 year = {2008},
 volume = {588},
 number = {2-3},
 pages = {198-206},
 doi = {10.1016/j.ejphar.2008.04.004},
 note = {\href{https://doi.org/10.1016/j.ejphar.2008.04.004}{DOI: 10.1016/j.ejphar.2008.04.004}}
}

@article{Goodwin2022,
 author = {Goodwin, Guy M. and Aaronson, Scott T. and Alvarez, Oscar and Arden, Peter C. and Baker, Annie and Bennett, James C. and others},
 title = {Single-Dose Psilocybin for a Treatment-Resistant Episode of Major Depression},
 journal = {New England Journal of Medicine},
 year = {2022},
 volume = {387},
 number = {18},
 pages = {1637-1648},
 doi = {10.1056/NEJMoa2206443},
 note = {\href{https://doi.org/10.1056/NEJMoa2206443}{DOI: 10.1056/NEJMoa2206443}}
}

@article{Muneta2020,
 author = {Muneta-Arrate, Itziar and Diez-Alarcia, Rebeca and Horrillo, Igor and Meana, J. Javier},
 title = {Pimavanserin exhibits serotonin 5-HT2A receptor inverse agonism for G$\alpha$i1- and neutral antagonism for G$\alpha$q/11-proteins in human brain cortex},
 journal = {European Neuropsychopharmacology},
 year = {2020},
 volume = {36},
 pages = {83-89},
 doi = {10.1016/j.euroneuro.2020.05.004},
 note = {\href{https://doi.org/10.1016/j.euroneuro.2020.05.004}{DOI: 10.1016/j.euroneuro.2020.05.004}}
}

@article{GonzalezMaeso2008,
 author = {González-Maeso, Javier and Ang, Rosalind L. and Yuen, Tony and Chan, Pokman and Weisstaub, Noelia V. and López-Giménez, Juan F. and others},
 title = {Identification of a serotonin/glutamate receptor complex implicated in psychosis},
 journal = {Nature},
 year = {2008},
 volume = {452},
 number = {7183},
 pages = {93-97},
 doi = {10.1038/nature06612},
 note = {\href{https://doi.org/10.1038/nature06612}{DOI: 10.1038/nature06612}}
}

@article{Schmid2008,
 author = {Schmid, Cullen L. and Raehal, Kirsten M. and Bohn, Laura M.},
 title = {Agonist-directed signaling of the serotonin 2A receptor depends on $\beta$-arrestin-2 interactions
                    in vivo},
 journal = {Proceedings of the National Academy of Sciences},
 year = {2008},
 volume = {105},
 number = {3},
 pages = {1079-1084},
 doi = {10.1073/pnas.0708862105},
 note = {\href{https://doi.org/10.1073/pnas.0708862105}{DOI: 10.1073/pnas.0708862105}}
}

@article{Englot2009,
 author = {Englot, D. J. and Modi, B. and Mishra, A. M. and DeSalvo, M. and Hyder, F. and Blumenfeld, H.},
 title = {Cortical deactivation induced by subcortical network dysfunction in limbic seizures},
 journal = {Journal of Neuroscience},
 year = {2009},
 volume = {29},
 number = {41},
 pages = {13006-13018},
 doi = {10.1523/JNEUROSCI.3846-09.2009},
 note = {\href{https://doi.org/10.1523/JNEUROSCI.3846-09.2009}{DOI: 10.1523/JNEUROSCI.3846-09.2009}}
}

@article{Englot2010,
 author = {Englot, D. J. and Yang, L. and Hamid, H. and Danielson, N. and Bai, X. and Marfeo, A. and others},
 title = {Impaired consciousness in temporal lobe seizures: role of cortical slow activity},
 journal = {Brain},
 year = {2010},
 volume = {133},
 number = {12},
 pages = {3764-3777},
 doi = {10.1093/brain/awq316},
 note = {\href{https://doi.org/10.1093/brain/awq316}{DOI: 10.1093/brain/awq316}}
}

@article{Vijayraghavan2007,
 author = {Vijayraghavan, S. and Wang, M. and Birnbaum, S. G. and Williams, G. V. and Arnsten, A. F. T.},
 title = {Inverted-{U} dopamine {D1} receptor actions on prefrontal neurons engaged in working memory},
 journal = {Nature Neuroscience},
 year = {2007},
 volume = {10},
 number = {3},
 pages = {376-384},
 doi = {10.1038/nn1846},
 note = {\href{https://doi.org/10.1038/nn1846}{DOI: 10.1038/nn1846}}
}

@article{Berridge2006,
 author = {Berridge, C. W. and Devilbiss, D. M. and Andrzejewski, M. E. and Arnsten, A. F. T. and Kelley, A. E. and Schmeichel, B. and others},
 title = {Methylphenidate preferentially increases catecholamine neurotransmission within the prefrontal cortex at low doses that enhance cognitive function},
 journal = {Biological Psychiatry},
 year = {2006},
 volume = {60},
 number = {11},
 pages = {1111-1120},
 doi = {10.1016/j.biopsych.2006.04.022},
 note = {\href{https://doi.org/10.1016/j.biopsych.2006.04.022}{DOI: 10.1016/j.biopsych.2006.04.022}}
}

@article{Kitagawa1990,
 author = {Kitagawa, K. and Matsumoto, M. and Tagaya, M. and Hata, R. and Ueda, H. and Niinobe, M. and others},
 title = {'Ischemic tolerance' phenomenon found in the brain},
 journal = {Brain Research},
 year = {1990},
 volume = {528},
 number = {1},
 pages = {21-24},
 doi = {10.1016/0006-8993(90)90189-I},
 note = {\href{https://doi.org/10.1016/0006-8993(90)90189-I}{DOI: 10.1016/0006-8993(90)90189-I}}
}

@article{Allen2005,
 author = {Allen, N. J. and Karadottir, R. and Attwell, D.},
 title = {A preferential role for glycolysis in preventing the anoxic depolarization of rat hippocampal area {CA1} pyramidal cells},
 journal = {Journal of Neuroscience},
 year = {2005},
 volume = {25},
 number = {4},
 pages = {848-859},
 doi = {10.1523/JNEUROSCI.4157-04.2005},
 note = {\href{https://doi.org/10.1523/JNEUROSCI.4157-04.2005}{DOI: 10.1523/JNEUROSCI.4157-04.2005}}
}

@article{VanHall2009,
 author = {van Hall, G. and Stromstad, M. and Rasmussen, P. and Jans, O. and Zaar, M. and Gam, C. and others},
 title = {Blood lactate is an important energy source for the human brain},
 journal = {Journal of Cerebral Blood Flow and Metabolism},
 year = {2009},
 volume = {29},
 number = {6},
 pages = {1121-1129},
 doi = {10.1038/jcbfm.2009.35},
 note = {\href{https://doi.org/10.1038/jcbfm.2009.35}{DOI: 10.1038/jcbfm.2009.35}}
}

@article{Owen1967,
 author = {Owen, O. E. and Morgan, A. P. and Kemp, H. G. and Sullivan, J. M. and Herrera, M. G. and Cahill, G. F.},
 title = {Brain metabolism during fasting},
 journal = {Journal of Clinical Investigation},
 year = {1967},
 volume = {46},
 number = {10},
 pages = {1589-1595},
 doi = {10.1172/JCI105650},
 note = {\href{https://doi.org/10.1172/JCI105650}{DOI: 10.1172/JCI105650}}
}

@article{McCullough1999,
 author = {McCullough, J. N. and Zhang, N. and Reich, D. L. and Juvonen, T. S. and Klein, J. J. and Spielvogel, D. and others},
 title = {Cerebral metabolic suppression during hypothermic circulatory arrest in humans},
 journal = {Annals of Thoracic Surgery},
 year = {1999},
 volume = {67},
 number = {6},
 pages = {1895-1899},
 doi = {10.1016/S0003-4975(99)00441-5},
 note = {\href{https://doi.org/10.1016/S0003-4975(99)00441-5}{DOI: 10.1016/S0003-4975(99)00441-5}}
}

@article{Beier2015,
 author={Beier, Kevin T. and Steinberg, Elizabeth E. and DeLoach, Katherine E. and Xie, Stanley and Miyamichi, Kazunari and Schwarz, Lindsay and Gao, Xiaojing J. and Kremer, Eric J. and Malenka, Robert C. and Luo, Liqun},
 title={Circuit Architecture of {VTA} Dopamine Neurons Revealed by Systematic Input-Output Mapping},
 journal={Cell}, year={2015}, volume={162}, number={3}, pages={622--634},
 doi={10.1016/j.cell.2015.07.015}
}

@article{Lerner2015,
 author={Lerner, Talia N. and Shilyansky, Carrie and Davidson, Thomas J. and Evans, Kathryn E. and Beier, Kevin T. and Zalocusky, Kelly A. and Crow, Ailey K. and Malenka, Robert C. and Luo, Liqun and Tomer, Raju and Deisseroth, Karl},
 title={Intact-Brain Analyses Reveal Distinct Information Carried by {SNc} Dopamine Subcircuits},
 journal={Cell}, year={2015}, volume={162}, number={3}, pages={635--647},
 doi={10.1016/j.cell.2015.07.014}
}

@article{Xie2013,
  author = {Xie, Lulu and Kang, Hongyi and Xu, Qiwu and Chen, Michael J. and Liao, Yonghong and Thiyagarajan, Meenakshisundaram and O'Donnell, John and Christensen, Daniel J. and Nicholson, Charles and Iliff, Jeffrey J. and Takano, Takahiro and Deane, Rashid and Nedergaard, Maiken},
  title = {Sleep drives metabolite clearance from the adult brain},
  journal = {Science},
  year = {2013},
  volume = {342},
  number = {6156},
  pages = {373--377},
  doi = {10.1126/science.1241224},
  url = {https://pmc.ncbi.nlm.nih.gov/articles/PMC3880190/}
}

@article{Miao2024,
  author = {Miao, Andawei and Luo, Tianyuan and Hsieh, Bryan and Edge, Christopher J. and Gridley, Morgan and Wong, Ryan Tak Chun and Constandinou, Timothy G. and Wisden, William and Franks, Nicholas P.},
  title = {Brain clearance is reduced during sleep and anesthesia},
  journal = {Nature Neuroscience},
  year = {2024},
  volume = {27},
  number = {6},
  pages = {1046--1050},
  doi = {10.1038/s41593-024-01638-y},
  url = {https://www.nature.com/articles/s41593-024-01638-y}
}

@article{Hauglund2025,
  author = {Hauglund, Natalie L. and Andersen, Mie and Tokarska, Klaudia and Radovanovic, Tessa and Kjaerby, Celia and S{\o}rensen, Frederikke L. and Bojarowska, Zuzanna and Untiet, Verena and Ballestero, Sheyla B. and Kolmos, Mie G. and Weikop, Pia and Hirase, Hajime and Nedergaard, Maiken},
  title = {Norepinephrine-mediated slow vasomotion drives glymphatic clearance during sleep},
  journal = {Cell},
  year = {2025},
  volume = {188},
  number = {3},
  pages = {606--622.e17},
  doi = {10.1016/j.cell.2024.11.027},
  url = {https://pubmed.ncbi.nlm.nih.gov/39788123/}
}

@article{Chouchani2014,
  author = {Chouchani, Edward T. and Pell, Victoria R. and Gaude, Edoardo and Aksentijevi{\'c}, Dunja and Sundier, Stephanie Y. and Robb, Ellen L. and Logan, Angela and Nadtochiy, Sergiy M. and Ord, Emily N. J. and Smith, Anthony C. and Eyassu, Filmon and Shirley, Rachel and Hu, Chou-Hui and Dare, Anna J. and James, Andrew M. and Rogatti, Sebastian and Hartley, Richard C. and Eaton, Simon and Costa, Ana S. H. and Brookes, Paul S. and Davidson, Sean M. and Duchen, Michael R. and Saeb-Parsy, Kourosh and Shattock, Michael J. and Robinson, Alan J. and Work, Lorraine M. and Frezza, Christian and Krieg, Thomas and Murphy, Michael P.},
  title = {Ischaemic accumulation of succinate controls reperfusion injury through mitochondrial {ROS}},
  journal = {Nature},
  year = {2014},
  volume = {515},
  number = {7527},
  pages = {431--435},
  doi = {10.1038/nature13909},
  url = {https://pubmed.ncbi.nlm.nih.gov/25383517/}
}

@article{SolmsPanksepp2012,
 author={Solms, Mark and Panksepp, Jaak},
 title={The ``Id'' Knows More than the ``Ego'' Admits: Neuropsychoanalytic and Primal Consciousness Perspectives on the Interface Between Affective and Cognitive Neuroscience},
 journal={Brain Sciences},year={2012},volume={2},number={2},pages={147--175},doi={10.3390/brainsci2020147}
}

@article{DehaeneChangeux2011,
 author={Dehaene, Stanislas and Changeux, Jean-Pierre},
 title={Experimental and theoretical approaches to conscious processing},
 journal={Neuron},year={2011},volume={70},number={2},pages={200--227},doi={10.1016/j.neuron.2011.03.018}
}

@article{Albantakis2023,
 author={Albantakis, Larissa and Barbosa, Leonardo and Findlay, Graham and Grasso, Matteo and Haun, Andrew M. and Marshall, William and others},
 title={Integrated information theory ({IIT}) 4.0: Formulating the properties of phenomenal existence in physical terms},
 journal={PLOS Computational Biology},year={2023},volume={19},number={10},pages={e1011465},doi={10.1371/journal.pcbi.1011465}
}

@article{CarhartHarris2014,
 author={Carhart-Harris, Robin L. and Leech, Robert and Hellyer, Peter J. and Shanahan, Murray and Feilding, Amanda and Tagliazucchi, Enzo and Chialvo, Dante R. and Nutt, David},
 title={The entropic brain: a theory of conscious states informed by neuroimaging research with psychedelic drugs},
 journal={Frontiers in Human Neuroscience},year={2014},volume={8},pages={20},doi={10.3389/fnhum.2014.00020}
}

@article{CarhartHarrisFriston2019,
 author={Carhart-Harris, Robin L. and Friston, Karl J.},
 title={{REBUS} and the Anarchic Brain: Toward a Unified Model of the Brain Action of Psychedelics},
 journal={Pharmacological Reviews},year={2019},volume={71},number={3},pages={316--344},doi={10.1124/pr.118.017160}
}

@article{Wailke2011,
 author={Wailke, S. and Herzog, J. and Witt, K. and Deuschl, G. and Volkmann, J.},
 title={Effect of controlled-release levodopa on the microstructure of sleep in {Parkinson's} disease},
 journal={European Journal of Neurology},year={2011},volume={18},number={4},pages={590--596},doi={10.1111/j.1468-1331.2010.03213.x}
}

@article{Portas1998,
 author={Portas, C. M. and Bjorvatn, B. and Fagerland, S. and Gr{\o}nli, J. and Mundal, V. and S{\o}rensen, E. and Ursin, R.},
 title={On-line detection of extracellular levels of serotonin in dorsal raphe nucleus and frontal cortex over the sleep/wake cycle in the freely moving rat},
 journal={Neuroscience},year={1998},volume={83},number={3},pages={807--814},doi={10.1016/S0306-4522(97)00438-7}
}

@article{Gallo2018,
 author={Gallo, E. F. and Meszaros, J. and Sherman, J. D. and Chohan, M. O. and Teboul, E. and Choi, C. S. and Moore, H. and Javitch, J. A. and Kellendonk, C.},
 title={Accumbens dopamine {D2} receptors increase motivation by decreasing inhibitory transmission to the ventral pallidum},
 journal={Nature Communications},year={2018},volume={9},pages={1086},doi={10.1038/s41467-018-03272-2}
}

@article{Bello2011,
 author={Bello, Estefan{\'i}a P. and Mateo, Yolanda and Gelman, Diego M. and Noa{\'i}n, Daniela and Shin, Jung Hoon and Low, Malcolm J. and Alvarez, Veronica A. and Lovinger, David M. and Rubinstein, Marcelo},
 title={Cocaine supersensitivity and enhanced motivation for reward in mice lacking dopamine {D2} autoreceptors},
 journal={Nature Neuroscience},year={2011},volume={14},pages={1033--1038},doi={10.1038/nn.2862}
}

@article{Power2015,
 author={Power, Robert A. and Steinberg, Stacy and Bjornsdottir, Gyda and others},
 title={Polygenic risk scores for schizophrenia and bipolar disorder predict creativity},
 journal={Nature Neuroscience},year={2015},volume={18},number={7},pages={953--955},doi={10.1038/nn.4040}
}

@article{Mullins2017,
 author={Mullins, Niamh and Ingason, Andr{\'e}s and Porter, Heather and others},
 title={Reproductive fitness and genetic risk of psychiatric disorders in the general population},
 journal={Nature Communications},year={2017},volume={8},pages={15833},doi={10.1038/ncomms15833}
}

@article{Pardinas2018,
 author={Pardi{\~n}as, Antonio F. and Holmans, Peter and Pocklington, Andrew J. and others},
 title={Common schizophrenia alleles are enriched in mutation-intolerant genes and in regions under strong background selection},
 journal={Nature Genetics},year={2018},volume={50},number={3},pages={381--389},doi={10.1038/s41588-018-0059-2}
}

@article{Jongsma2018,
 author={Jongsma, Hannah E. and Gayer-Anderson, Charlotte and Lasalvia, Antonio and others},
 title={Treated Incidence of Psychotic Disorders in the Multinational {EU-GEI} Study},
 journal={JAMA Psychiatry},year={2018},volume={75},number={1},pages={36--46},doi={10.1001/jamapsychiatry.2017.3554}
}

@article{Zoetmulder2016,
 author={Zoetmulder, Marielle and Nikolic, Miki and Biernat, Heidi and Korbo, Lise and Friberg, Lars and Jennum, Poul},
 title={Increased Motor Activity During {REM} Sleep Is Linked with Dopamine Function in Idiopathic {REM} Sleep Behavior Disorder and {Parkinson} Disease},
 journal={Journal of Clinical Sleep Medicine},year={2016},volume={12},number={6},pages={895--903},doi={10.5664/jcsm.5896}
}

@article{EbanRothschild2016,
  author = {Eban-Rothschild, Ada and Rothschild, Gideon and Giardino, William J. and Jones, Jeff R. and de Lecea, Luis},
  title = {{VTA} dopaminergic neurons regulate ethologically relevant sleep--wake behaviors},
  journal = {Nature Neuroscience},
  year = {2016},
  volume = {19},
  pages = {1356--1366},
  doi = {10.1038/nn.4377}
}

@article{Solms2000,
  author = {Solms, Mark},
  title = {Dreaming and {REM} sleep are controlled by different brain mechanisms},
  journal = {Behavioral and Brain Sciences},
  year = {2000},
  volume = {23},
  number = {6},
  pages = {843--850},
  doi = {10.1017/S0140525X00003988}
}

@article{HowesKapur2009,
  author = {Howes, Oliver D. and Kapur, Shitij},
  title = {The dopamine hypothesis of schizophrenia: version {III}, the final common pathway},
  journal = {Schizophrenia Bulletin},
  year = {2009},
  volume = {35},
  number = {3},
  pages = {549--562},
  doi = {10.1093/schbul/sbp006}
}

@article{Kapur2003,
  author = {Kapur, Shitij},
  title = {Psychosis as a state of aberrant salience: a framework linking biology, phenomenology, and pharmacology in schizophrenia},
  journal = {American Journal of Psychiatry},
  year = {2003},
  volume = {160},
  number = {1},
  pages = {13--23},
  doi = {10.1176/appi.ajp.160.1.13}
}

@article{Cummings2014,
  author = {Cummings, Jeffrey and Isaacson, Stuart and Mills, Roger and Williams, Hilde and Chi-Burris, Kathy and Corbett, Anne and Dhall, Rohit and Ballard, Clive},
  title = {Pimavanserin for patients with {Parkinson's} disease psychosis: a randomised, placebo-controlled phase 3 trial},
  journal = {The Lancet},
  year = {2014},
  volume = {383},
  number = {9916},
  pages = {533--540},
  doi = {10.1016/S0140-6736(13)62106-6}
}
\endgroup
\clearpage
\appendix
\section{Simulation specification and mathematical checks}\label{app:simulation}
Model outputs are displayed to three decimal places, while calculations use full precision.
Empirical estimates retain the precision needed to show their intervals and statistical results.
Small regional score differences use scientific notation with three significant figures.
Calculations and plotted coordinates retain full precision.

The conceptual drawings are schematic illustrations, unless an equation-generated curve is explicitly stated.
Their numerical values are not fits to participant data.
The source package includes the plotting inputs and numerical checks.
The TikZ sources contain the coordinates for each schematic drawing.

The state curve and its numerical checks use the common parameter set below.
The reference is quiet wakefulness: $x=1$, $C(1)=1$, with reserve and injury factors equal to one.
The low-signal and upper-collapse factors remain active.
Their reference values equal one at displayed precision.

Let the relative amplitudes be $a=(0.22,0.70,1.00,0.95)$.
Set $Z_C=G_0(1)V_0(1)\sum_i a_i\sig((1-\mu_i)/\tau_i)$ and $\alpha_i=a_i/Z_C$.
This fixed normalization sets the output unit without changing the dopamine input scale.
With the parameters below, $Z_C=1.696$ and $C'(1)=0.649$.
The reference lies within the increasing cumulative response, before the upper collapse transition.

\begin{table}[H]
\centering\small
\caption{Parameters for the normalized state-curve example and other numerical checks. Figure~\ref{fig:continuum} uses this state curve. These parameters are mathematical examples, not empirical estimates.}
\begin{tabularx}{\textwidth}{@{}p{3.5cm}Y@{}}
\toprule
Component & Values \\
\midrule
Recruitment midpoints & $\mu=(0.20,0.50,0.80,2.50)$ \\
Recruitment amplitudes & $\alpha=(0.22,0.70,1.00,0.95)/Z_C$ \\
Transition scales & $\tau=(0.07,0.08,0.16,0.18)$ \\
Collapse gate & $x_c=3.20$, $s_c=0.15$, with $x_c=2.8$ and $3.6$ for sensitivity \\
Low-signal boundary & $x_\ell=0.10$ \\
Regional comparison & $p=(0.5,0.5)$, $\kappa=4$ \\
RPR adaptation & Priming interval 0 to 2. Equilibrium densities 1.5 and 1.3. Adaptation times 1 and 2. \\
RPR recovery & Both equilibrium densities return to 1. Both recovery times are 3. Residual blockade decays at rate 2 after priming. \\
Seizure state example & Same normalized state curve, with reference serotonergic context and reserve, zero injury, and $C_{\mathrm{acc}}=0.50$. Uses W0 regional weights. \\
RPR gain curves & $A=1$, $u_A\in\{1,1.5,2,3\}$, with variable $u_N$ \\
\bottomrule
\end{tabularx}
\end{table}

The numerical checks verify soft-min normalization, bounds, derivatives, and the small-$\kappa$ moment expansion.
They verify the collapse values, onset maximum, and PAC maximum against their analytical expressions.
They also verify continuity of the developmental function and the RPR gain inequality.

The normalized example has one maximum at $x=2.958$, with $C=1.649$, and no interior valleys.
The upper viability gate produces the decline near collapse.
The cumulative core remains nondecreasing in 640,000 ordered-input comparisons across 5,000 positive parameter sets.
Its analytic derivative agrees with complex-step differentiation, with maximum scaled error below $10^{-14}$.
The single maximum describes this parameter set, rather than every possible collapse transition.
These checks establish internal consistency, not biological validity.

\begin{table}[H]\centering\small
\caption{Numerical checks of the equations and the normalized state example. The state example separates increasing recruitment from the upper collapse gate.}
\begin{tabularx}{\textwidth}{@{}p{5.1cm}Y@{}}\toprule
Check & Result \\\midrule
Soft-min normalization and bounds & Passed all 1,000 random configurations. \\
Soft-min derivative & Maximum absolute numerical error below $10^{-8}$. \\
Collapse gate at $x=3.0$ and $3.1$ & 0.958 and 0.179. \\
Onset maximum & Numerical derivative below $10^{-8}$ at the analytic maximum. \\
Waking-reference state curve & $C(1)=1.000$, $C'(1)=0.649$. One upper-collapse maximum and no interior valleys. \\
Cumulative core & No decreases in 640,000 ordered-input comparisons. Analytic derivative error below $10^{-14}$. \\
RPR gain inequality & Passed all 1,000 random comparisons. \\
\bottomrule\end{tabularx}\end{table}

The equal-mean redistribution and common-shift identities pass 1,000 random numerical comparisons, with maximum absolute error below $10^{-10}$.
Numerical integration of the memory equation reproduces the gain-phase identity with maximum absolute error below $10^{-8}$.
For $a=1$, $b=3$, and $c=1$, the feedback equation has three stable zero-input equilibria.
They occur at $z=0$ and $z\approx\pm1.618$, separated by unstable equilibria near $\pm0.618$.
These additional checks concern the restrictions in the shared-parameter tests.

\subsection{Optional peak in the receptor response}\label{app:receptorpeak}
A peak in the ligand-to-signal response is an optional property of the receptor model.
The upper viability gate supplies a separate high-input decline in the signal-to-state function.
For a two-pool reduction, absorb the regional scale into $a$ and set
\begin{samepage}
\begin{equation}
N(L)=\frac{aL}{K_N+L},\qquad A(L)=\frac{bL}{K_A+L},\qquad
D(L)=\frac{aL(K_A+L)}{(K_N+L)\{K_A+(1+b)L\}}.
\label{eq:twopool}
\end{equation}
\begin{eqvars}
\item $L$ (\textbf{Ligand concentration}): Ligand concentration.
\item $N(L)$ (\textbf{Activating contribution}): Activating contribution, with positive gain $a$.
\item $A(L)$ (\textbf{Autoreceptor feedback}): Autoreceptor feedback, with positive gain $b$.
\item $K_N,K_A$ (\textbf{Occupancy constants}): Positive half-occupancy concentrations for the two pools.
\item $D(L)$ (\textbf{Effective signal}): Effective signal, with the same regional scale absorbed into $a$.
\end{eqvars}
\end{samepage}

A unique maximum at positive concentration exists exactly when
\begin{samepage}
\begin{equation}
bK_A>(1+b)K_N,\qquad
L_* = \frac{K_A}{\sqrt{b(K_A/K_N-1)}-1}.
\label{eq:receptorpeak}
\end{equation}
\begin{eqvars}
\item $L_*$ (\textbf{Peak concentration}): Concentration at the peak, defined only when the stated inequality holds.
\end{eqvars}
\end{samepage}

The signal then decreases toward the positive plateau $a/(1+b)$.
Equal occupancy constants give a monotone saturating curve.
Independent receptor measurements must determine whether the peak condition applies.

Direct differentiation gives
\begin{samepage}
\begin{equation}
\frac{dD}{dL}
=\frac{a\{-\delta L^2+2K_NK_A L+K_NK_A^2\}}
{(K_N+L)^2\{K_A+(1+b)L\}^2}.
\label{eq:receptorderivative}
\end{equation}
\begin{eqvars}
\item $\delta=bK_A-(1+b)K_N$ (\textbf{Peak condition}): Parameter combination that determines whether a positive-concentration peak exists.
\end{eqvars}
\end{samepage}
The denominator is positive for $L\geq0$.
If $\delta\leq0$, the numerator is strictly positive at every finite nonnegative concentration.
If $\delta>0$, its quadratic has one positive root and one negative root.
The derivative changes from positive to negative at the positive root in Equation~\ref{eq:receptorpeak}.
This proves both necessity and sufficiency of the peak condition for this reduction.

The signal remains positive for every finite $L>0$ and approaches $a/(1+b)$ as $L$ increases without bound.
The separate state and viability equations determine how this positive signal affects conscious intensity and injury.

Algebraic and numerical checks verify the derivative, peak location, limiting plateau, and equal-affinity reduction.
An independent numerical check uses 1,000 positive parameter sets and complex-step differentiation.
The largest scaled derivative error is below $2.3\times10^{-15}$.
The sample uses log-uniform parameters from 0.01 to 100 and random seed 20260906.
It tests the algebra across parameter regimes.
Its parameter frequencies have no biological interpretation.

\paragraph{Conditions for a deficit relative to reference:}
Subtracting the high-ligand plateau gives
\begin{samepage}
\begin{equation}
D(L)-\frac{a}{1+b}
=\frac{a(\delta L-K_NK_A)}
{(1+b)(K_N+L)\{K_A+(1+b)L\}}.
\label{eq:plateaudifference}
\end{equation}
\end{samepage}
All denominator factors are positive.
Thus, the high-ligand limit lies below a positive reference signal exactly when
\begin{samepage}
\begin{equation}
D(\infty)<D(L_{\mathrm{ref}})
\quad\Longleftrightarrow\quad
\delta>0,\qquad L_{\mathrm{ref}}>\frac{K_NK_A}{\delta}.
\label{eq:receptorbaseline}
\end{equation}
\begin{eqvars}
\item $L_{\mathrm{ref}}$ (\textbf{Reference concentration}): Specifies the independently measured ligand concentration used to normalize regional signaling.
\end{eqvars}
\end{samepage}
A peak can exist while the reference signal remains below the final plateau.
Such a system does not develop a reference-relative deficit at arbitrarily high ligand.
When the reference lies on the rising branch above that plateau, the second equal-signal concentration is
\begin{samepage}
\begin{equation}
L_{\mathrm{other}}
=\frac{K_NK_A(K_A+L_{\mathrm{ref}})}
{\delta L_{\mathrm{ref}}-K_NK_A}.
\label{eq:receptorsecondcrossing}
\end{equation}
\begin{eqvars}
\item $L_{\mathrm{other}}$ (\textbf{Second crossing}): Gives the other concentration with the reference response when the denominator is positive.
\end{eqvars}
\end{samepage}
A further rise beyond this second crossing produces the deficit.
If the reference is already on the falling branch, further ligand increases reduce signaling immediately in the static response.
The normalized limiting signal is $a/\{(1+b)D(L_{\mathrm{ref}})\}>0$.
Temporal filtering still limits the speed of the regional response.

\section{Empirical methods and sensitivity analyses}\label{app:empiricalmethods}
\subsection{Sleep reconstruction and complete timing family}\label{app:sleepmethods}
The following details complement Sections~\ref{sec:sleepresults} and~\ref{sec:statisticalmodels}.
The six-test sustained-entry family appears in Table~\ref{tab:mechanismresults}.
Table~\ref{tab:strictfamily} below gives the complete eight-test conventional timing family.
Both families include the calendar and participant-trend specifications.

\paragraph{Participant-deletion procedure:}
The sustained-entry influence analysis refits both temporal specifications after removing each participant in turn.
All remaining nights retain the original eligibility, outcome, and covariate rules.
The full-sample residualized exposure ranks information contributors separately for each specification.
Nested refits remove the highest-ranked one, two, and three participants, without reranking after deletion.
Each retained sample uses the original estimator and participant-cluster interval, including its finite-sample correction.
The two- and three-participant deletions retain 34 and 33 participants with formulation variation, respectively.

This descriptive analysis assesses participant influence separately from the six-test family.
Its 137 coefficients per specification are overlapping refits, not additional independent tests.

\paragraph{Medication reconstruction and clock convention:}
Exact duplicate medication intervals have identical participant, treatment, LEDD, start-month, and stop-month fields.
Their removal changed recorded dose on 469 nights from four participants in the broad sample.
Unresolved nonnumeric doses affected 2,410 nights from 14 participants.
The exclusion analyses below use the unresolved-dose flag from this reconstruction.

Dose components use the recorded LEDD field, divided by 100 for regression.
The formulation category uses product labels containing CR, ER, XL, XR, MODUTAB, RYTARY, EXTENDED, PROLONG, RETARD, DUODOPA, or DUOPA.
The last two labels identify intestinal delivery products within this operational category.
Thus, the contrast combines formulation and delivery characteristics.
An unrecorded component receives zero exposure, which does not establish biological absence.
The retained timing samples require positive recorded total levodopa LEDD.

Sleep dates use UTC timestamps, grouped after subtraction of 12 hours.
Onset sine and cosine describe UTC clock time, rather than verified local circadian phase.
Monthly prescription records do not establish administration time or adherence.
Duration and coverage adjustment can also condition on variables affected by treatment.

\paragraph{Conventional outcomes and eligibility:}
Reconstruction removes duplicate timestamps and retains unknown stages and gaps.
Sleep onset requires ten consecutive sleep epochs with exact timestamp continuity.
The interval ends after its last valid sleep epoch.
Eligible intervals last 4 to 12 hours and have at least 80\% scored coverage.

REM phase is the mean midpoint of REM epochs relative to the interval, expressed on a 0 to 100 scale.
REM latency runs from sleep onset to the first valid REM epoch.
Eligibility requires some REM, without a minimum episode count or a 5-to-240-minute latency restriction.
The REM sample contains 10,582 nights from 137 participants.
Deep-sleep phase and the direct stage contrast use 9,421 nights from 136 participants.
Each comparison therefore retains its outcome-specific sample.

\paragraph{Estimation and multiplicity:}
The regression removes participant intercepts and, in the trend specification, participant-specific linear slopes.
Calendar month and weekday enter as categorical effects.
Participant-cluster standard errors include the finite-sample correction for absorbed parameters.
Pointwise intervals use a $t$ reference with one fewer degree of freedom than participants.

Null-imposed wild cluster tests use Rademacher weights and 9,999 draws, with random seed 20260905.
Nuisance effects are re-estimated through projections within each draw.
Bootstrap tests compare absolute studentized statistics, with one added to the numerator and denominator counts.
Holm adjustment uses these bootstrap $p$ values within each declared family.
The pointwise intervals are not inversions of these tests.
Thus, an interval excluding zero can accompany a bootstrap $p$ value above 0.05.

\begin{table}[H]\centering\footnotesize
\caption{Complete strict formulation family. Effects are per 100 mg/day LEDD. Intervals are pointwise 95\% cluster-$t$ intervals. The last column adjusts wild-bootstrap $p$ values over all eight rows.}
\label{tab:strictfamily}\begin{tabular}{@{}llrrrlr@{}}\toprule
Model & Outcome & Nights & People & Effect & 95\% interval & $p_{\rm Holm}$ \\\midrule
Calendar & REM phase (pp) & 10,582 & 137 & $0.402$ & $0.112$ to $0.691$ & 0.130 \\
Calendar & REM latency (min) & 10,582 & 137 & $2.103$ & $0.713$ to $3.493$ & 0.130 \\
Calendar & Deep phase (pp) & 9,421 & 136 & $-0.080$ & $-0.418$ to $0.259$ & 1.000 \\
Calendar & REM minus deep (pp) & 9,421 & 136 & $0.523$ & $0.015$ to $1.031$ & 0.607 \\
Trends & REM phase (pp) & 10,582 & 137 & $0.331$ & $-0.021$ to $0.683$ & 0.694 \\
Trends & REM latency (min) & 10,582 & 137 & $1.410$ & $-0.284$ to $3.104$ & 0.846 \\
Trends & Deep phase (pp) & 9,421 & 136 & $-0.065$ & $-0.354$ to $0.225$ & 1.000 \\
Trends & REM minus deep (pp) & 9,421 & 136 & $0.424$ & $-0.140$ to $0.988$ & 0.874 \\\bottomrule\end{tabular}\end{table}

The calendar REM-phase and latency bootstrap $p$ values are 0.0165 and 0.0162.
They become 0.139 and 0.211 under participant trends.
The direct calendar REM-minus-deep phase contrast gives $p=0.101$, before Holm adjustment.
This direct contrast assesses stage specificity.
The minimum adjusted $p$ value across all eight rows is 0.130.

\paragraph{Exposure information and medication sensitivity:}
The formulation component varies within 36 participants, while total levodopa LEDD varies within 70.
For conventional timing, the largest participant supplies 20.6\% of residualized formulation information in the calendar model.
Inverse concentration indices are 10.6 participants for calendar adjustment and 8.9 with participant trends.
These indices describe information concentration, rather than effective degrees of freedom.

Exclusion of unresolved doses retains 9,926 conventional REM nights from 134 participants.
The trend-adjusted phase contrast is $+0.360$ percentage points (95\% interval $-0.022$ to $0.741$).
The latency contrast is $+1.80$ minutes ($0.01$ to $3.59$).
These sensitivity intervals are pointwise, without another wild-bootstrap test family.

For sustained entry, the largest participant supplies 21.1\% and 24.5\% of formulation information under the two temporal specifications.
Exclusion of unresolved doses retains 9,881 waiting-time nights and 9,798 persistence nights from 134 participants.
Calendar and trend waiting contrasts are $+1.313$ minutes ($0.461$ to $2.165$) and $+0.964$ minutes ($-0.095$ to $2.023$).
Persistence contrasts are $-0.181$ minutes ($-0.494$ to $0.131$) and $-0.263$ minutes ($-0.691$ to $0.165$).
These are descriptive sensitivity intervals, without another multiplicity family.

The initial coverage rule permits six minutes without classification within the first 120 minutes.
True entry during an unknown interval can remain unobserved.
Persistence conditions on a qualifying episode and excludes early-censored intervals.
Its horizon and conditional sample differ from the waiting outcome.
No equivalence margin was specified for persistence.
An interval spanning zero therefore does not establish preserved maintenance.

\subsection{Regional eligibility and ordinal estimation}\label{app:regionalmethods}
\paragraph{Pairing and image linkage:}
Each participant, event, and examination month must contain exactly one OFF record and one ON record.
Conflicting examination-page and medication-state labels are excluded.
All 14 lateral items must have integer values from 0 to 4 in each record.
Both records require medication use and an explicit absence of DBS.
Any known DBS or HIFU at or before that month excludes the pair.

OFF occurs 6 to 72 hours after medication, and ON occurs 1 to 6 hours after medication.
The ON clock follows the OFF clock by 0.5 to 6 hours, without an explicit ON-first order.
These monthly records cannot verify a common examination day.

Image linkage uses the nearest eligible scan month, from the examination month through three months earlier.
Identical scan copies are collapsed.
An ambiguous nearest scan month is excluded, without substitution of an older scan.
Of the 640 paired visits, 572 have same-month image links with unknown within-month order.
The sample contains 640 unique scans and 181 participants with repeated pairs.
HIFU status is missing in both records for 549 pairs, and legacy DBS codes add uncertainty.

\paragraph{Earlier baseline:}
The earlier examination is the nearest uniquely recorded eligible OFF month, from one to 12 months before the current pair.
The ordinal comparison additionally requires an explicit OFF state and recorded medication use at that earlier examination.
One first eligible triplet per participant supplies 42 item values.
The 254 participants therefore supply 10,668 item values.
The median separation is seven months.
Earlier dose timing is incomplete and does not enter the ordinal model.

\paragraph{Complete latent specification:}
All symbols in the following specification belong to the clinical observation model.
They do not redefine the theory's regional signaling variables.
For each participant, the latent specification is
\begin{equation}
\begin{aligned}
B_s&=\beta_m\widetilde m_r+s\beta_d\widetilde d_r+\sigma_b b+\sigma_a a_s,\\
B_{s,\mathrm{prior}}&=B_s-g(\delta_0+s\delta_d\widetilde d_r)+\sigma_{\mathrm{prog}}\sqrt{g}\,z_s,\\
U&=\tau_0+\tau_m\widetilde m_r+\tau_b\sigma_b b
  +\tau_{\mathrm{on}}\widetilde h_{\mathrm{on}}+\tau_{\mathrm{off}}\widetilde h_{\mathrm{off}}+\sigma_u c.
\end{aligned}
\label{eq:ordinalspecification}
\end{equation}
\begin{eqvars}
\item $B_s,B_{s,\mathrm{prior}}$ (\textbf{Latent burdens}): Current and earlier OFF burdens on body side $s$.
\item $s$ (\textbf{Body-side index}): $+1$ for left and $-1$ for right.
\item $\widetilde m_r,\widetilde d_r$ (\textbf{Imaging predictors}): Scaled mean and signed half-difference of putamen SBR.
\item $g$ (\textbf{Examination gap}): Earlier-to-current separation in years.
\item $\widetilde h_{\mathrm{on}},\widetilde h_{\mathrm{off}}$ (\textbf{Current dose timing}): Scaled logarithms of one plus current ON and OFF hours since medication.
\item $b,a_s,z_s,c$ (\textbf{Random effects}): Mutually independent standard normal variables. The same $b$ and $c$ apply to both sides.
\item $\sigma_b,\sigma_a,\sigma_{\mathrm{prog}},\sigma_u$ (\textbf{Random-effect scales}): Positive scales for common burden, side burden, progression, and common response.
\item $\beta_m,\beta_d,\delta_0,\delta_d$ (\textbf{Burden coefficients}): Common and regional imaging associations with baseline burden and progression.
\item $U,\tau_0,\tau_m,\tau_b,\tau_{\mathrm{on}},\tau_{\mathrm{off}}$ (\textbf{Common response}): Shared latent medication response and its regression coefficients.
\end{eqvars}

Equation~\ref{eq:ordinalresponse} adds only the regional response coefficient $\gamma$ to this common model.
The progression term models the examination gap.
The timing predictors describe current dose-to-examination intervals, without reconstructing treatment changes between examinations.

The cumulative probit link has four ordered finite thresholds for each item.
Thresholds are shared across states and sides, but fitted separately in each competing model.
Each item loading and the probit residual variance equal one, with no baseline intercept.
The common model has 41 parameters, and the regional model has 42.

\paragraph{Scaling and regularization:}
Training participants determine all centers and scales.
Mean SBR and both log timing variables are centered and divided by their training standard deviations.
The signed SBR half-difference is divided by its training root mean square, without centering.
This preserves side-swap symmetry.

Finite threshold spacings equal a softplus-transformed parameter plus 0.01.
Fixed Gaussian penalties have standard deviation 3 for burden, progression, and response coefficients, including $\gamma$.
Log standard deviations have penalty center $\log(0.5)$ and standard deviation 1.5.
First thresholds have penalty center zero and standard deviation 5.
Raw spacing parameters have penalty center zero and standard deviation 3.
These fixed regularizers do not provide posterior intervals.

The objective is mean negative joint log likelihood plus total penalty divided by the training participant count.
L-BFGS-B uses at most 250 iterations, objective tolerance $10^{-9}$, and gradient tolerance $10^{-4}$.
Line search permits 30 steps and retains ten correction vectors.
Bounds are $[-8,5]$ for first thresholds and $[-5,4]$ for raw spacings.
Coefficient bounds are $[-6,6]$, and log standard deviation bounds are $[-5,1.5]$.

The common fit starts from empirical category thresholds and small latent variances.
Regional fits start from the fitted common model, with $\gamma=0$, $+0.3$, and $-0.3$.
Selection retains the converged start with the lowest penalized objective.
This numerical selection does not add outcome-specific predictors.

\paragraph{Prediction and numerical checks:}
One shuffled five-fold partition uses random seed 260906.
Each participant contributes once to evaluation, with all transformations fitted in the corresponding training sample.
Training uses seven Gauss--Hermite nodes per scalar latent integral, with evaluation at 7, 11, and 15 nodes.
The conditional score subtracts the 28-item OFF marginal log probability from the 42-item joint log probability.
Reserved ON observations enter only the evaluation score.
Intervals use 10,000 bootstrap resamples of paired participant scores from the fitted models, with seed 260906.

All five common fits and 15 regional starts report optimizer convergence, without parameters at computational bounds.
The largest selected unprojected gradient is $5.30\times10^{-4}$.
Some fits meet the objective-change criterion before the stricter gradient criterion.
Probability, gradient, symmetry, conditional-score, and direct-integration checks pass.

Between 11 and 15 evaluation nodes, the mean absolute score change is at most $1.59\times10^{-3}$ per ON item.
The near-zero mean gain changes sign across integration orders.
An 11-node first-fold refit changes the 15-node mean gain by $1.45\times10^{-6}$ log units.
This checks numerical stability in one fold, rather than another complete five-fold evaluation.

At 15 nodes, the secondary joint gain is $-9.18\times10^{-4}$, with 95\% interval $[-4.50,2.79]\times10^{-3}$.
The primary conditional comparison is reported in Section~\ref{sec:regionalmean}.
Both intervals condition on fitted models and omit complete-refitting uncertainty.

Four synthetic train/test splits assess recovery in samples of 254 people.
Two planted large regional effects yield reserved ON gains of 0.793 and 0.586 log units per triplet.
Two common-response controls yield $5.37\times10^{-4}$ and $-0.0139$.
These checks demonstrate recovery of large planted effects, without estimating power or establishing equivalence for small biological effects.
They are numerical identification checks, rather than participant evidence for either mechanism.

\begin{landscape}
\section{Phenomenological catalogue}\label{app:phenomenology}
\begin{figure}[H]\centering
\fitlandart{\begin{tikzpicture}[
    yscale=1,
    xscale=1.85,
    state/.style={
        rectangle,
        rounded corners=4pt,
        fill=white,
        draw=#1!70,
        line width=1.5pt,
        inner sep=4pt,
        font=\small\sffamily,
        align=center,
        minimum width=2cm,
        minimum height=0.8cm,
        drop shadow={opacity=0.15, shadow xshift=1pt, shadow yshift=-1pt}
    },
    keystate/.style={
        rectangle,
        rounded corners=6pt,
        fill=#1!10,
        draw=#1!90,
        line width=2pt,
        inner sep=6pt,
        font=\footnotesize\sffamily\bfseries,
        align=center,
        minimum width=2.4cm,
        minimum height=1cm,
        drop shadow={opacity=0.25, shadow xshift=2pt, shadow yshift=-2pt}
    },
    danger/.style={
        rectangle,
        rounded corners=6pt,
        fill=red!25,
        draw=red!85,
        line width=2pt,
        inner sep=5pt,
        font=\footnotesize\sffamily\bfseries,
        align=center,
        text=red!90
    },
    quadrant/.style={
        font=\large\bfseries\sffamily,
        text=#1!65!black
    },
    axis/.style={
        font=\normalsize\bfseries\sffamily,
        fill=white,
        inner sep=3pt
    },
    path/.style={
        line width=2.5pt,
        opacity=0.5,
        ->,
        >=stealth,
        #1!70
    }
]
\fill[consciouscolor!6] (-7.5,0) rectangle (0,6.5);
\fill[flowcolor!6] (0,0) rectangle (7.5,6.5);
\fill[sleepcolor!6] (-7.5,-6.5) rectangle (0,0);
\fill[dopaminecolor!6] (0,-6.5) rectangle (7.5,0);
\fill[gray!18] (-7.5,-6.5) rectangle (-5.95,6.5);
\draw[gray!60,line width=1.5pt,dashed] (-5.95,-6.5)--(-5.95,6.5);
\node[font=\small\bfseries,text=gray!70!black,rotate=90] at (-6.72,2.8) {DEATH};
\fill[red!10] (5.9,-6.5) rectangle (7.9,6.5);
\draw[red!70,line width=2pt,dashed] (5.9,-6.5) -- (5.9,6.5);
\node[font=\small\bfseries,text=red!80,rotate=90] at (7,2.8)
 {COLLAPSE AND INJURY};
\draw[->,line width=2pt,gray!55] (-5.95,0) -- (7.8,0)
 node[axis,right,align=center,fill=none,xshift=6pt] {Effective DA\\{\scriptsize Intensity axis}};
\draw[->,line width=2pt,gray!55] (0,-6.8) -- (0,6.8)
 node[axis,above,align=center,fill=none] {Effective 5-HT\textsubscript{2A}\\{\scriptsize Complexity axis}};
\foreach \y/\val in {-5/Low,5/High} {
 \draw[gray!45] (-.1,\y)--(.1,\y);
 \node[text=gray!70,font=\small,rotate=90] at (.35,\y) {\val};
}
\foreach \x/\val in {-5.1/Very low,-3/Low,3/High,5/Very high} {
 \draw[gray!45] (\x,-.1)--(\x,.1);
 \node[text=gray!70,font=\small] at (\x,-.5) {\val};
}
\node[quadrant=consciouscolor,font=\normalsize\bfseries\sffamily] at (-3.8,6.95)
 {DREAMING \& MEDITATIVE};
\node[quadrant=flowcolor,font=\normalsize\bfseries\sffamily] at (2.7,6.95)
 {FLEXIBLE \& ASSOCIATIVE};
\node[quadrant=sleepcolor,font=\normalsize\bfseries\sffamily] at (-3.8,-6.95)
 {LOW ACCESS \& WITHDRAWAL};
\node[quadrant=dopaminecolor,font=\normalsize\bfseries\sffamily] at (2.7,-6.95)
 {FOCUSED \& GOAL-DIRECTED};
\draw[path=flowcolor] (0,0) .. controls (2,1) and (4,3.5) .. (4.85,4.2);
\draw[path=consciouscolor] (0,0) .. controls (-.4,.4) and (-.7,.8) .. (-.7,1.1);
\draw[path=consciouscolor,dashed] (-.7,2) .. controls (-1.6,2.6) and (-1.6,3.9) .. (-.7,4.2);
\node[keystate=gray,fill=yellow!5] (baseline) at (0,0) {BASELINE\\Normal Wake};
\node[keystate=consciouscolor] (meditation) at (-0.7,4.75) {Deep Meditation\\Vivid Presence\\Ego Dissolution};
\node[keystate=consciouscolor] (rem) at (-2.6,3.1) {REM Dreams\\Vivid Imagery};
\node[state=consciouscolor] (psilocybin) at (-2.6,4.75) {Psilocybin\\Low Dose};
\node[state=consciouscolor] (cannabis) at (-2.6,1.55) {Cannabis\\Relaxed Imagery};
\node[state=consciouscolor] (lucid) at (-0.7,3.1) {Lucid\\Dreaming};
\node[state=consciouscolor] (mindfulness) at (-0.7,1.55) {Mindfulness\\Open Awareness};
\node[keystate=flowcolor] (flow) at (1.25,3.1) {Creative Flow\\Associations};
\node[keystate=flowcolor] (psychedelic) at (4.85,4.75) {LSD Peak\\Expanded Content};
\node[state=flowcolor] (eureka) at (3.1,4.75) {Eureka\\Moment};
\node[state=flowcolor] (artistic) at (3.1,3.1) {Artistic Expression\\Rich Imagery};
\node[state=flowcolor] (mdma) at (1.25,4.75) {MDMA\\Vivid Empathy};
\node[state=flowcolor] (runner) at (3.1,1.55) {Runner's High\\Vivid Presence};
\node[keystate=sleepcolor] (sleep) at (-5.05,-3.9) {Deep Sleep\\No Experience};
\node[keystate=sleepcolor] (depression) at (-1.3,-3.1) {Depression\\Dull Experience};
\node[state=sleepcolor] (anesthesia) at (-4.9,-5.65) {General Anesthesia\\No Experience};
\node[state=sleepcolor] (onset) at (-3.4,-1.55) {Sleep Onset\\Fading Experience};
\node[state=sleepcolor] (fatigue) at (-1.3,-1.55) {Fatigue\\Dull Experience};
\node[state=sleepcolor] (alcohol) at (-3.4,-3.1) {Alcohol\\Stupor};
\node[keystate=dopaminecolor] (focus) at (1.3,-3.1) {Focused Work\\Clear Presence};
\node[keystate=dopaminecolor] (athletic) at (3.1,-3.1) {Athletic Peak\\Motor Flow};
\node[state=dopaminecolor] (learning) at (1.3,-1.55) {Learning\\Focused Attention};
\node[state=dopaminecolor] (problem) at (3.1,-1.55) {Problem Solving\\Absorbed Attention};
\node[state=dopaminecolor] (stimulant) at (4.85,-3.1) {Stimulant\\Narrow Focus};
\node[state=dopaminecolor] (competition) at (4.85,-1.55) {Competition\\Clear Alertness};
\node[danger] (psychosis) at (5,3) {Psychosis \\(Stimulants)\\Rich Content};
\node[danger] (regionalpsychosis) at (-1.7,6.05) {Regional-Deficit Psychosis};
\node[danger] (mania) at (5,1.35) {Mania\\Racing Thoughts};
\node[danger,text width=2.2cm] (terminal) at (6.9,-4.1) {Irreversible\\Injury\\$J\geq J_{\mathrm{irr}}$};
\node[danger] (seiz) at (-3.1,-5.0) {After Seizure\\No Experience};

\draw[draw=sleepcolor!75,line width=2.4pt,-{Stealth[length=5pt,width=5pt]},shorten >=2pt,shorten <=2pt]
 (baseline.west) .. controls (-3.5,0) and (-5.05,-1.1) .. (sleep.north);
\draw[draw=sleepcolor!75,line width=2.4pt,-{Stealth[length=5pt,width=5pt]},shorten >=2pt,shorten <=2pt]
 (sleep.west) .. controls (-5.9,-1.3) and (-5.1,3.1) .. (rem.west);
\draw[draw=sleepcolor!75,line width=2.1pt,-{Stealth[length=5pt,width=5pt]},shorten >=2pt,shorten <=2pt]
 (baseline.north west) .. controls (-2.1,.3) and (-1.45,2.35) .. (rem.south east);
\node[keystate=gray,text width=2.1cm,minimum width=0cm] (death) at (-6.72,-3.9)
 {Death\\Absent\\Experience};
\node[font=\tiny\sffamily,text=gray!80,align=center,text width=2.15cm] at (-6.72,-5.45)
 {Absorbing outcome.\\Not a DA threshold.};
\node[font=\tiny\sffamily,text=red!75,align=center,text width=2.6cm] at (6.9,-5.25) {Separate criterion.\\Outside the map.};
\node[font=\footnotesize\sffamily,text=gray!65,align=center] at (0,-7.6)
 {No-experience labels specify selected episodes, not all sleep, anesthesia, or seizures.};
\node[fill=white,draw=sleepcolor!70,rounded corners=3pt,
 font=\small\bfseries,inner sep=4pt] at (-4.5,-8.35) {Sleep Transition};
\node[fill=white,draw=consciouscolor!70,rounded corners=3pt,
 font=\small\bfseries,inner sep=4pt] at (0,-8.35) {Meditation Scenarios};
\node[fill=white,draw=flowcolor!70,rounded corners=3pt,
 font=\small\bfseries,inner sep=4pt] at (4.5,-8.35) {Psychedelic Transition};
\node[fill=white,draw=gray!50,rounded corners=6pt,drop shadow,
 inner sep=8pt,font=\footnotesize\sffamily,align=left] at (0,-10.55) {
 \textbf{Illustrative ordering within the theory}\\
 Right: greater felt intensity and higher proposed effective dopamine signaling.\\
 Up: richer experience and stronger proposed 5-HT\textsubscript{2A} signaling.\\
 Labels describe selected experiences. Spacing does not quantify transmitter levels.\\
 Sleep retains very low proposed signaling. Death is a separate absorbing outcome.
};
\end{tikzpicture}}
\caption{Experiential examples on an ordinal hypothesis map. Each label describes a selected presentation, not all responses to that condition or drug. Distances and cross-drug ordering are not fitted receptor measurements. Two psychosis labels are shown: upper right for the high-input route, and upper left for the regional-deficit route below waking intensity (Section~\ref{sec:psychosispatterns}). This catalogue motivates questions. The quantitative tests are the commitments in Table~\ref{tab:commitments}.}
\label{fig:extendedmap}
\end{figure}
\end{landscape}

\section{Technical glossary}
The glossary fixes the meaning of the main technical terms within this article.

\begin{table}[H]\centering
\begin{tabularx}{\textwidth}{@{}p{3.8cm}Y@{}}
\toprule
Term & Meaning in this article \\
\midrule
Effective signaling & Net receptor and circuit response represented by the regional model. \\
Intensity & Latent strength with which experience is subjectively present, pending an operational scale. \\
Dynamic complexity & Felt complexity and personal significance of experience, assessed with separate ratings and a fixed summary score. \\
Structural capacity & Proposed substrate property that bounds available dynamic complexity. \\
Soft minimum & Normalized smooth approximation to a minimum of regional signals. \\
Hysteresis & Dependence of a transition on input history that persists in the relevant slow-input limit. \\
LEDD & Levodopa equivalent daily dose, expressed in mg/day equivalent units. \\
REM phase & Mean timing of REM within the sleep interval, on a 0 to 100 scale. \\
Receptor priming & Proposed adaptation before a later receptor-directed input. \\
\bottomrule
\end{tabularx}
\end{table}

\clearpage
\section{Symbol index}\label{app:symbols}
The index links symbols to their local definitions.
Grouped symbols follow the definitions beside the equations.
Repeated symbols have separate rows when their meanings differ.
Roman and Greek symbols are ordered by their names.

\begin{table}[H]\centering\small
\setlength{\tabcolsep}{4pt}
\renewcommand{\arraystretch}{1.0}
\begin{tabularx}{\textwidth}{@{}p{4.25cm}Y p{3.05cm}@{}}
\toprule
Symbol & Meaning in this context & Definition \\
\midrule
$\mathbf 1$ & All-ones vector & Eq.~\ref{eq:redistribution} \\
$\mathbf 1(x<x_c)$ & Boundary indicator & Eq.~\ref{eq:collapse} \\
$a$ & Age & Eq.~\ref{eq:development} \\
$A$ & Autoreceptor feedback in the two-pool reduction & Eq.~\ref{eq:twopool} \\
$A$ & Baseline feedback & Eq.~\ref{eq:rprgain} \\
$A$ & Baseline state label in sensitivity transfer & Eq.~\ref{eq:ratiotransfer} \\
$a$ & Common shift & Eq.~\ref{eq:redistribution} \\
$A$ & Efficacy & Eq.~\ref{eq:devdose} \\
$a_0,a_1,a_2$ & EEG coefficients & Eq.~\ref{eq:eeg} \\
$a_{\mathrm{ad}}$ & Protective adaptation & Eq.~\ref{eq:injuryhistory} \\
$a_{\mathrm{all}},b_{\mathrm{all}}$ & Allometric reference & Eq.~\ref{eq:anatomicalmetrics} \\
$A,B$ & Baseline states & Eq.~\ref{eq:ratiotransfer} \\
$a,b$ & Positive gains in the two receptor pools & Eq.~\ref{eq:twopool} \\
$a,b,c$ & Feedback coefficients & Eq.~\ref{eq:transition}, Eq.~\ref{eq:foldcondition} \\
$a_D,a_H$ & Relaxation rates & Eq.~\ref{eq:coupling} \\
$a_i$ & Individual baseline & Eq.~\ref{eq:formulation} \\
$A(L)$ & Autoreceptor feedback & Eq.~\ref{eq:twopool} \\
$\alpha_{\mathrm{EEG}}$ & Spectral exponent & Eq.~\ref{eq:eeg} \\
$\alpha_E,\epsilon_E$ & Feedback parameters & Eq.~\ref{eq:reservefeedback} \\
$\alpha_i(h)$ & Response amplitude & Eq.~\ref{eq:multistate} \\
$A_r$ & Autoreceptor feedback & Eq.~\ref{eq:deff} \\
$B$ & Baseline state label in sensitivity transfer & Eq.~\ref{eq:ratiotransfer} \\
$B$ & Metabolite mapping & Eq.~\ref{eq:measurements} \\
$b,a_s,z_s,c$ & Random effects & Eq.~\ref{eq:ordinalspecification} \\
$b_{\mathrm{brain}}$ & Brain mass fraction & Eq.~\ref{eq:anatomicalmetrics} \\
$b_{\mathrm{clearance}}$ & Clearance contribution & Eq.~\ref{eq:measurements} \\
$\beta_m,\beta_d,\delta_0,\delta_d$ & Burden coefficients & Eq.~\ref{eq:ordinalspecification} \\
$\beta_T,\beta_E$ & Dose coefficients & Eq.~\ref{eq:formulation} \\
$b_{\mathrm{ex}}$ & Rare-trait advantage & Eq.~\ref{eq:evolutionfrequency} \\
$B_j$ & Antagonist input & Eq.~\ref{eq:rprdynamic} \\
$b_r^{(A)}$ & Local outcome slope & Eq.~\ref{eq:ratiotransfer} \\
$B_s$ & Latent baseline burden & Eq.~\ref{eq:ordinalresponse} \\
$B_s,B_{s,\mathrm{prior}}$ & Latent burdens & Eq.~\ref{eq:ordinalspecification} \\
$C$ & Conscious intensity & Section~\ref{sec:definitions} \\
$C_{\mathrm{acc}}$ & Experience-related boundary & Eq.~\ref{eq:seizureexample} \\
$c_{\mathrm{ex}}$ & Frequency-dependent cost & Eq.~\ref{eq:evolutionfrequency} \\
$C_{\mathrm{ex}}$ & Illustrative state output & Eq.~\ref{eq:seizureexample} \\
$c_Q(h,E)$ & Demand scale & Eq.~\ref{eq:regionaldemand} \\
$c_R$ & Repair energy cost & Eq.~\ref{eq:injuryhistory} \\
$C_{\mathrm{total}}$ & State output & Eq.~\ref{eq:ceiling} \\
$d$ & Common mean & Eq.~\ref{eq:redistribution} \\
$d$ & Common starting signal & Eq.~\ref{eq:weightedspread} \\
$\overline D$ & Mean signal & Eq.~\ref{eq:softmin} \\
$d$ & Remaining reference signal & Eq.~\ref{eq:excessceiling} \\
\bottomrule
\end{tabularx}
\end{table}

\clearpage
\noindent\textbf{Symbol index (continued):}
\begin{table}[H]\centering\small
\setlength{\tabcolsep}{4pt}
\renewcommand{\arraystretch}{1.0}
\begin{tabularx}{\textwidth}{@{}p{4.25cm}Y p{3.05cm}@{}}
\toprule
Symbol & Meaning in this context & Definition \\
\midrule
$d$ & Remaining signal & Eq.~\ref{eq:deficitfloor} \\
$\mathbf D$ & Signal vector & Eq.~\ref{eq:redistribution} \\
$D_1,D_{2,\mathrm{post}},D_{2,\mathrm{auto}}$ & Receptor compartments & Eq.~\ref{eq:deff} \\
$\deff(r,t)$ & Effective dopamine signal & Eq.~\ref{eq:deff} \\
$\delta$ & Peak condition & Eq.~\ref{eq:receptorderivative} \\
$\delta$ & Redistribution size & Eq.~\ref{eq:weightedspread} \\
$\delta$ & Regional displacement & Eq.~\ref{eq:redistribution} \\
$\Delta_j$ & Entry duration & Eq.~\ref{eq:rementryrule} \\
$\mathcal D_H$ & Normalized pattern diversity & Eq.~\ref{eq:signalentropy} \\
$D_\kappa$ & Bottleneck signal & Eq.~\ref{eq:softmin} \\
$\partial D_\kappa/\partial D_r$ & Current regional influence & Eq.~\ref{eq:softweight} \\
$D(L)$ & Effective signal & Eq.~\ref{eq:twopool} \\
$D_{\min}$ & Critical minimum & Eq.~\ref{eq:softmin} \\
$\widetilde d_r$ & Scaled imaging contrast & Eq.~\ref{eq:ordinalresponse} \\
$\dot{\widetilde D}_r$ & Signal change & Eq.~\ref{eq:memory} \\
$D_r^{(A)}$ & Baseline regional signal & Eq.~\ref{eq:ratiotransfer} \\
$D_r-D_s$ & Signal difference & Eq.~\ref{eq:rescueratio} \\
$\widetilde D_r(t)$ & Integrated regional signal & Eq.~\ref{eq:memory} \\
$D_r(t)$ & Regional dopamine signal & Section~\ref{sec:definitions} \\
$D_\star$ & Target coordinate & Eq.~\ref{eq:reachability} \\
$e$ & Relative reserve & Eq.~\ref{eq:reservefeedback} \\
$E_{it}$ & Formulation dose & Eq.~\ref{eq:formulation} \\
$e_{it}$ & Residual & Eq.~\ref{eq:formulation} \\
$E,J$ & Reserve and injury & Eq.~\ref{eq:multistate}, Eq.~\ref{eq:injury}, Eq.~\ref{eq:regionalinjury} \\
$E,J,P,Q_{\mathrm{tot}}$ & Reserve, injury, supply, and demand & Eq.~\ref{eq:extendedinjury} \\
$\ell$ & Lower signal & Eq.~\ref{eq:deficitfloor} \\
$\epsilon_s,\epsilon_r,\epsilon_c$ & Observation errors & Eq.~\ref{eq:measurements} \\
$\mathrm{EQ}$ & Encephalization quotient & Eq.~\ref{eq:anatomicalmetrics} \\
$\eta_{\mathrm{ad}}$ & Protection strength & Eq.~\ref{eq:injuryhistory} \\
$\eta_r$ & Relative transporter clearance & Eq.~\ref{eq:staticdat} \\
$\eta_{s,\mathrm{ON}}^{(0)},\eta_{s,\mathrm{ON}}^{(1)}$ & Predicted latent states & Eq.~\ref{eq:ordinalresponse} \\
$F$ & Capacity mapping & Eq.~\ref{eq:substrate} \\
$\mathcal F$ & Reduced-signal set & Eq.~\ref{eq:deficitfloor} \\
$f_{\mathrm{const}}$ & 5-HT2A constitutive activity & Eq.~\ref{eq:htoccupancy} \\
$\partial F_C/\partial x$ & Dopamine response slope & Eq.~\ref{eq:intensityslope} \\
$F_C(x,h)$ & Cumulative intensity response & Eq.~\ref{eq:multistate} \\
$f_{\mathrm{ex}}$ & Exploratory trait frequency & Eq.~\ref{eq:evolutionfrequency} \\
$f_j(L,t)$ & Pathway activity & Eq.~\ref{eq:pathways} \\
$f_{\mathrm{met}}$ & Brain metabolic fraction & Eq.~\ref{eq:anatomicalmetrics} \\
$g$ & Examination gap & Eq.~\ref{eq:ordinalspecification} \\
$g$ & Generational time & Eq.~\ref{eq:evolutionfrequency} \\
$G_0(x)$ & Low-signal factor & Eq.~\ref{eq:lowgate} \\
$\gamma$ & Dopamine attenuation & Eq.~\ref{eq:htoccupancy} \\
$\gamma$ & Regional response coefficient & Eq.~\ref{eq:ordinalresponse} \\
\bottomrule
\end{tabularx}
\end{table}

\clearpage
\noindent\textbf{Symbol index (continued):}
\begin{table}[H]\centering\small
\setlength{\tabcolsep}{4pt}
\renewcommand{\arraystretch}{1.0}
\begin{tabularx}{\textwidth}{@{}p{4.25cm}Y p{3.05cm}@{}}
\toprule
Symbol & Meaning in this context & Definition \\
\midrule
$G_{\mathrm{glc}}$ & Local glucose & Eq.~\ref{eq:injury} \\
$g,m$ & Observation functions & Eq.~\ref{eq:measurements} \\
$G(\omega)$ & Amplitude gain & Eq.~\ref{eq:gainphase} \\
$G(q)$ & Constitutive gain & Eq.~\ref{eq:remconstitutive} \\
$G_{\mathrm{RPR}}$ & Response gain & Eq.~\ref{eq:rprgain} \\
$\mathcal H$ & Raised-signal set & Eq.~\ref{eq:excessceiling} \\
$h$ & Serotonergic context & Eq.~\ref{eq:multistate} \\
$\mathcal H_{2A}$ & Target-related hazard & Eq.~\ref{eq:hazard} \\
$\widetilde h_{\mathrm{on}},\widetilde h_{\mathrm{off}}$ & Current dose timing & Eq.~\ref{eq:ordinalspecification} \\
$\mathcal H_{\mathrm{other}}$ & Other hazard & Eq.~\ref{eq:hazard} \\
$H_r(t)$ & Regional serotonin signal & Section~\ref{sec:definitions} \\
$H_{\mathrm{sig}}$ & Pattern entropy & Eq.~\ref{eq:signalentropy} \\
$\mathcal H(t)$ & Total hazard & Eq.~\ref{eq:hazard} \\
$\mathcal I_j$ & Entry range & Eq.~\ref{eq:rementryrule} \\
$i,n$ & Recruitment components & Eq.~\ref{eq:multistate} \\
$i,t$ & Observation indices & Eq.~\ref{eq:formulation} \\
$j$ & Pathway index & Eq.~\ref{eq:pathways} \\
$j$ & Receptor index & Eq.~\ref{eq:deff} \\
$K$ & Occupancy constant & Eq.~\ref{eq:devdose} \\
$\kappa$ & Bottleneck strength & Eq.~\ref{eq:softmin} \\
$\kappa$ & Fixed bottleneck strength & Section~\ref{sec:fixedbenchmark} \\
$\kappa_{DH}$ & Serotonin-to-dopamine coupling & Eq.~\ref{eq:coupling} \\
$\kappa_{HD}$ & Dopamine-to-serotonin coupling & Eq.~\ref{eq:coupling} \\
$k_{B,j}$ & Antagonist clearance & Eq.~\ref{eq:rprdynamic} \\
$k_{\mathrm{DAT},r}$ & Transporter clearance & Eq.~\ref{eq:occupancy} \\
$k_J,k_T$ & Injury accumulation & Eq.~\ref{eq:injury} \\
$K_{jr}$ & Occupancy constant & Eq.~\ref{eq:occupancy} \\
$K_N,K_A$ & Occupancy constants & Eq.~\ref{eq:twopool} \\
$k_{\mathrm{other},r}$ & Other clearance & Eq.~\ref{eq:occupancy} \\
$K_{\mathrm{pat}}$ & Pattern alphabet & Eq.~\ref{eq:signalentropy} \\
$k_r$ & Regional gain & Eq.~\ref{eq:deff} \\
$k_{\mathrm{ref},r}$ & Reference clearance & Eq.~\ref{eq:staticdat} \\
$k_W,k_Z$ & Additional injury coefficients & Eq.~\ref{eq:extendedinjury} \\
$L$ & Ligand concentration & Eq.~\ref{eq:twopool} \\
$L_*$ & Peak concentration & Eq.~\ref{eq:receptorpeak} \\
$L^*$ & Required concentration & Eq.~\ref{eq:devdose} \\
$\lambda_c,\lambda_L$ & Signaling gains & Eq.~\ref{eq:htoccupancy} \\
$\lambda_j$ & Pathway weight & Eq.~\ref{eq:pathways} \\
$\lambda_W,\lambda_Z$ & Removal rates & Eq.~\ref{eq:burdenbalance} \\
$L_H,K_H$ & 5-HT2A ligand and occupancy constant & Eq.~\ref{eq:htoccupancy} \\
$L_{it}$ & Total dose & Eq.~\ref{eq:formulation} \\
$L_{\mathrm{other}}$ & Second crossing & Eq.~\ref{eq:receptorsecondcrossing} \\
$\mathcal L_Q$ & Regional load index & Eq.~\ref{eq:regionaldemand} \\
$L_r$ & Ligand concentration & Eq.~\ref{eq:occupancy} \\
\bottomrule
\end{tabularx}
\end{table}

\clearpage
\noindent\textbf{Symbol index (continued):}
\begin{table}[H]\centering\small
\setlength{\tabcolsep}{4pt}
\renewcommand{\arraystretch}{1.0}
\begin{tabularx}{\textwidth}{@{}p{4.25cm}Y p{3.05cm}@{}}
\toprule
Symbol & Meaning in this context & Definition \\
\midrule
$L_{\mathrm{ref}}$ & Reference concentration & Eq.~\ref{eq:receptorbaseline} \\
$M_{\mathrm{brain}},M_{\mathrm{body}}$ & Brain and body masses & Eq.~\ref{eq:anatomicalmetrics} \\
$M_E(e)$ & Reserve demand factor & Eq.~\ref{eq:reservefeedback} \\
$\mathbf m_{\mathrm{met}}$ & Metabolic context & Eq.~\ref{eq:injury} \\
$M(\Phi)$ & Complexity ceiling & Eq.~\ref{eq:substrate} \\
$\widetilde m_r,\widetilde d_r$ & Imaging predictors & Eq.~\ref{eq:ordinalspecification} \\
$m,s$ & Curve location and width & Eq.~\ref{eq:pac} \\
$M(t)$ & Exposure context & Eq.~\ref{eq:hazard} \\
$\mu_i(h)$ & Recruitment location & Eq.~\ref{eq:multistate} \\
$N(L)$ & Activating contribution & Eq.~\ref{eq:twopool} \\
$N_r$ & Postsynaptic contribution & Eq.~\ref{eq:deff} \\
$N_{\mathcal S},V_{\mathcal S}$ & Matched count and volume & Eq.~\ref{eq:anatomicalmetrics} \\
$O_2$ & Local oxygen & Eq.~\ref{eq:injury} \\
$O_H$ & 5-HT2A occupancy & Eq.~\ref{eq:htoccupancy} \\
$O_{jr}$ & Receptor occupancy & Eq.~\ref{eq:deff} \\
$O(\kappa^2)$ & Approximation remainder & Eq.~\ref{eq:moments} \\
$\omega$ & Input frequency & Eq.~\ref{eq:gainphase} \\
$\omega_{\Psi}$ & Observation weight & Eq.~\ref{eq:contentscore} \\
$\omega(r)$ & Regional weight density in the continuous field & Eq.~\ref{eq:softmin} \\
$P$ & Energy supply & Eq.~\ref{eq:injury} \\
$p$ & Sleep-pressure input to the transition model & Eq.~\ref{eq:transition} \\
$p$ & Statistical significance probability & Section~\ref{sec:statisticalmodels} \\
$p_0$ & Coupling amplitude & Eq.~\ref{eq:pac} \\
$\mathrm{PAC}$ & Phase-amplitude coupling & Eq.~\ref{eq:pac} \\
$P_{\mathrm{brain}},P_{\mathrm{body}}$ & Metabolic powers & Eq.~\ref{eq:anatomicalmetrics} \\
$p_{\mathcal F}$ & Affected weight & Eq.~\ref{eq:deficitfloor} \\
$p_{\mathcal H}$ & Raised weight & Eq.~\ref{eq:excessceiling} \\
$\Phi$ & Structural capacity & Eq.~\ref{eq:ceiling} \\
$\pi_k$ & Pattern probability & Eq.~\ref{eq:signalentropy} \\
$P,k_J,k_T,T_c$ & Supply and injury terms & Eq.~\ref{eq:regionalinjury} \\
$p_{\mathrm{motor}}$ & Motor cortical weight & Section~\ref{sec:fixedbenchmark} \\
$p_{\mathrm{PFC}}$ & Prefrontal weight & Section~\ref{sec:fixedbenchmark} \\
$p_r$ & Prior regional weight & Eq.~\ref{eq:softmin} \\
$p_r/p_s$ & Prior-weight ratio & Eq.~\ref{eq:rescueratio} \\
$\Psi$ & Experiential complexity & Section~\ref{sec:definitions} \\
$\Psi_0$ & Reference complexity & Eq.~\ref{eq:htoccupancy} \\
$\Psi_{\mathrm{eff}}$ & Effective complexity & Eq.~\ref{eq:ceiling} \\
$\Psi_{\max}(\Phi)$ & Complexity ceiling & Eq.~\ref{eq:ceiling} \\
$\Psi_{\mathrm{REM}}$ & REM complexity drive & Eq.~\ref{eq:remconstitutive} \\
$p_{\mathrm{SN}}$ & Nigral weight & Section~\ref{sec:fixedbenchmark} \\
$p_{\mathrm{striatum}}$ & Striatal weight & Section~\ref{sec:fixedbenchmark} \\
$p_{\mathrm{VTA}}$ & VTA weight & Section~\ref{sec:fixedbenchmark} \\
$q$ & State-control process & Eq.~\ref{eq:remconstitutive} \\
$Q_b(h,E)$ & Background demand & Eq.~\ref{eq:regionaldemand} \\
\bottomrule
\end{tabularx}
\end{table}

\clearpage
\noindent\textbf{Symbol index (continued):}
\begin{table}[H]\centering\small
\setlength{\tabcolsep}{4pt}
\renewcommand{\arraystretch}{1.0}
\begin{tabularx}{\textwidth}{@{}p{4.25cm}Y p{3.05cm}@{}}
\toprule
Symbol & Meaning in this context & Definition \\
\midrule
$Q_{\mathrm{ref}}(\mathbf D,h)$ & Reference-reserve demand & Eq.~\ref{eq:reservefeedback} \\
$Q_{\mathrm{reg}}$ & Total regional demand & Eq.~\ref{eq:regionaldemand} \\
$q_r(t)$ & Release rate & Eq.~\ref{eq:occupancy} \\
$Q_{\mathrm{tot}}$ & Total required demand & Eq.~\ref{eq:injuryhistory} \\
$Q(x,h,E)$ & Energy demand & Eq.~\ref{eq:injury} \\
$r$ & Regional index & Section~\ref{sec:definitions} \\
$\mathcal R_{\mathrm{crit}}$ & Critical set & Eq.~\ref{eq:softmin} \\
$r_E(E)$ & Recovery capacity & Eq.~\ref{eq:injuryhistory} \\
$\rho$ & Relative density & Eq.~\ref{eq:devdose} \\
$\rho_{\mathrm{demo}}(a)$ & Density trajectory & Eq.~\ref{eq:development} \\
$\rho_j$ & Relative receptor density & Eq.~\ref{eq:rprdynamic} \\
$\rho_J$ & Repair rate & Eq.~\ref{eq:injuryhistory} \\
$\rho_{j,\mathrm{eq}}(B_j)$ & Adapted equilibrium & Eq.~\ref{eq:rprdynamic} \\
$\rho_{jr}$ & Relative receptor density & Eq.~\ref{eq:deff} \\
$\rho_N$ & Neuronal density & Eq.~\ref{eq:anatomicalmetrics} \\
$\mathcal R,\mathcal I,\mathcal D$ & Substrate measures & Eq.~\ref{eq:substrate} \\
$r,s$ & Demand comparison pair & Eq.~\ref{eq:demandrestrictions} \\
$r,s$ & Perturbed regions & Eq.~\ref{eq:weightedspread} \\
$r,s$ & Region pair & Eq.~\ref{eq:rescueratio} \\
$R(t)$ & 5-HT2A temporal response & Eq.~\ref{eq:htoccupancy} \\
$s$ & Body-side index & Eq.~\ref{eq:ordinalresponse}, Eq.~\ref{eq:ordinalspecification} \\
$s$ & PAC curve width & Eq.~\ref{eq:pac} \\
$s$ & Second region in a local sensitivity comparison & Eq.~\ref{eq:rescueratio} \\
$s$ & Summation index & Eq.~\ref{eq:softweight} \\
$S^*$ & Target response & Eq.~\ref{eq:devdose} \\
$s_c$ & Collapse width & Eq.~\ref{eq:collapse} \\
$s_i$ & Recruitment fraction & Eq.~\ref{eq:intensityslope} \\
$\sigma$ & Normalized supply deficit & Eq.~\ref{eq:injuryhistory} \\
$\sigma_b,\sigma_a,\sigma_{\mathrm{prog}},\sigma_u$ & Random-effect scales & Eq.~\ref{eq:ordinalspecification} \\
$\sig(z)$ & Activation function & Eq.~\ref{eq:multistate} \\
$S_{jr}$ & Signaling efficacy & Eq.~\ref{eq:deff} \\
$t^*$ & Response maximum & Eq.~\ref{eq:onset} \\
$t$ & Time & Section~\ref{sec:definitions} \\
$T$ & Tissue temperature & Eq.~\ref{eq:injury} \\
$t_0$ & NREM starting time & Eq.~\ref{eq:rementryrule} \\
$\tau_{\mathrm{ad}}$ & Adaptation time & Eq.~\ref{eq:injuryhistory} \\
$\tau_{\mathrm{des}}$ & Decay time & Eq.~\ref{eq:onset} \\
$\tau_i$ & Recruitment scale & Eq.~\ref{eq:multistate} \\
$\tau_m$ & Memory time & Eq.~\ref{eq:memory}, Eq.~\ref{eq:gainphase} \\
$\tau_{\mathrm{on}}$ & Onset time & Eq.~\ref{eq:onset} \\
$\tau_{\rho,j}$ & Adaptation time & Eq.~\ref{eq:rprdynamic} \\
$\tau_z$ & Transition time & Eq.~\ref{eq:transition} \\
$T_c$ & Temperature threshold & Eq.~\ref{eq:injury} \\
$\theta_r$ & Postsynaptic D2 circuit gain & Eq.~\ref{eq:deff} \\
\bottomrule
\end{tabularx}
\end{table}

\clearpage
\noindent\textbf{Symbol index (continued):}
\begin{table}[H]\centering\small
\setlength{\tabcolsep}{4pt}
\renewcommand{\arraystretch}{1.0}
\begin{tabularx}{\textwidth}{@{}p{4.25cm}Y p{3.05cm}@{}}
\toprule
Symbol & Meaning in this context & Definition \\
\midrule
$T_j(t)$ & Residence time & Eq.~\ref{eq:rementryrule} \\
$t_{\mathrm{REM}},t_{\mathrm W}$ & Candidate entry times & Eq.~\ref{eq:rementryrule} \\
$U$ & Common drug response & Eq.~\ref{eq:ordinalresponse} \\
$u$ & Input age & Eq.~\ref{eq:memory} \\
$u_A$ & Feedback change & Eq.~\ref{eq:rprgain} \\
$u_{B,j}(t)$ & Input rate & Eq.~\ref{eq:rprdynamic} \\
$u_D,u_H$ & External inputs & Eq.~\ref{eq:coupling} \\
$u_N$ & Postsynaptic change & Eq.~\ref{eq:rprgain} \\
$U,\tau_0,\tau_m,\tau_b,\tau_{\mathrm{on}},\tau_{\mathrm{off}}$ & Common response & Eq.~\ref{eq:ordinalspecification} \\
$u_W,u_Z$ & Formation rates & Eq.~\ref{eq:burdenbalance} \\
$u(x,h,p)$ & Transition input & Eq.~\ref{eq:transition} \\
$V_0$ & Viability gate & Eq.~\ref{eq:collapse} \\
$\operatorname{Var}_p(D)$ & Regional dispersion & Eq.~\ref{eq:moments} \\
$\operatorname{Var}_p(\mathbf D)$ & Weighted signal variance & Eq.~\ref{eq:demandrestrictions} \\
$\varphi(\omega)$ & Phase lag & Eq.~\ref{eq:gainphase} \\
$v_E(E),v_J(J)$ & Reserve and injury factors & Eq.~\ref{eq:injury} \\
$V(x,E,J)$ & Viability factor & Eq.~\ref{eq:multistate} \\
$W$ & Selected solute & Eq.~\ref{eq:burdenbalance} \\
$\mathrm{W0}$ & Fixed theoretical benchmark & Section~\ref{sec:fixedbenchmark} \\
$W_c,Z_c$ & Burden boundaries & Eq.~\ref{eq:extendedinjury} \\
$w_{jr},v_{jr}$ & Receptor weights & Eq.~\ref{eq:deff} \\
$w_r^Q$ & Demand weight & Eq.~\ref{eq:regionaldemand} \\
$x$ & Coordinated input & Eq.~\ref{eq:rementryrule} \\
$x$ & Normalized dopamine input & Eq.~\ref{eq:multistate}, Eq.~\ref{eq:pac} \\
$\mathrm{X0}$ & Regional demand candidate with fixed weights and exponent & Section~\ref{sec:regionalexcess} \\
$x_0,h_0$ & Reference coordinates & Eq.~\ref{eq:coupling} \\
$\mathrm{X1}$ & Declared reserve-feedback candidate & Section~\ref{sec:reservefeedback} \\
$x_c$ & Collapse location & Eq.~\ref{eq:anchors}, Eq.~\ref{eq:collapse} \\
$x_\ell$ & Low-signal scale & Eq.~\ref{eq:lowgate} \\
$\dot x,\dot h$ & Coordinate rates & Eq.~\ref{eq:coupling} \\
$x_{\mathrm{PAC}}^*$ & Peak input & Eq.~\ref{eq:pacpeak} \\
$x_{\mathrm{REM}},x_{\mathrm W}$ & Sleep-transition boundaries & Eq.~\ref{eq:rementryrule} \\
$x_s$ & Attenuation scale & Eq.~\ref{eq:htoccupancy} \\
$Y_{\mathrm{cx}},Y_{\mathrm{sig}}$ & Content ratings & Eq.~\ref{eq:contentscore} \\
$Y_{it}$ & Sleep outcome & Eq.~\ref{eq:formulation} \\
$Y_{\Psi}$ & Content score & Eq.~\ref{eq:contentscore} \\
$Y_{\mathrm{sleep}},Y_{\mathrm{report}},Y_{\mathrm{CSF}}$ & Observed outcomes & Eq.~\ref{eq:measurements} \\
$[z]_+$ & Positive part & Eq.~\ref{eq:injury} \\
$Z$ & Selected oxidant & Eq.~\ref{eq:burdenbalance} \\
$z$ & Transition state & Eq.~\ref{eq:transition}, Eq.~\ref{eq:foldcondition} \\
$Z_{it},\boldsymbol\gamma$ & Covariates & Eq.~\ref{eq:formulation} \\
$z_{\mathrm{turnover}}$ & Transmitter turnover & Eq.~\ref{eq:measurements} \\
\bottomrule
\end{tabularx}
\end{table}

\clearpage
\section{Results and interpretation}
\begin{table}[H]
\centering\small
\caption{Empirical results and their meaning. All analyses are exploratory. Intervals are 95\% intervals with the limitations stated in the methods.}
\label{tab:testinterpretation}
\begin{tabularx}{\textwidth}{@{}p{3.0cm}p{5.0cm}Y@{}}
\toprule
Question & Result & Scientific meaning \\
\midrule
Sustained REM entry & Waiting contrast: $+1.220$ min ($0.425$ to $2.014$), six-test Holm $p=0.0546$. With trends: $+0.905$ ($-0.037$ to $1.847$). & Compatible with delayed entry. The coefficient stays positive with trends, with an interval that spans zero. \\
Participant-deletion robustness & With trends, removing the largest contributor gives $+0.482$ min ($-0.308$ to $1.272$). Removing the largest two or three gives $+0.728$ or $+0.878$ min. & Coefficients remain positive, but joint-deletion intervals include zero. Exposure information is concentrated, and influence checks do not remove confounding. \\
Persistence and stage specificity & Persistence: $-0.185$ min ($-0.471$ to $0.101$). Direct REM-minus-deep waiting: $+1.077$ ($-0.308$ to $2.462$). & Intervals span zero for persistence and stage differences. The direct stage contrast is secondary. \\
Conventional REM timing & Phase: $+0.402$ pp. Latency: $+2.10$ min. Trend intervals include zero. Minimum eight-test Holm $p=0.130$. & The timing pattern is consistent across outcomes. It does not identify the global dopamine crossing, and the observations overlap. \\
Regional medication response & Conditional ordinal log-score gain: $-7.38\times10^{-5}$ (95\% interval $[-3.72,3.68]\times10^{-3}$), in 254 participants. & The added regional term has uncertain predictive value after modeling baseline error. The comparison concerns a clinical response proxy. \\
\bottomrule
\end{tabularx}
\par\smallskip
\parbox{\textwidth}{\footnotesize Formulation effects are per 100 mg/day LEDD substitution at fixed total levodopa dose. Positive regional gains favor the added regional term. Intervals for prediction gains use saved participant losses and omit full refitting uncertainty.}
\end{table}

\begin{table}[H]
\centering\small
\caption{Mathematical status and untested theoretical predictions. Internal consistency and observational evidence answer different questions.}
\label{tab:theorystatus}
\begin{tabularx}{\textwidth}{@{}p{3.0cm}p{4.9cm}Y@{}}
\toprule
Component & Result & Scientific meaning \\
\midrule
Mathematical consistency & Weighted redistribution, sensitivity transfer, gain-phase relations, response maxima, and conditional receptor gain pass numerical checks. & The checked equations are internally consistent. Shared parameters provide stronger experimental constraints than isolated dose associations. \\
W0 and state mapping & Isolated prefrontal floor: $0.245$. The common normalized response gives $C_{\mathrm{ex}}\approx0.072$ under distributed suppression, versus $1.000$ at reference. & Regional inputs and state parameters jointly determine the outcome. The example permits preserved or impaired states under a specified observation boundary. \\
Sleep transitions and content & Dopamine magnitude and residence time select REM or waking entry. Effective \HT{} drive sets the proposed felt complexity and significance. & Tests entry type and timing separately from content. Requires regional signaling, validated sleep stages, and experience measurements. \\
Evolutionary persistence & Frequency-dependent selection permits a stable nonzero exploratory trait frequency. & Predicts frequency-dependent fitness returns and an environment-sensitive clinical tail, rather than a universal prevalence. \\
Comparative capacity & Density, body scaling, metabolic allocation, and signal entropy have separate operational definitions. & Differentiated recurrent responses must predict a complexity ceiling across species and receptor perturbations. Random diversity alone is insufficient. \\
Untested predictions & W0, X0/X1, sensitivity transfer, injury history, state boundaries, regional rescue, REM signaling, comparative predictions, safety, and RPR await tests. & These are explicit hypotheses with proposed failure criteria. The available PPMI observations do not validate them. \\
Waste and oxidative injury & Explicit formation and removal balances add optional injury routes. & No PPMI evaluation. Independent burden and removal measurements must distinguish these routes from energy deficit and heat. \\
\bottomrule
\end{tabularx}
\end{table}

\end{document}